%% file: ijcai-arxiv.tex
\documentclass{article}
\usepackage{ijcai25}

\usepackage{times}
\usepackage{soul}
\usepackage{url}
\usepackage[hidelinks]{hyperref}
\usepackage[utf8]{inputenc}
\usepackage[small]{caption}
\usepackage{graphicx}
\usepackage{amsmath}
\usepackage{amsthm}
\usepackage{booktabs}
\usepackage{algorithm}
\usepackage{algorithmic}
\usepackage[switch]{lineno}
\usepackage{multirow}

\newtheorem{example}{Example}
\newtheorem{theorem}{Theorem}
\newtheorem{definition}{Definition}

\usepackage{amssymb}
\usepackage{listings}
\usepackage{tcolorbox}
\usepackage{hyperref}
\usepackage{bookmark}
\usepackage{chapterbib}  % 让每个 \include 文件都有自己的 bib

\tcbset{
  mycodebox/.style={
    colback=gray!5,
    colframe=blue!75!black,
    sharp corners,
    boxrule=1pt,
  }
}

\title{Automated Strategy Invention for Confluence of Term Rewrite Systems}

\author{
Liao Zhang$^{1,2}$
\and
Fabian Mitterwallner$^1$ 
\and
Jan Jakubův$^3$ \And
Cezary Kaliszyk$^{4,1}$\\
\affiliations
$^1$University of Innsbruck\\
$^2$Shanghai Jiao Tong University\\
$^3$Czech Technical University in Prague\\
$^4$University of Melbourne\\
\emails
zhangliao714@gmail.com,
fabian.mitterwallner@uibk.ac.at,
\{jakubuv, cezarykaliszyk\}@gmail.com
}
\begin{document}

\maketitle

\begin{abstract}

Term rewriting plays a crucial role in software verification and compiler optimization. 
With dozens of highly parameterizable techniques developed to prove various system properties,
automatic term rewriting tools work 
in an extensive parameter space. This complexity exceeds human capacity
for parameter selection, motivating an investigation into automated strategy invention.
In this paper, we focus on confluence
of term rewrite systems, and apply AI techniques to invent strategies for automatic confluence proving.
Moreover, we randomly generate a large dataset to analyze confluence for term rewrite systems. 
We improve the state-of-the-art automatic confluence prover CSI: When equipped with our invented strategies,
it surpasses its human-designed strategies both on the augmented dataset
and on the original human-created benchmark dataset ARI-COPS, proving/disproving
the confluence of several term rewrite systems for which no automated proofs
were known before.

\end{abstract}

\section{Introduction}
\label{sec:intro}
Term rewriting studies 
%the question of
substituting subterms of a formula with other terms~\cite{baader1998term}, playing an important role in automated reasoning~\cite{bachmair1994rewrite}, software verification~\cite{meseguer2003software}, and compiler optimization~\cite{willsey2021egg}.
Mathematicians have developed various techniques to analyze the properties of term rewrite systems (TRSs).
However, many properties are undecidable~\cite{baader1998term}, implying that no technique can consistently prove a particular property.
To navigate this undecidability, modern term rewriting provers typically employ complicated strategies, incorporating wide arrays of rewriting analysis techniques, with the hope that one will be effective. 
Each technique often accompanies several flags to control its behavior. The diversity of techniques and their controlling flags result in a vast parameter space for modern automated term rewriting provers.

Manually optimizing strategies for undecidable problems is beyond human capacity given the extensive parameter space. This inspires us to apply AI techniques to search for appropriate strategies automatically.
In this paper, we focus on confluence, an important property of term rewriting, and discuss automated strategy invention for the state-of-the-art confluence prover CSI~\cite{nagele2017csi}. We modify Grackle~\cite{huula2022targeted}, an automatic tool to generate a strategy portfolio, %for a solver, 
encoding strategies that require transformations and complex schedules such as parallelism.

Directly using a tool like Grackle to randomly generate parameters for CSI may produce unsound results. 
This is a unique challenge compared to previous applications of Grackle~\cite{huula2022targeted,aleksandrova2024prover9}. 
The solvers to which Grackle was previously applied always produce sound results, while CSI's users need to carefully specify their strategies to ensure soundness.

We also augment the human-built confluence problems database (ARI-COPS)\footnote{https://ari-cops.uibk.ac.at/}, a representative benchmark for the annual confluence competition (CoCo)\footnote{https://project-coco.uibk.ac.at/}.
Before 2024, CoCo used the COPS database as the benchmark. 
An unpublished duplicate checker is executed to remove duplicated problems in COPS, resulting in the ARI-COPS database, which is used in CoCo 2024.
As ARI-COPS has been created manually, it includes only 566 TRSs. They are of high quality, but the relatively small number is still inadequate for data-driven AI techniques that require large amounts of training data.
To handle this problem, we generate a large number of TRSs randomly, but ensure that they are interesting enough to analyze. For this, we develop a procedure to confirm a relative balance in the number of TRSs most quickly solved by different confluence analysis techniques within the dataset.

We evaluate our strategy invention approach in ARI-COPS and the augmented dataset. On both of the datasets, the invented strategies surpass CSI's competition strategy.  
In particular, we prove (non-)confluence for several TRSs that have not been proved by any automatic confluence provers in the history of the CoCo competition.

As an example, our invented strategy is able to disprove confluence for the ARI-COPS problem \texttt{846.ari} (\texttt{991.trs} in COPS), never proved by any participant in CoCo.
The key is the application of the redundant rule technique~\cite{nagele2015improving} with non-standard arguments.
CSI's competition strategy performs \texttt{redundant -narrowfwd -narrowbwd -size 7} prior to performing non-confluence analysis. The flags \texttt{narrowfwd} and \texttt{narrowbwd} determine the categories of redundant rules to generate. 
Our tool automatically discovered that by changing the original redundant rule transformation to \texttt{redundant -development 6 -size 7}, we can prove this problem.
A larger value for the flag \texttt{development} causes a larger number of development redundant rules to be added. 
We notice that the value six is crucial as small values below three are ineffective for \texttt{846.ari}. 
% Although, with values above three, CSI can discover proof for \texttt{991.trs}, the time consumption would be significantly larger. \textbf{???}
This is only one of the several TRSs which our new strategies can solve as discussed in the later sections.

The main reason why it is difficult to discover new proofs in CoCo, is because CSI's competition strategy developed rewriting experts is very complicated, for which a comprehensive explanation is presented in the technical appendix.
For example the competition strategy includes the development redundant rule technique~\cite{nagele2015improving}.
The original evaluation of it shows no improvement over other redundant rule techniques in COPS at that time. 
Thus, CSI's developers decided not to use it in the competition strategy. 
As COPS grows, it becomes helpful in some new TRSs such as \texttt{846.ari}.
However, the default strategy has only slightly changed over the past years, and the development redundant rule technique has never been tried. 
One reason for this could be that choosing sound parameters is challenging even for rewriting experts. 
Meanwhile, competition strategy is highly complicated and has a prohibitively large configuration space both in the number of parameters and structures of the strategy itself.
We leverage Grackle to do the tedious strategy search. It can automatically optimize the strategies better than experts as the dataset grows.
Other rewriting tools do not discover the proof perhaps because they do not implement the essential techniques for solving the problems.

\paragraph{Contributions.}
    First, to our best knowledge, our work is the first application of AI techniques to automatic confluence provers. 
    We automatically generate a lot of strategies for the state-of-the-art confluence prover CSI and combine them as a unified strategy.
    Second, we carefully design the parameter search space for CSI to confirm the soundness of strategy invention. 
    Third, we build a large dataset for confluence analysis, comprising randomly generated TRSs and problems in the ARI-COPS dataset.
    Finally, empirical results show that our strategy invention approach surpasses CSI's competition strategy both in ARI-COPS and the augmented datasets. Notably, we discover several proofs for (non-)confluence that have never been discovered by any automatic confluence provers in the annual confluence competition.

\section{Background}
\subsection{Term Rewriting}
\label{sec:rewriting}
We informally define some theoretical properties of term rewriting in this section, hoping to ease the understanding of the behavior underlining automatic confluence provers. A formal description can be found in the technical appendix.

We assume a disjoint set of \textit{variable} symbols and a finite signature of \textit{function} symbols.
\textit{Constants} are function symbols with zero arity.
% $\mathcal{F}$. 
% consisting of at least one constant symbol. % FM we usually don't restrict signatures like this. For confluence this would be OK if necessary, since adding constants does not change the confluence of a TRS.
The set of \textit{terms} is built up from variables and function symbols.
% $\mathcal{V}$ and $\mathcal{F}$. 
The set of variables occurring in a term $t$ is denoted by $\mathit{Var}(t)$.
A term rewrite system (TRS) consists of a set of rewrite rules $l \rightarrow r$ where  $l, r \in \mathit{terms}$, $l \notin \mathit{variables}$, and $\mathit{Var}(r) \subseteq \mathit{Var}(l)$.
We write $t_1 \rightarrow^* t_n$ to denote $t_1 \rightarrow t_2 \rightarrow ... \rightarrow t_n$ where $n$ can be one. 
% Joinability is denoted by $t \downarrow u$, meaning that $\exists v, t \rightarrow^* v \land u \rightarrow^* v$.
A TRS is \textit{confluent} if and only if $\forall s, t, u \in \mathit{terms} (s\rightarrow^* t \land s\rightarrow^* u \Rightarrow \exists v \in \mathit{terms}(t \rightarrow^* v \land u \rightarrow^* v))$. 
Consider the TRS of $\{f(g(x),h(x)) \rightarrow a, g(b) \rightarrow d, h(c) \rightarrow d\}$~\cite{gramlich1996confluence}.
It is not confluent since $f(d,h(b)) \leftarrow f(g(b),h(b)) \rightarrow a$, and no rules are applicable to $f(d,h(b))$ and $a$. 
% A term is called \textit{linear} if no variable multiply occurs in it.
A rewrite rule $l \rightarrow r$ is called \textit{left-linear} if no variable occurs multiple times in $l$.
% $l$ is linear.
A TRS is called left-linear if all its rules are left-linear.
Left-linearity is crucial for confluence analysis since most existing confluence techniques only apply to such systems.
%decision procedures depend on it to determine the confluence of TRSs.
In this paper, a term is called \textit{complex} if it is neither a variable nor a constant.

\subsection{CSI}
\label{sec:csi}
CSI is one of the state-of-the-art automatic confluence provers that participates in CoCo.
It ranked first in five categories of competitions in CoCo 2024.
To show (non-)confluence of TRSs, CSI automatically executes a range of techniques, %decision procedures,
scheduled by a complicated configuration document written by experts in confluence analysis.
Subsequently, CSI either outputs \texttt{YES}, \texttt{NO}, or \texttt{MAYBE} indicating confluence, non-confluence, or indetermination, respectively.

CSI implements many techniques applicable to the analysis of TRSs (many of them parametrized or transforming the system into one that can be analyzed by other techniques) and utilizes a complicated strategy language to control them. In CSI, these techniques are called \textit{processors}. They are designed to prove the properties of TRSs, perform various transformations, and check the satisfiability of certain conditions. 
The strategy language can flexibly combine the execution of processors such as specifying parallel or sequential applications, disregarding unexpected results, assigning time limits, and designating repeated applications.
The details of the strategy language are presented in the technical appendix.
% supplementary material.

Since the generated proofs are almost always large and difficult to check manually, CSI relies on an external certifier CeTA~\cite{thiemann2009certification} to verify its 
proofs. To utilize CeTA, CSI outputs a certificate of its proof in the certification problem format~\cite{sternagel2014certification}. 
Given a certificate, CeTA will either answer \texttt{CERTIFIED} or present a reason to reject it.
Not all processors implemented in CSI are verifiable because CSI cannot produce certificates for all processors, and CeTA does not implement the verification procedures for all processors.

\subsection{Grackle}

\begin{algorithm}[tb]
\caption{GrackleLoop: an outline of the strategy portfolio invention loop.}
\label{alg:grackle}
\textbf{Input}: initial strategies $\mathcal{S}$, benchmark problems $\mathcal{P}$,   \\
\phantom{\textbf{Input}:} hyperparameters $\beta$ \\
% \phantom{\textbf{Input}:} $\mathit{left}$, whether the term is on the rewrite rule's left side \\
% \phantom{\textbf{Input}:} $linear$, whether to construct a linear term \\
\textbf{Output}: a strategy portfolio $\Phi$
% encapsulating current state. It includes all strategies ($\Phi_{start}$) invented so far.   
\begin{algorithmic}[1]
\STATE $\Phi_{strat} \gets \mathcal{S}$    
\WHILE{termination criteria is not satisfied}
\STATE Evaluate($\mathcal{P}, \Phi, \beta$)   
\STATE $\Phi_{cur} \gets \text{Reduce}(\mathcal{P}, \Phi, \beta)$
\STATE $s \gets \text{Select}(\mathcal{P}, \Phi, \beta)$
% \STATE if $s$ is None then return $\Phi$
% \IF {$s$ is None} return $\Phi$
% \IF {$s$ is None}  \STATE return $\Phi$ \ENDIF
\STATE \textbf{if} $s$ is None \textbf{then return} $\Phi$
\STATE $s_0 \gets \text{Specialize}(s, \mathcal{P}, \Phi, \beta)$
\STATE $\Phi_{strat} \gets \Phi_{strat} \cup s_0$
\ENDWHILE
\end{algorithmic}
\end{algorithm}

Grackle~\cite{huula2022targeted}
%A comprehensive explanation of Grackle is presented in its original paper
is a strategy optimization system designed to automate the generation of various effective strategies for a given solver based on benchmark problems. 
Such solvers receive a problem and decide the satisfiability of a particular property of the problem.
It was originally designed for automated reasoning tools and has been applied to various provers such as Prover9~\cite{prover9-mace4} and Lash~\cite{brown2022lash}.
% problems~\cite{de2011satisfiability} and for proving satisfiability modulo theories problems~\cite{de2011satisfiability}.
We choose Grackle for our research, as it is highly adaptable and we are not aware of any strategy invention program that would allow the kinds of strategies needed for automatic rewriting tools. Additionally, Grackle has achieved good results with the solvers it was previously applied to. 
% \textbf{???? remove applied to sat, no tool for rewriting}
The strategy invention problem of Grackle is formally defined below.

\begin{definition}[Strategy Invention Problem]
Assume a set of initial strategies $\mathcal{S}$. 
% Each strategy $s \in S$ is employed with a set of parameters $\theta$. 
In the benchmark of examples $\mathcal{P}$, the problem is to invent a bounded set of complementary strategies $\mathcal{S'}$ that can prove the largest number of problems in $\mathcal{P}$. 
% (likely larger than $S$) 
% via a genetic algorithm and parameter tuning. 
% Each strategy $s' \in S'$ is either the same as an initial strategy $s$ or derived from $s$ with different parameters $\theta'$, which is heuristically discovered by an optimization algorithm with randomness. 
% The invented strategies are expected to be complementary in solving different problems. This means that for each $s_i' \in S'$, $s$ should be most efficient on solving a subset of problems $E_i' \subseteq E$, such that  $\forall i \neq j$, $s_j' \in S'$ cannot solve any problem in $E_i'$ quicker than $s_i'$.
Complementary strategies means that $\forall s_i' \in \mathcal{S'}$, $s_i'$ should \textit{master} a subset of problems $\mathcal{P}_i' \subseteq \mathcal{P}$, such that  $\forall i \neq j$, $s_j' \in \mathcal{S'}$ cannot solve any problem in $ \mathcal{P}_i'$ quicker than $s_i'$.  
\end{definition}
% Given a set of problems as input, Grackle automatically invents a large number of strategies and selects those most complementary in problem solving. Here, complementation first means that each selected strategy must be most efficient for solving a number of unique problems. Moreover, it indicates that within a limit of the number of selected strategies, the combination of all selected strategies can maximize the number of provable problems.

Algorithm~\ref{alg:grackle} outlines the strategy portfolio invention loop of Grackle, which
invents strategies via a genetic algorithm and parameter tuning with randomness. 
% Grackle is based on a genetic algorithm to invent strategies.
% It repeatedly executes a dual-phase process to generate complementary strategies. 
The variable $\Phi$ denotes the current state, including information like all invented strategies $\Phi_{strat}$, and the current generation of strategies $\Phi_{cur}$. 
The first phase is \textit{generation evaluation (evaluate)}. 
% followed by \textit{strategy invention}. 
% Grackle users need to provide a number of initial strategies before the execution.
In this phase, Grackle evaluates all strategies $ \Phi_{strat} $ in its portfolio on the benchmark $\mathcal{P}$. 
The evaluation results are stored in $\Phi$ to avoid duplicated execution.
% In the first loop iteration, $ \Phi_{strat} $ only comprises the initial strategies $\mathcal{S}$.
% After the evaluation,

Next, Grackle performs \textit{generation reduction (reduce)}.
It assigns scores to every strategy in $ \Phi_{strat} $ based on the evaluation results in the previous phase. 
A configurable number of strategies with the highest scores becomes the current generation of strategies $\Phi_{cur}$.
% uses a scoring function to discover the strategies that are most \textit{complementary} in solving problems.
% This means that for each $s_i' \in S'$, $s$ should be most efficient on solving a subset of problems $E_i' \subseteq E$, such that  $\forall i \neq j$, $s_j' \in S'$ cannot solve any problem in $E_i'$ quicker than $s_i'$
% Only the strategies with the best evaluation performance on $E$ are kept in the portfolio.

The third phase is \textit{strategy selection (select)}. It selects a strategy $s$ from the current generation of strategies $\Phi_{cur}$ based on certain criteria, 
which is then used to invent new strategies. If no strategy can be selected, the algorithm terminates.

Finally, \textit{strategy specialization (specialize)} invents a new strategy $s_0$ via specializing
% based on the selected strategy $s$.
$s$ over its best-performing problems $\mathcal{P}_{s}$ in $\mathcal{P}$. 
Grackle then executes external parameter tuning programs such as ParamILS~\cite{hutter2009paramils} or SMAC3~\cite{lindauer2022smac3}, tuning parameters for the selected strategy $s$ with randomness.
The goal is to invent a new strategy $s_0$ such that it performs better than $s$ ion $\mathcal{P}_{s}$.
The new strategy $s_0$  will be added to the portfolio $\Phi$.
% New strategies are provided by the invention phase.
% and the strategies in the portfolio are updated based on the evaluation results. 
% This evaluation clusters the set of benchmark problems into subsets $P_S$ for each strategy $s$, where $P_s$ contains the problems where the strategy $S$ performs best.
% Only the strategies mastering at least a given number of unique problems remain in the portfolio. 
% In the invention phase, Grackle invents new strategies using the best-performing strategy $S$ and its corresponding benchmark problems $P_S$. New strategies are invented and tested on $P_S$ via external parameter tuning programs ParamILS~\cite{hutter2009paramils} or SMAC3~\cite{lindauer2022smac3}. 
% Then, Grackle integrates the newly invented strategies into the portfolio and repeats the evaluation phase.

Grackle employs the same approach to describe its parameter search space as ParamILS.
The space is described by a set of available parameters, each of which is associated with a default value and several disjoint potential values.
Grackle users need to input the potential values based on their domain-specific experiences on the particular solvers.
We refer to~\cite{huula2022targeted} for a comprehensive explanation of Grackle.
\section{Strategy Invention and Combination}
To generate a better strategy for CSI, we first 
%apply Grackle to
invent a large set of complementary strategies, and then appropriately combine a subset of the invented strategies into a single strategy.

\subsection{Strategy Invention\label{sec:strat_invent}}

To find new strategies for CSI, we first need to represent the parameter space in a meaningful way. The parameter space needs to be designed with precision to guarantee soundness.
% Directly using a tool like Grackle to randomly combine parameters may produce unsound results. 
% This is a unique challenge compared to previous applications of strategy invention. 
% The solvers to which Grackle was previously applied always produce sound results, while CSI's users need to carefully specify their strategies to ensure soundness. 

% \textbf{the details will be explained later, change order, transform}

There are three reasons why CSI may produce unsound results given an entirely random strategy. 
First, some processors are not intended for confluence analysis.
They may intend to prove other properties of TRSs, such as termination~\cite{baader1998term}.  
% CSI can also contain processors which should be applied to termination analysis or complexity analysis
Second, even for the same processor, it may be designed to prove different properties of TRSs with different flags.
% some arguments of processors are not intended to prove confluence problems.
% They may be used for termination analysis or complexity analysis.
Third, some transformation processors may change the goal of CSI to prove another property of TRSs, which is different from confluence such as relative termination~\cite{zantema2004relative}. 
% transform the confluent problem into 
% proving a differ
% an unexpected problem.
% For example, they can transform it to a relative termination problem~\cite{zantema2004relative}.
% for which processors for conflucene produce wrong results.
% For a relative termination problem, the processors for confluence may produce entrily misl

% \textcolor{blue}{DELETE??
% For CoCo, CSI utilizes a complicated configuration document written by term rewriting experts. 
% A comprehensive explanation of the competition strategy is presented in the technical appendix.}

We separate CSI's competition strategy into 23 sub-strategies, which, along with CSI's competition strategy, also serve as the initial strategies for Grackle.
Among the 23 sub-strategies, nine are mainly used to show confluence, and 14 are used to show non-confluence.
A comprehensive explanation of the division is shown in the technical appendix. 
% and one is capable of showing both.

We maintain the structure used in CSI's competition strategy during the strategy invention because CSI relies on certain combinations of processors to (dis)prove confluence.
There are papers proving theorems for confluence analysis, stating that if some properties of a TRS can be proved, then it is (non-)confluent.
Such a theorem can be implemented as a single processor, which checks whether the given TRS satisfies the properties required by the theorem. 
However, not all such theorems are implemented as a processor. 
To utilize such theorems, we need to combine CSI's strategy language and processors to perform transformations on the original TRS and prove the necessary properties of the transformed problem.
If we generate strategies randomly, it will be difficult to generate such useful structures and may produce unsound strategies due to inappropriate transformations.

We search for three categories of parameters. 
First, we search for \textit{processor flags} which do not violate the soundness guarantee. 
For instance, \texttt{-development 6} in Section~\ref{sec:intro} is a processor flag for the \texttt{redundant} processor.
To ensure soundness, we only search for flags of processors existing in CSI's competition strategy. 
Second, we include \textit{iteration parameters}, such as time limits or repeated numbers of execution, to regulate the running of a certain sub-strategy. 
These parameters are defined in CSI's strategy language.
Moreover, we add a \textit{boolean execution-controlling parameter} for some parallel or sequential executed sub-strategies, indicating whether to run the particular sub-strategies in confluence analysis. 
% Each of them is a boolean parameter.
Assume a strategy \texttt{A||B}, where \texttt{||} denotes a parallel execution.
The boolean parameters for \texttt{A} and \texttt{B} can represent whether to run one, both, or neither of them.

We need to construct a strategy for CSI using the parameters searched by Grackle.
To achieve this, we start with CSI's competition strategy, replacing the processor flags and iteration parameters with relevant invented parameters.
Then, we disable sub-strategies according to the boolean execution-controlling parameters.

The most challenging part of our work is the proper definition of the parameter space to confirm CSI's soundness. 
As the exact definition is quite technical and verbose, we present the explanation of the parameter space and show an invented strategy in the technical appendix.

\subsection{Strategy Combination}
After inventing several complementary strategies, we want to combine them into a single strategy and compare it with the competition strategy of CSI. 
The combination is performed by choosing a few strategies from Grackle's final portfolio and appropriately assigning a time limit to each of them.

To effectively divide the time, we split the whole one minute into several time splits.
Next, we greedily allocate a strategy to each time split in the sequence by order. Each newly chosen strategy aims at proving the largest number of remaining benchmark problems that have not been proved by the previously chosen strategies. We shuffle the sequence 100 times and greedily select strategies for each shuffled sequence, resulting in 
strategy schedules comprising sequences of pairs of strategies and time splits.
To use a strategy schedule, CSI executes each strategy in it by order for a duration of the relevant time split.
We split the one-minute duration into many sequences and perform the greedy strategy selection for each. 
We finally choose the strategy schedule that maximizes the number of provable problems.
The details of the strategy combination are explained in the technical appendix.

\section{Dataset Augmentation}
Although ARI-COPS is meticulously built by term rewriting experts, it is unsuitable for AI techniques. 
First, it is relatively small which is insufficient for contemporary AI techniques. 
% Second, there may exist an imbalance within ARI-COPS because certain groups may submit many similar problems upon discovering an interesting property, despite the rarity of TRSs with such properties.
Second, there may be an imbalance in ARI-COPS because the problems come from rewriting literature. The examples are often of theoretical interest and are constructed to illustrate specific confluence analysis techniques. 
% Third, the TRSs in ARI-COPS are often in their most general format to demonstrate a certain theoretical property. 
% Nevertheless, TRSs encountered in practical applications can contain redundant rules irrelevant to showing a certain property.
However, TRSs encountered in practical applications can contain redundant rules that are irrelevant to illustrating a certain property.

\subsection{TRS Generation Procedure}
We develop a program to randomly generate a large dataset of TRSs, receiving multiple parameters to control the overall generation procedure.
First, the maximum number of available function symbols $F$, constants $C$, variables $V$, and rules $R$ establish the upper bound of the respective quantities of symbols and rules.
For each of $F$, $C$, and $V$, a value is randomly selected between zero and the specified maximum, determining the actual number of available symbols.
%While the actual number of rules is randomly chosen between 1 and $R$. % yan: something seems missing here, fixed below.
The actual number of rules is randomly chosen between one and $R$.
Second, we define a parameter $M$, used during the initialization of function symbols. For each function symbol, an arity is randomly chosen between one and $M$
% arities are randomly chosen for each function symbol, ranging from one an a specified maximum arity parameter.

Another important parameter is the probability of generating a left-linear TRS $L$, which is associated with the likelihood of producing provably confluent TRSs.
The majority of contemporary techniques for proving confluence are merely effective for left-linear TRSs. 
Without regulating the ratios of left-linearity, randomly generated TRSs rarely exhibit left-linearity, making it theoretically difficult to show confluence for them. 
We also notice that, in practice, CSI can merely prove confluence of very few generated TRSs if the ratios of left-linearity are not controlled.
By default, we force 60\% of generated TRSs to be left-linear. 

Moreover, for a rule $l \rightarrow r$, there is a parameter called $CT$ 
% regulating the iprobability of generating $l$ and $r$ that are constants or variables. 
related to the probability of generating $l$ and $r$ that are complex terms. 
We need it because we prefer complex terms, whereas constants and variables are quite simple.
% The value of $CT$ can be larger than one, as it indicates the maximum probability. 
% For each TRS, a value is randomly chosen between zero and $CT$.
% determining the inverse likelihood of generating terms  constants or variables.

Algorithm~\ref{alg:mk_term} presents the generation procedure of a single term.
% The inputs contain lists of available constants, variables, and functions initialized based on the parameters $F$, $V$, and $C$.
% Two conditions need to be considered while choosing the root symbol.
While choosing the root symbol, we first randomly sample a value between zero and one and compare it with $comp$ to determine whether to only use $\mathit{funs}$ as candidates for the root symbol.
Here, $comp$ is a value randomly chosen between zero and $CT$ during the initialization stage of the generation of a TRS.
If the $comp$ is larger than one, we can only generate complex terms.
Meanwhile, according to the definition of rewrite rules in Section~\ref{sec:rewriting}, the left term $l$ in $l \rightarrow r$ cannot be a variable.
After choosing a root symbol for the term $t$, we continuously choose new symbols for undefined 
function arguments until all of them are defined.
After selecting a new variable, we need to remove it from the set of available variables if we are generating a left-linear TRS.
The size of the terms generated by us is at most 15, where the \textit{size} of a term is defined as the number of symbols in it. 
We choose 15 as the maximum value because the sizes of most terms in ARI-COPS are smaller than 15. 

To generate a rule $l \rightarrow r$, we first execute Algorithm~\ref{alg:mk_term} to generate $l$ and then execute it again to generate $r$.
We extract all used variables in $l$ and mark them as available variables for the generation of $r$, thereby $\mathit{Var(r)} \subseteq \mathit{Var(l)}$, as required by the definition of rewrite rules in Section~\ref{sec:rewriting}.

We repeatedly generate rewrite rules until they reach the expected number and then return the newly generated TRS.

\begin{algorithm}[tb]
\caption{Term Generation}
\label{alg:mk_term}
\textbf{Input}: $consts, vars, \mathit{funs} $ \\
\phantom{\textbf{Input}:} $comp$, the likelihood of making a complex term \\
\phantom{\textbf{Input}:} $\mathit{left}$, whether the term is on the rewrite rule's left side \\
\phantom{\textbf{Input}:} $linear$, whether to construct a linear term \\
\textbf{Output}: a term $t$ 
% and the variables $us$ in $t$
\begin{algorithmic}[1] %[1] enables line numbers
% \STATE Let $t=0$.
% \WHILE{condition}
% \STATE Do some action.
\IF {$random(0, 1) < comp$}
    \STATE $root\_symbols \gets \mathit{funs}$
\ELSIF{left}
    \STATE $root\_symbols \gets \mathit{funs} + consts$
\ELSE
    \STATE $root\_symbols \gets \mathit{funs} + consts + vars$
\ENDIF
\STATE $t \gets random\_choose\_one(root\_symbols)$
\STATE $\mathit{undefs} \gets $ undefined function arguments in $t$
\WHILE{$\mathit{undefs}$ is not empty}
  \FORALL{$\mathit{undef} \in \mathit{undefs}$}
     \STATE $sym \gets  random\_choose\_one(\mathit{funs} + consts + vars)$
     \STATE replace the undefined function argument corresponding to $\mathit{undef}$ in $t$ with $sym$
     \IF{$linear$ and $is\_var(sym)$ and $\mathit{left}$}
       \STATE remove $sym$ from $vars$ 
     \ENDIF
  \ENDFOR
  \STATE $\mathit{undefs} \gets $ undefined function arguments of $t$
\ENDWHILE
% \STATE $used\_vars \gets$ variables in $t$
\STATE \textbf{return} $t$
% , $used\_vars$
\end{algorithmic}
\end{algorithm}

\subsection{Dataset Generation}
\label{sec:dat_gen}
We utilize the program explained in this section to construct a large dataset, facilitating the application of AI techniques to confluence analysis. 
First, we randomly generate 100,000 TRSs with the parameters of the maximum number of available function symbols $F=12$, constants $C=5$, variables $V=8$, and rules $R=15$. Other parameters include the maximum arity of function symbols $M=8$, the probability of generating left-linear TRSs $L = 0.6$, and the value related to the possibility of generating complex terms $CT = 1.6$.

However, the randomly generated dataset can be imbalanced.
First, there may be significant differences in the number of confluent, non-confluent, and indeterminate TRSs.
Second, the number of TRSs mastered by different confluence analysis techniques may vary considerably. 

We develop a multi-step procedure to build a relatively balanced dataset. 
First, we execute CSI's competition strategy on all generated TRSs for one minute using a single CPU. CSI outputs \texttt{NO}, \texttt{YES}, and \texttt{MAYBE} for 69317, 25012, and 5671 TRSs, respectively.

Second, we randomly choose 5,000 problems from each set of problems classified as \texttt{NO}, \texttt{YES}, and \texttt{MAYBE} by CSI. 

Third, we execute the duplicate checker used in CoCo 2024 to remove the duplications in the 15,000 chosen TRSs and 566 ARI-COPS TRSs. 
% It first uses a special fingerprint function to check if the syntactical structure is the same.
It checks the equivalence of syntactical structures between TRSs modulo renaming of variables and a special renaming on function symbols of their signatures.
If TRSs of an equivalence class occur both in the randomly generated dataset and ARI-COPS, we only remove those randomly generated TRSs.

Fourth, we want to mitigate the imbalance in the number of problems mastered by different confluence techniques.
%decision procedures.
We execute 26 strategies for all TRSs, aiming at labeling each of them with the most effective strategy.
The labeling strategies contain all initial strategies for Grackle, which are explained in Section~\ref{sec:strat_invent}. 
The other two that are used to prove confluence are extracted from two complicated initial strategies, both consisting of many sub-strategies and integrated with transformation techniques that potentially simplify the search for proofs.
Specifically, the two complicated initial strategies parallelly execute two important confluence analysis techniques, 
development closedness~\cite{van1997developing} and decreasing diagrams~\cite{van1994confluence}, not used by the other initial sub-strategies.
If we do not use them for labeling, we will not be able to understand whether a TRS is mastered by one of the two important confluence analysis techniques.
The details of the two new labeling strategies are explained in the technical appendix.
% them or the combination of multiple strategies and transformation techniques. 
The time limit for using CSI's competition strategy as a labeling strategy is one minute.
The time limit for other labeling strategies is 30 seconds, smaller than one minute because the execution of decomposed sub-strategies is more efficient.
We calculate the number of problems most quickly solved by each labeling strategy.
The details of labeling strategies are presented in the technical appendix.
The randomly generated dataset is quite imbalanced, four strategies master more than 1,000 problems; however, 16 strategies master 
less than 250 problems.
To address the imbalance, we randomly choose at most 300 problems for a strategy from its set of mastered problems.
We also randomly add 1,200 problems that cannot be solved by any labeling strategy to the dataset.

Finally, we obtain a dataset of 5,267 TRSs. 
% Within this dataset, 1,569 TRSs were determined to be confluent, 1,936 were identified as non-confluent, and 1,687 remained indeterminate when evaluated by CSI using a single CPU within a one-minute time limit.
Within this dataset, 1,647 TRSs are classified as confluent, 1,910 as non-confluent, and 1,710 as indeterminate when evaluated by CSI using a single CPU within a one-minute time limit.

\begin{figure}[t]
\centering
\includegraphics[width=\columnwidth]{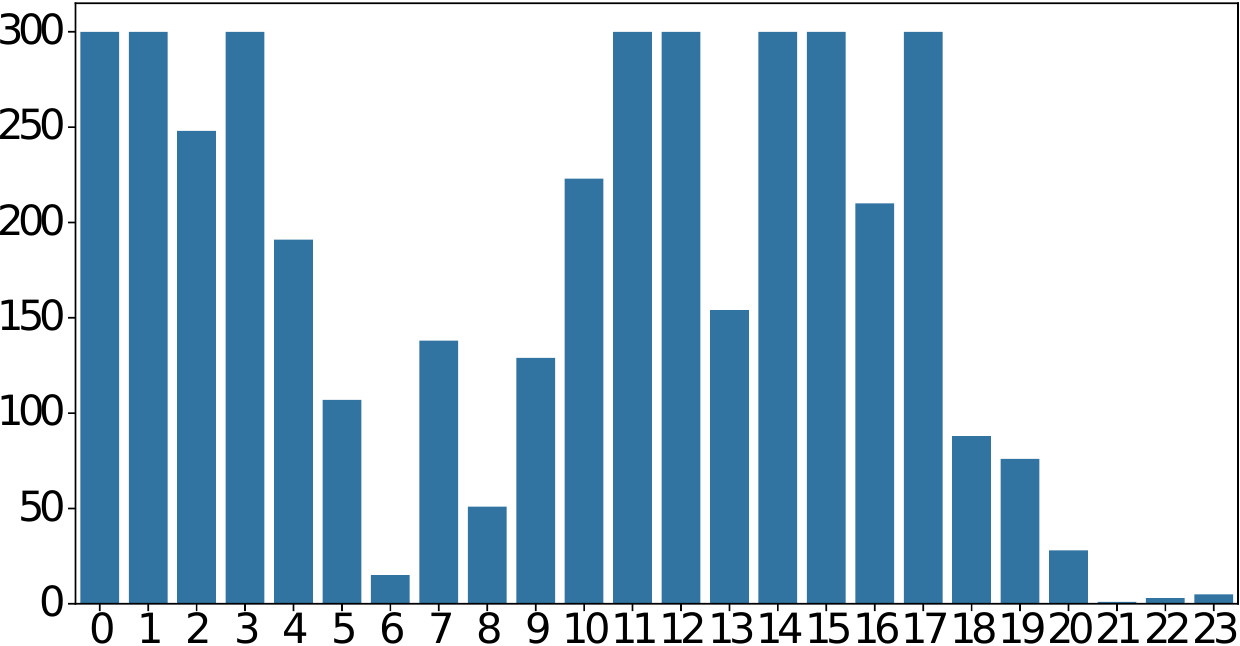} % Reduce the figure size so that it is slightly narrower than the column. Don't use precise values for figure width.This setup will avoid overfull boxes.
\caption{
The number of TRSs solved most quickly (y-axis) for each labeling strategy (x-axis).
%Number of TRSs most quickly solved by each labeling strategy. 
Two labeling strategies that do not master any problems are ignored in the x-axis.}
\label{fig:distri}
\end{figure}

Figure~\ref{fig:distri} shows the final distribution of the number of problems mastered by each labeling strategy.
% It is not perfectly balanced; however, we consider it relatively balanced because some strategies can only master the problems satisfying particular properties.
It is not perfectly balanced; however, we consider it relatively balanced, given that certain strategies can only master problems that satisfy particular properties.
Such properties can be uncommon in randomly generated TRSs and practical applications.

There are infinitely many strategies that can be chosen as labeling strategies, such as strategies obtained by changing processor flags. 
We do not choose other labeling strategies as we have already decomposed CSI's competition strategy, enabling us to label problems with all categories of confluence analysis techniques implemented in CSI. 
% Labeling the dataset based on the categories of confluence analysis techniques
Further decomposition or modification of processor flags may allocate problems to different labeling strategies that only slightly differ.

% The other two are extracted from Grackle's 23 initial strategies.
% Two of Grackle's initial strategies for proving confluence are combinations of several smaller sub-strategies and are integrated with a transformation technique called redundant rule generation.\textbf{cite}
% While redundant rule generation is not utilized in the other seven sub-strategies for proving confluence.
% In particular, the two complicated sub-strategies contain two smaller sub-strategies that do not exist in the other seven sub-strategies.
% We additionally execute the two smaller sub-strategies for all TRSs since 
% are complicated and can be divided into smaller sub-strategies.

% As the initial strategies for Grackle, we divide th
% Using the program explained before, we construct a dataset consisting of 5... TRSs 

\begin{table}[t]
\centering
\begin{tabular}{|c|c|c|c|c|}
% \begin{tabular}{c c c c c}
\hline
& \multicolumn{2}{c|}{ARI-COPS} & \multicolumn{2}{c|}{augment} \\
\hline
CPU & 1 & 4 & 1 & 4 \\
\hline
 init & 475 & 477 & 846 & 852\\
 total & 479 & 484 & 873 & 871 \\
 confs & 73 & 93 & 92 & 104 \\
 both in final & 6 & 2 & 22 & 12 \\

\hline
\end{tabular}
\caption{Statistics of Grackle's training procedure. The rows \textit{init} and \textit{total} denote the number of problems solved by Grackle's initial strategy and the number of problems solved by strategies in Grackle's final portfolio, respectively. The row \textit{confs} denotes the number of strategies that remains in Grackle's final portfolio.  The row \emph{both in final} represents the number of strategies in the final portfolio that master both confluence and non-confluence of TRSs.
% We both try executing a strategy with a single CPU and four CPUs.
}
\label{tab:grackle_train}
\end{table}

\section{Experiments}
% \textcolor{blue}{
We evaluate our strategy invention method on ARI-COPS and a combination of the randomly generated TRSs and ARI-COPS datasets.
In both datasets, CSI with invented strategies outperforms CSI with the competition strategy, the state-of-the-art approach in confluence analysis for TRSs. 

\subsection{Experimental Settings}
%Two datasets are employed in the experiments.
The ARI-COPS 2024 dataset comprises a total of 1,613 problems of which 566 are TRS problems. 
We focus on evaluating our approach on TRS problems since they are standard term rewriting problems for confluence analysis and represent the major category in ARI-COPS.
Another evaluation dataset consists of data from both ARI-COPS and our randomly generated datasets in Section~\ref{sec:dat_gen}.
For training purposes, we arbitrarily select 283 examples from ARI-COPS and 800 examples from the randomly generated dataset.
To build the test dataset, we exclude the examples in the training dataset, subsequently randomly selecting 800 examples from the randomly generated dataset and the remaining 283 examples from ARI-COPS.

%We meticulously configure Grackle's parameters to optimize the strategy invention procedures for CSI.
The Grackle time limit for proving a TRS is 30 seconds, employed both in the evaluation and the strategy specialization phases.
During the specialization phase, Grackle launches ParamILS for parameter tuning.
% which tries to solve the problems with various strategies and outputs the best strategy found.
The overall time limit for one strategy specialization phase is 45 minutes.
% Grackle interleaves several evaluation and invention phases.
% We set 2,700 seconds as the time limit for one iteration of the strategy invention phase.
The total execution time of Grackle is two days.
Grackle performs parallel execution in both the evaluation and specialization phases; thus, we also limit the number of CPUs it can use.
For each dataset, we perform two Grackle runs, configuring the numbers of available CPUs for a single strategy run to be either one or four.
When it is set to one and four, the total number of available CPUs for Grackle is set to 52 and 66, respectively.
Here, a CPU denotes a core of the AMD EPYC 7513 32-core processor.
Grackle's portfolio stores at most 200 of the best strategies.

The use of four CPUs has been selected to match the results of CSI's competition strategy in CoCo 2024 on the competition setup. Given exactly the same problems solved by CSI in our own setup described above with four CPUs and in the  CoCo competition in their Starexec~\cite{stump2014starexec} setup we consider the further comparisons in the paper fair.

\begin{table}[t]
\centering
% \begin{tabular}{c c c c c c c c}
\begin{tabular}{|c|c|c|c|c|c|c|c|}
\hline
& \multicolumn{2}{c|}{comp} & \multicolumn{2}{c|}{total} & \multicolumn{2}{c|}{combine} &
% & \multicolumn{2}{c}{comp} & \multicolumn{2}{c}{total} & \multicolumn{2}{c}{combine} &
\multirow{2}{*}{
% \begin{tabular}{@{}c@{}} CoCo \\ 2023\end{tabular}
CoCo
}
\\
% \midrule{1-7}
\cline{1-7}
CPU & 1 & 4 & 1 & 4 & 1 & 4 & \\
% \midrule
\hline
 yes & 266  & 272 & 271  & 277 & 271 & 276 & 272 \\
 no & 203  & 205 &  208 & 207 & 207 & 207  & 205 \\
 % maybe & ? & 90 & ? & ? & ? & 87 & 90\\
 solved & 469  & 477 & 479 & 484 & 478 & 483 & 477 \\
% \bottomrule
\hline
\end{tabular}
\caption{Numbers of solved TRSs on ARI-COPS.
The column \textit{comp} represents CSI's competition strategy, \textit{total} shows the total number of problems proved by all invented strategies,  and \textit{combine} denotes combining invented strategies as a single strategy.
CoCo denotes the results obtained by CSI in CoCo 2024. }
\label{tab:coco2023}
\end{table}

\subsection{Experimental Results}
\paragraph{Performance on ARI-COPS.} 
Table~\ref{tab:grackle_train} depicts the statistics of Grackle's training procedure.
The value \textit{total} shows the number of solved TRSs after the training, while
\textit{init} is the number solved by the initial strategies.
% The difference between the number of problems proved by \textit{total} and those solved by \textit{init} is insignificant compared to previous Grackle papers~\cite{huula2022targeted, aleksandrova2024prover9}.
% The hypothesis is that unsolvable problems in ARI-COPS for CSI often contain certain theoretical properties beyond modern confluence analysis techniques.
% Enumerating the parameter space can discover more efficient strategies but is not very helpful for theoretically unsolvable problems.
% Moreover, the problems submitted by rewriting experts are often in their most general format, again decreasing the effectiveness of efficient strategies.
When using four CPUs, Grackle's final portfolio contains more strategies than those in the final portfolio generated using one CPU. A probable reason is that executing with four CPUs can discover some strategies that are only effective with enough computation resources.
% \textcolor{blue}{
The final augmented portfolios contain more strategies that master both confluence and non-confluence of TRSs. 
The likely reason is that a larger dataset makes training slower, and it is more difficult for Grackle to find optimal strategies for particular theoretical properties of TRSs.
% }
% exg (A||B)|C vs (A|C)

% smaller compared to previous Grackle papers~\cite{huula2022targeted, aleksandrova2024prover9}, where the number of problems proved by \textit{total} is typically 10\% larger than that of \textit{init}.
% The hypothesis is that rewriting tools work on undecidable problems, and the manual-built dataset contains diverse problems 

\begin{table}[t]
\centering
\begin{tabular}{|c|c|c|c|c|c|c|}
% \hline
\cline{1-7}
& \multicolumn{3}{c|}{never by CSI} & \multicolumn{3}{c|}{never in CoCo} 
% \multirow{2}{*}{
% \begin{tabular}{@{}c@{}} CoCo \\ 2023\end{tabular}}
\\
\hline
% \cline{1-7}
CPU & yes & no & solved & yes & no & solved \\
\hline
 1 & 2 & 3 & 5 & 1 & 3 & 4 \\
 % CeTA-1 & &  & & & &  \\
4 & 4 & 2 & 6 & 1 & 2 & 3 \\
 % CeTA-4 &  &  &  & & &   \\ 
 1\&4 & 6 & 3 & 9 & 2 & 3 & 5 \\
 1-CeTA & 0 & 3 & 3 & 0 & 3 & 3 \\ 
 4-CeTA & 1 & 0 & 1 & 0 & 0 & 0 \\ 
 1\&4-CeTA & 1 & 3 & 4 & 0 & 3 & 3 \\ 
\hline
\end{tabular}
\caption{Numbers of TRSs solved by all strategies in Grackle's final portfolio that have never been solved by all versions of CSI or any tool in CoCo. The suffix CeTA denotes the proofs can be certified by CeTA. The notion 1\&4 means the union of all strategies invented by employing one CPU and four CPUs per strategy execution. 
% If different strategies are invented for the same TRS, we regard the proof for the TRS as verifiable if any of the strategies is verifiable.  
}
\label{tab:newly-solved}
\end{table}

Table~\ref{tab:coco2023} compares the invented strategies with CSI's competition strategy.
With a single CPU per each strategy evaluation, Grackle's final portfolio proves ten more problems than CSI's competition strategy. 
With four CPUs, \textit{total} proves seven more problems than \textit{comp}.

The invented strategies additionally (dis)prove several TRSs that have never been proved by different versions of CSI or all CoCo's participants, as depicted in Table~\ref{tab:newly-solved}. 
In total, we show (non-)confluence for nine TRSs that could not be solved by any versions of CSI. 
Five of the nine new proofs have never been proven by all CoCo's participants.
% If we compare the invented strategies with CSI's results in CoCo 2024, 10 more problems are proved. 
% However, three of them can be proved by previous versions of CSI that employ a simpler much faster strategy.
% Furthermore, the repeated execution of some sub-strategies contributes to some of the newly solved problems.
% with four of these proofs formally certified.

% where three are non-confluent and four are confluent.}

We combine the invented strategies as a single strategy to compare it with CSI's competition strategy.
The number of time splits and the exact time assigned for each invented strategy are presented in the technical appendix. 
With single and four CPUs, \textit{combine} proves nine and six more problems than the competition strategy, respectively.
% Particular, the results of \text

When using one CPU, we gain more improvements over CSI's competition strategy compared to using four CPUs.
A likely reason is that our strategy invention approach is particularly good at generating efficient strategies.
With four CPUs, CSI can run several processors in parallelly, %decreasing the requirement for efficient strategies.
effectively reducing the runtime.

% The number of problems solved by Grackle's init strategy is presented in Table~\ref{tab:grackle_init}, which are not very big compare with 

% We perform three verification procedures to verify the soundness of the invented strategies.
\paragraph{Certification.} 
First, we check whether the answers found by the invented strategies are consistent with the answers discovered in CoCo.
% CSI's competition strategy.
Second, we execute CeTA to verify the proofs for the newly solved problems.
% produced by our strategies.
Table~\ref{tab:newly-solved} depicts the number of newly solved problems certifiable by CeTA.
If we cannot certify the proofs due to the limitation of CeTA and CSI as explained in Section~\ref{sec:csi}, we analyze the related strategies. 
We aim to understand what changes they perform to the original strategy lead to the proofs.
From the analysis, we either slightly modify the sub-strategy defined in the competition strategy or directly use some existing sub-strategies to produce the same answers as the invented strategies. 
These modifications that lead to the answers are employed in the corresponding invented strategies, which are small and sound according to our knowledge of term rewriting.
We also check the certification errors output by CeTA to figure out whether they are indeed errors or just caused by limitations of CSI and CeTA.
Third, for each strategy in Grackle's final portfolio, we run CSI on its mastered problems and apply CeTA to verify the proofs. 
Only 234 and 226 proofs can be verified when one and four CPUs are employed for strategy invention, respectively. 
We manually check the proofs that cannot be verified by CeTA.
The details of our certification procedures are shown in the technical appendix. 
% Second, we verify that the results obtained by the combined strategies and each strategy in Grackle's final portfolio in ARI-COPS are consistent with known labels in ARI-COPS. \textbf{labels from all provers?}
% Finally, we verify that the results of each strategy in Grackle's final portfolio are consistent with ARI-COPS' labels. \textbf{??}

% Table~\ref{tab:coco2024} presents the results of combined strategies on the benchmark of CoCo 2024.
% The duplicate checker explained in Section~\ref{sec:dat_gen} is applied to ARI-COPS in CoCo 2024, reducing the number of TRSs to 566 after removing the duplications and generating a new database ARI-COPS\footnote{https://ari-cops.uibk.ac.at/ARI/}. 
% We do not invent strategies on ARI-COPS since it was recently published.
% The results are consistent with Table~\ref{tab:coco2023}, albeit our improvements become a little smaller after removing the duplications.

\paragraph{Performance on the augmented dataset.}
Table~\ref{tab:grackle_train} also summarizes Grackle's training procedure in the augmented dataset.
Compared to the training in ARI-COPS, Grackle's final portfolios consist of more strategies.
The likely reason is that the augmentation dataset comprises more examples, necessitating more diverse strategies to cover them. 
We notice that with one CPU, the invented strategies prove more problems than those invented with four CPUs. This is probably caused by the randomness in the strategy invention. 

The results of the evaluation in the test dataset are presented in Table~\ref{tab:argu_test}.
With one and four CPUs, \textit{combine} respectively proves 60 and 13 more problems than \textit{comp}.
Notice that here the training examples are disjoint from the testing examples, whereas in the evaluation for ARI-COPS, they are the same.  
%From this, we can conclude that our invented strategies can be well generalized for unencountered data.
From this, we can conclude that our invented strategies generalize well to unseen data.
With four CPUs, the unified strategy proves more problems than using one CPU. 
The likely reason is that the invented strategies with four CPUs can discover proofs more quickly, leading to a stronger unified strategy within the one-minute time limit.

\begin{table}[t]
\centering
\begin{tabular}{|c|c|c|c|c|}
\hline
& \multicolumn{2}{c|}{comp} & \multicolumn{2}{c|}{combine}  \\
\hline
CPU & 1 & 4 & 1 & 4 \\
\hline
 yes & 403 & 412 & 412 & 418 \\
 no  & 399 & 442 & 450 & 449 \\
 % maybe?  & 285 & 240 & 235 & 215 \\
 solved & 802 & 854 & 862 & 867 \\
\hline
\end{tabular}
\caption{Numbers of solved TRSs on the testing examples of the augmented dataset. 
}
\label{tab:argu_test}
\end{table}

\section{Examples}
Besides the example in Section~\ref{sec:intro}, we present two more examples of the invented strategies that (dis)prove problems unprovable by any participant in CoCo.

The core structure of the first example is \texttt{AT}. 
It proves confluence for \texttt{794.ari} in ARI-COPS (\texttt{939.trs} in COPS).
The sub-strategy \texttt{AT}, denoting Aoto-Toyama criteria~\cite{aoto2012reduction}, is defined in CSI's competition configuration document.
% Here, \texttt{REDUNDANT\_DEL} denotes a category of redundant rule technique different from that in Section~\ref{sec:intro}. 
% It removes some redundant rules of the TRS to improve efficiency.
% The symbols \texttt{;} and \texttt{?} together denote a sequential execution.
CSI's competition strategy executes \texttt{AT} in parallel with many other sub-strategies, reducing the computational resources allocated to it and failing to find a proof.

Another example is similar to that in Section~\ref{sec:intro}, we discover that if CSI employs \texttt{redundant -development 6} to generate redundant rules in the competition strategy, it can disprove confluence for \texttt{852.ari}
 (\texttt{997.trs} in COPS), and the proof can be certified by CeTA.

\section{Related Work}
% \paragraph{Term rewriting}
There have been several attempts to apply machine learning to rewriting; however, none have been applied to automatic confluence provers.
% Prior work investigates~\cite{winkler2019smarter} feature characterization of term rewrite systems, but it does not build any learning models based on the features.
While~\cite{winkler2019smarter} investigate feature characterization of term rewrite systems, they do not build any learning models based on the features.
% It utilizes rewrite rules to simplify optimization problems but does not use automatic term rewriting tools nor study the theoretical properties of term rewrite systems, such as confluence.
There are works analyzing the termination of programs using neural networks to learn from the execution traces of the program~\cite{giacobbe2022neural,abate2021learning}. Nevertheless, they do not transform programs to term rewrite systems and apply machine learning to guide automatic term rewriting tools in termination analysis.
% Aurora% the choice of the application of the next rewrite rules
MCTS-GEB~\cite{he2023mcts} applies reinforcement learning to build equivalence graphs for E-graph rewriting, but it focuses on optimization problems, not on confluence.

There has been extensive research on parameter tuning and strategy portfolio optimization in automated reasoning. 
Hydra~\cite{xu2010hydra} employs a boosting algorithm~\cite{freund1997decision} to select complementary strategies for SAT solvers. 
\cite{7814606} propose an evolutionary algorithm for strategy generation in the SMT solver Z3~\cite{de2008z3}. 
A comprehensive review of these approaches is provided by~\cite{kerschke2019automated}.
% \cite{piotrowski2019can} 

\section{Conclusion and Future Work}
We have proposed an approach to automatically invent strategies for the state-of-the-art confluence analysis prover CSI.
%The invented strategies are then combined into a unified strategy.
We have performed data augmentation by randomly generating a large number of term rewrite systems and mixing these with the human-built dataset ARI-COPS.
We have evaluated the invented combined strategy both on the original ARI-COPS dataset and the augmented dataset. 
The invented strategies discover significantly more proofs than CSI's competition strategy on both datasets.
% The combined strategy obtains the state-of-the-art results 
Notably, five of the human-written problems have never been proved by any automatic confluence provers in the annual confluence competitions.

Future work includes applying machine learning to individual term-rewriting techniques, for example those that perform search in a large space. Prioritizing the more promising parts of the search space could improve the individual techniques.
Our strategy invention approach could also be extended to other automatic term rewriting provers.
It would also be possible to apply neural networks to directly predict appropriate strategies for automatic term rewriting tools, however, soundness of proofs generated using such an approach remains a major challenge.
%Since some term rewriting analysis techniques require a large search space, automatic term rewriting provers employ external automated reasoning solvers to discover proof in the search space.

\section*{Acknowledgements}
This research was supported by the ERC PoC project \emph{FormalWeb3} no. 101156734, the University of Innsbruck doctoral scholarship \emph{promotion of young talent}, the National Natural Science Foundation of China 92370201, and the Czech Science Foundation project no. 24-12759S.

\bibliographystyle{named}
% \bibliography{ijcai25}

\clearpage

\appendix
\include{arxiv-appendix}
\bibliography{arxiv}
\end{document}

%% file: arxiv-appendix.tex
\section{Term Rewriting}
\label{sec:rewriting}
% \subsection{Basic Concepts}
We explain the essential concepts of term rewriting in this section.

% Depending on the formalizations of objects, there exist different kinds of rewrite systems. The simplest one is the abstract rewrite system (ARS).
Various types of rewrite systems exist based on the formalization of objects, with the simplest being the abstract rewrite system (ARS).
\begin{definition}
An ARS is a pair $\mathcal{A}=(A, \rightarrow)$ of a set $A$ and a binary relation $\rightarrow$ on $A$.
\end{definition}

A (possibly infinite) \textit{rewrite sequence} is a sequence $a_0 \rightarrow a_1 \rightarrow \cdots $ such that $a_i \in A $.
We write $a \rightarrow^* b $ if there is a rewrite sequence $a \rightarrow \cdots \rightarrow b$.

\begin{definition}
An ARS $(A, \rightarrow)$ is \textit{terminating} if $\forall a \in A$, there are no infinite rewrite sequences starting from $a$.
\end{definition}

The notation $ a \downarrow b $ denotes that $a$ and $b$ are \textit{joinable}, meaning that there exists an element $c \in A$ such that $ a \rightarrow^* c $ and $ b \rightarrow^* c $.

\begin{definition}
An ARS $(A, \rightarrow)$ is \textit{confluent} if $\forall a, b,c \in A$ with $ b \leftarrow^* a \rightarrow^*c $, we have $b \downarrow c$.
\end{definition}

Consider an abstract reduction system (ARS) $\mathcal{E} = (E, \rightarrow)$, where $E = \{a, b, c, d\}$ and $\rightarrow = \{(a, b), (b, d), (c, b), (d, c)\}$. 
% This system produces the ARS illustrated in Figure~\ref{intro:fig-abs}. 
The ARS $\mathcal{E}$ is non-terminating as it admits an infinite rewrite sequence: $c \rightarrow b \rightarrow d \rightarrow c \rightarrow \cdots$.

% A key limitation of ARSs is their reliance on concrete instances alone, which restricts their applicability to computations that involve contexts and variables, such as logical inference rules in ATP. In contrast, 
Term rewrite systems (TRSs) extend ARS by incorporating first-order variables and employing first-order terms.

We then define the notions of rewriting terms using contexts and holes.

\begin{definition}
A \textit{hole} is defined as a special symbol $\Box \notin \mathcal{F}$, and a \textit{context} $C$ is a term that contains exactly one hole.
The notion $C[t]$ denotes the \textit{application of the term $t$ to the context $C$}, which is defined as follows:
\[
C[t] =
\begin{cases}
t & \text{if } C = \Box \\
f(t_1,\ldots,C'[t],\ldots, t_n) & \text{if } C = f(t_1, \ldots ,C', \ldots,t_n)
\end{cases}
\]
\end{definition}

\begin{definition}
The \textit{set of variables in a term} $t$ is defined as
\[
\mathit{Var}(t) =
\begin{cases}
\{t\} & \text{ if } t \text{ is a variable} \\
\varnothing& \text{ if } t \text{ is a constant} \\
\bigcup\limits_{i=1}^{n} Var(t_i) & \text{ if } t = f(t_1, \ldots ,t_n)
\end{cases}
\]
\end{definition}

\begin{definition}
A \textit{rewrite rule} for terms $l$ and $r$ is written as $l \rightarrow r$ where $l \notin \mathcal{V}$ and $Var(r) \subseteq Var(l) $.
A \textit{term rewrite system} $\mathcal{R}$ consists of a set of rewrite rules.
Consider the TRS $\mathcal{R}$, we write the \textit{rewrite relation} $t \rightarrow_{\mathcal{R}} u$ for terms $t, u$ if there exists a rewrite rule $l \rightarrow r \in \mathcal{R}$, a context $C$, and a substitution $\sigma$ such that $t = C[l\sigma]$ and $u = C[r\sigma]$.
\end{definition}

We write $\rightarrow_{\mathcal{R}}^*$ to denote the transitive-reflexive closure of $\rightarrow_{\mathcal{R}}$.
Similar to ARSs, we obtain the definitions of rewrite sequences and $\downarrow_\mathcal{R}$ for TRSs.
We drop the subscript $\mathcal{R}$ for the relations on terms in the subsequent sections if it is contextually inferrable.

\begin{definition}
A TRS $\mathcal{R}$ is \textit{terminating} if $\forall t \in \mathcal{T}(\mathcal{F}, \mathcal{V}) $, there is not any infinite rewrite sequence $t \rightarrow t_1 \rightarrow \cdots $ starting from $t$.

\end{definition}

The TRS $\mathcal{A} = \{ f (x) \rightarrow g(f(x)), g(y) \rightarrow f(g(y)) \}$ is not terminating, as it allows the infinite rewrite sequence $f(x) \rightarrow g(f(x)) \rightarrow f(g(f(x))) \rightarrow \cdots$.
This sequence is infinite because the term $f(x)$ within $g(f(x))$ can be rewritten back to $g(f(x))$, resulting in an infinite loop.

\begin{definition}[confluence]
A TRS $\mathcal{R}$ is \textit{confluent} if and only if $\forall s, t, u \in \mathcal{T(F,V)}, s\rightarrow_\mathcal{R}^* t \land s\rightarrow_\mathcal{R}^* u \Rightarrow t \downarrow_\mathcal{R} u$.
\end{definition}
The TRS $\mathcal{B} = \{ f(x,x) \rightarrow a, f(x,g(x)) \rightarrow b, c \rightarrow g(c) \}$ is not confluent, as it permits the following rewrite sequences:
$ a \leftarrow f(c,c) \rightarrow f(c, g(c)) \rightarrow b $.
Since no rules can be applied to $a$ and $b$, convergence between them is not achievable. 

A term is called \textit{linear} if no variable multiply occurs in it.
A rewrite rule $l \rightarrow r$ is called \textit{left-linear} if $l$ is linear.
A TRS is called left-linear if all its rules are left-linear.
Left-linearity is crucial for confluence analysis since most existing confluence analysis techniques depend on it to determine the confluence of TRSs.
In this paper, a term is called \textit{compositional} if it is neither a variable nor a constant.

We will also explain some rewriting concepts that are important for the understanding of the parameter space in Section~\ref{sec:param}. However, our parameter involves an extensive number of rewriting techniques. 
Explaining all basic rewriting concepts requires extremely large amount of work, which is beyond the scope of our paper. 
Moreover, a lot of definitions or theorems rely on previous definitions and theorems. Presenting all the dependent also necessiate too much work for us. 

We recommend readers to read some textbooks~\cite
{baader1998term,bezem2003term} for the concepts that they cannot understand.

You may skip the following definitons and theorems if you can get a feeling of our parameter space.

\subsection{Basic Termination Techniques}

To prove termination, many techniques try to discover a well-founded monotone algebra that is compatible with the given TRS.
The crucial part is the discovery of interpretations. Depending on their formats, there are integer interpretations, polynomial interpretations, matrix interpretations, etc.

\begin{definition}[interpretation]
Let $\mathcal{F}$ be a signature. An $\mathcal{F}$-\textit{algebra} $\mathcal{A}$ is a set A equipped with operations $f_\mathcal{A} : A^n \rightarrow A$ for every n-ary function symbol $f \in \mathcal{F}$. The underlying set $A$ is called the \textit{carrier} of $\mathcal{A}$ and $f_\mathcal{A}$ is called the interpretation of $f$.
\end{definition}

\begin{definition}
Let $\mathcal{A}$ be an arbitrary algebra. We inductively define a mapping $[\cdot]_\mathcal{A}$
from the set of ground terms to $\mathcal{A}$ as follows: $[f (t_1 , . . . , t_n )]_\mathcal{A} = f_\mathcal{A} ([t_1]_\mathcal{A}, \ldots, [t_n ]\mathcal{A} )$. In particular, if $t$ is a constant then $[t]_\mathcal{A} = t_\mathcal{A}$ .    
\end{definition}

\begin{definition}[well-founed relation]
Let $R$ be a relation on a set $A$.
A relation $R$ is called \textit{well-founded} if there are no infinite descending sequences $a_1 R a_2 R a_3 R \cdots$ of elements of $A$.
    
\end{definition}

\begin{definition}[monotone algebra]
A monotone $\mathcal{F}$-algebra $(\mathcal{A}, >)$ consists of a non-empty $\mathcal{F}$-algebra $\mathcal{A}$
and a proper order $>$ on the carrier $\mathcal{A}$ of $\mathcal{A}$ such that every algebra operation is strictly monotone in all its coordinates, i.e., if $f \in F$ has arity $n \geq 1$ then
$f_\mathcal{A} (a_1 , \ldots , a_i , \ldots , an ) > f_\mathcal{A} (a_1 , \ldots , b, \ldots , a_n )$
for all $a_1 , \ldots , a_n , b \in A$ and $i \in {1, \ldots, n}$ with $a_i > b$. We call a monotone $\mathcal{F}$-algebra $(\mathcal{A}, >)$ well-founded if $>$ is well-founded.    
\end{definition}

\begin{theorem}
A TRS is terminating if and only if it is compatible with a well-founded
monotone algebra.    
\end{theorem}

% \paragraph{Termination Techniques}
% We explain some basic termination techniques.
% \begin{theorem}
% A TRS $\mathcal{R}$ is \textit{terminating} if and only if there exists a well-founded order $>$ on terms such that $\rightarrow_{\mathcal{R}} \subseteq >$.
% \end{theorem}

\begin{example}
Consider the TRS $\mathcal{R}_1$ consisting of the single rewrite rule
$f(f(x, y), z) \rightarrow f(x, f(y, z))$
Let $(\mathcal{A}, >) = (\mathbb{N}, >_\mathbb{N} )$, the set of natural numbers equipped with the usual order, and define $f_\mathcal{A} (x, y) = 2x + y + 1$ for all $x, y \in \mathbb{N}$. The operation $f_\mathcal{A}$ is strictly monotone in both
coordinates: if $x >_\mathbb{N} x'$ and $y >_\mathbb{N} y'$ then $2x+y+1 >_\mathbb{N} 2x' +y+1$ and $2x+y+1 >_\mathbb{N} 2x+y' +1$.
We have
$f_\mathcal{A} (f_\mathcal{A} (x, y), z) = 4x + 2y + z + 3 >\mathbb{N} 2x + 2y + z + 2 = f_\mathcal{A} (x, f_\mathcal{A} (y, z))$
for all $x, y, z \in \mathbb{N}$. Hence $[\alpha]_\mathcal{A} (f(f(x, y), z)) >_\mathbb{N} [\alpha]_\mathcal{A} (f(x, f(y, z)))$ for every assignment $\alpha$,
yielding the termination of $\mathcal{R}_1$
\end{example}

\subsection{Basic Confluence Techniques}
One typical way of proving confluence is first proving termination and then proving local confluence.

\begin{definition}[local confluence]
Let $\mathcal{R}$ be a TRS. An element $a \in \mathcal{T}$ is \textit{locally confluent} if
for all elements $b, c \in \mathcal{T}$ with $b \rightarrow a \rightarrow c$ we have $b \downarrow c$. The TRS $\mathcal{T}$ is locally confluent if all its elements are locally confluent.
\end{definition}

\begin{theorem}[Newman’s Lemma]
Every terminating and locally confluent TRS is confluent.
\end{theorem}

% Another typical way to prove confluent is based on Knuth-Bendix completion.

% There are other techniques. Due to the extreme amount of work, we recommend readers to learn them from textbooks.

\subsection{Basic Non-confluence Techniques}
We want to introduce critical pairs since they are crucial for non-confluence analysis.

\begin{definition}[substitution]
A \textit{substitution} is a mapping $\sigma$ from $\mathcal{V}$ to $\mathcal{T}(\mathcal{F}, \mathcal{V})$.
The \textit{application of the substitution $\sigma$ to the term $t$} is defined as:
\[
t \sigma =
\begin{cases}
\sigma(t) & \text{if } t \in \mathcal{V} \\
f(t_1\sigma, \ldots ,t_n\sigma) & \text{if } t= f(t_1, \ldots ,t_n)
\end{cases}
\]
\end{definition}

% \begin{definition}[generalization]
% A term $u$ is called a \textit{generalization} of $s$ and $t$ if there exist substitutions $\sigma_1$ and $\sigma_2$ such that $u\sigma_1 = s \land u \sigma_2 = t$.
% \end{definition}

\begin{definition}[unifiability]
Two terms $s$ and $t$ are unifiable if there exists a substitution $\sigma$ such that $ \sigma s = \sigma t$.
\end{definition}

\begin{definition}[variant]
A \textit{variable substitution} is a substitution from $\mathbf{V}$ to $\mathcal{V}$. 
A \textit{renaming} is a bijective variable substitution. A term $s$ is a variant of a term $t$ if $s = t\sigma$ for some
renaming $\sigma$.
\end{definition}

\begin{definition}[Position]
\[
 \mathcal{P}os(t) =
\begin{cases}
\{ \epsilon \}  \text{ if } t \text{ is a variable} \\
\{ \epsilon \} \cup \{ip | 1 \leq i \leq n \text{ and } p \in Pos(t_i)\} \\
  \text{ \ \ if } t = f (t_1 , \ldots , t_n ) \\
% \varnothing& \text{ if } t \text{ is a constant} \\
\end{cases}
\]

Let $p \in \mathcal{P}os(t)$. The subterm of $t$ at position $p$ is denoted by $t|_p$ , i.e.,
\[
t|_p = 
\begin{cases}
t \text{ if } p = \epsilon \\
t_{i}|{q} \text{ if } t = f (t_1 , \ldots, t_n ) \text{ and } p = iq
\end{cases}
\]
The symbol $t(p)$ at position $p$ in $t$ is defined as $t(p) = root(t|_p)$. We partition the set $\mathcal{P}os(t)$ into $\mathcal{P}os_{\mathcal{V}}(t) = \{p \in \mathcal{P}os(t) | t|_p \in\mathcal{V} \}$ and $\mathcal{P}os_{\mathcal{F}} (t) = \mathcal{P}os(t) \setminus Pos_\mathcal{V}(t)$.

\end{definition}
$\mathcal{P}os_{\mathcal{V}}(t)$ denotes the positions of variables in the term $t$.
$\mathcal{P}os_{\mathcal{F}}(t)$ denotes the positions of function symbols in the term $t$.

\begin{definition}[Overlap]
 An \textit{overlap} of a TRS ($\mathcal{F}, \mathcal{R}$) is a triple $\langle l_1 \rightarrow  r_1 , p, l_2  \rightarrow r_2 \rangle$ satisfying
the following properties:
\begin{enumerate}
    \item $l_1 \rightarrow r_1$ and $l_2 \rightarrow r_2$ are variants of rewrite rules of $\mathcal{R}$ without common variables,
    \item $p \in \mathcal{P}os_{\mathcal{F}(l_2)}$
    \item $l_1$ and $l_2|_p$ are unifiable,
    \item if $p = \epsilon$ then $l_1 \rightarrow r_1$ and $l_2 \rightarrow r_2$ are not variants.
\end{enumerate}

\end{definition}
% Definition 5.1.2.
\begin{definition}[Critical Pair]
Suppose $\langle l_1 \rightarrow r_1 , p, l_2 \rightarrow r_2 \rangle$ is an overlap of a TRS $\mathcal{R}$. Let $\sigma$ be a most general unifier of $l_1$ and $l_2 |_p$ . The term $l_2 \sigma[l_1 \sigma]p = l_2 \sigma$ can be rewritten in two different ways:
$ l_2 \sigma[r_1 \sigma]p \xleftarrow[p]{l_1 \rightarrow r_1} l_2 \sigma[l_1 \sigma]p = l_2 \sigma \xrightarrow[\epsilon]{l_2 \rightarrow r_2} r_2 \sigma $
We call the quadruple $(l_2 \sigma[r_1 \sigma]p , p, l_2 \sigma, r_2 \sigma)$ a \textit{critical peak} and the equation $l_2 \sigma[r_1 \sigma]p \approx r_2 \sigma$ a \textit{critical pair} of $\mathcal{R}$, obtained from the overlap $\langle l_1 \rightarrow r_1 , p, l_2 \rightarrow r_2 \rangle$.
\end{definition}

\begin{example}
Let the TRS $\mathcal{R}$ have two rules $f(a, g(x)) \rightarrow f(x, x)$ and $g(b) \rightarrow c$.
We have an overlap $\langle g(b) \rightarrow c, 2, f(a, g(x)) \rightarrow f(x, x) \rangle$.
It gives rise to the critical peak $f(a, c) \xleftarrow[]{2} f(a, g(b)) \xrightarrow[]{\varepsilon}
 f(b, b)$ and the critical pair $f(a, c) \approx f(b, b)$.    
\end{example}

To disprove confluence of a TRS $\mathcal{R}$, we consider peaks of the form$$t  \leftarrow^{\leq m} t_1 \leftarrow s \rightarrow u_1 \rightarrow^{\leq n} u$$
such that $t_1 = s[r_1 \sigma]p \leftarrow s[t_1 \sigma]p = s = s[t_2 \sigma]q \rightarrow s[r_2 \sigma]q = u_1$ with $t_1 \rightarrow r_1$,
$t_2 \rightarrow r_2 \in R, q \leq p$, and $p \in \mathcal{P}os(s[t_2 ]q )$.
The basic idea is to show non-joinability of $t$ and $u$.
In order to test non-joinability of $t$ and $u$ we consider ground instances of $t$ and $u$. Here, ground instances mean substituting all variables with constants.
Let $c_x$ be a fresh constant for every variable $x$ and let $\hat{t}$
denote the result of replacing every variable in a term $t$ with the corresponding
constant. Since for terms $s$ and $w$ we have $s \rightarrow_\mathcal{R} w$ if and only if $ \hat{s} \rightarrow_R  \hat{w}$, it follows that terms $t$ and $u$ are joinable if and only if $\hat{t}$ and $\hat{u}$ are joinable. 
To test non-joinability of $\hat{t}$ and $\hat{u}$ we overapproximate the sets of reducts
for $\hat{t}$ and $\hat{u}$ and check if the intersection of these sets is empty.

\section{CSI Strategy Language}
\label{sec:csi-grammar}
Besides the technical appendix, \cite{nagele2017csi} also explains CSI's 
strategy language.

\paragraph{Overall Grammar}
A strategy is defined by the grammar
\begin{verbatim}
s ::= m | (s) | c | i | e    
e ::= s% | s! | s[f] | {s}o
i ::= s? | s* | s+ | sn* | s[f]*
c ::= s;s | s|s | s||s 
      | if p then s else s
\end{verbatim}
where \texttt{s} expresses the possible strategies of CSI, \texttt{m} denotes the name of 
any available processor, \texttt{p} denotes the name of any available predicate, and 
\texttt{c}, \texttt{i}, and \texttt{e} define the available combinators, iterators, and specifiers. 
Here combinators are used to combine two strategies whereas iterators are used to repeat a given strategy a designated number of times. In contrast, specifiers are used to control the behavior of strategies. 
A strategy works on a confluence problem.
Whenever CSI executes a strategy, internally, a so-called proof object is constructed which represents the actual proof. Depending on the shape of the resulting proof object after applying a 
strategy \texttt{s}, we say that \texttt{s} succeeded or  \texttt{s} failed. This should not be confused with the possible answers of the prover: \texttt{YES}, \texttt{NO}, and \texttt{MAYBE}. 
Here \texttt{YES} means that confluence could be proved, \texttt{NO} indicates a successful non-confluence proof, and \texttt{MAYBE} refers to the case when confluence could neither be proved nor disproved. On the success of a strategy \texttt{s}, it depends on 
the internal proof object whether the final answer is \texttt{YES} or \texttt{NO}. On 
failure, the answer is always \texttt{MAYBE}. Based on the two possibilities success 
or failure, the semantics of the strategy operators is as follows.

\paragraph{Combinators}
\begin{itemize}
\item \texttt{s;s'}: First applies \texttt{s} to the given problem. If this fails, then \texttt{s;s'} fails. Otherwise \texttt{s'} is applied to the resulting problems.
\item \texttt{s|s'}: Applies \texttt{s} to the given problem. If this succeeds, its result is returned. Otherwise \texttt{s'} is applied to the given problem.
\item \texttt{s||s'}: Runs \texttt{s} and \texttt{s'} in parallel on the given problem. As soon as at least one of \texttt{s} and \texttt{s'} succeeds, the resulting problem is returned.
\item \texttt{if p then s else s'}: Applies \texttt{s} to the given problem if \texttt{p} is satisfied by the underlying problem. Otherwise \texttt{s'} is applied.
\end{itemize}

\paragraph{Iterators}
\begin{itemize}
\item \texttt{s?}: Applies \texttt{s} to the given problem. On success, its result is returned. Otherwise, the original problem is returned and unmodified.
\item \texttt{s*}: Applies \texttt{s} recursively to the given problem until it cannot be 
      modified anymore. Note that \texttt{s*} is always successful.
\item \texttt{s+}: Applies \texttt{s} recursively to the given problem until it cannot be modified anymore. I.e., \texttt{s+} is successful if it can prove or disprove termination or confluence of the given problem. Otherwise, it fails. Note that \texttt{s+ = 
s*;s} but \texttt{s+} is not equivalent to \texttt{s;s*}.
\item \texttt{sn*}: Applies \texttt{s} recursively to the given problem until it cannot be 
      modified anymore or \texttt{s} has been applied $n$ times. Note that \texttt{sn*} 
      is always successful.
\item \texttt{s[f]*}: Applies \texttt{s} recursively to the given problem until it cannot be modified anymore or \texttt{f} seconds are elapsed. Note that \texttt{s[f]*} is always successful.
\end{itemize}

\paragraph{Specifiers}
\begin{itemize}
\item \texttt{s\%}: Applies \texttt{s} to the given problem. If \texttt{s} fails, the computation is aborted and \texttt{s\%} fails. Otherwise, it succeeds.
\item \texttt{s!}: Applies \texttt{s} to the given problem. If \texttt{s} proves or disproves termination or confluence of the given problem, \texttt{s!} is successful. Otherwise, it fails.
\item \texttt{s[f]}: Tries to modify a given problem via \texttt{s} for at most \texttt{f} seconds. If \texttt{s} does not succeed or fail within \texttt{f} seconds, \texttt{s[f]} fails. Otherwise \texttt{s[f]} returns the resulting problem. Hence it succeeds (fails) if \texttt{s} succeeds (fails).
\item \texttt{\{s\}o}: Applies \texttt{s} to the given problem. If \texttt{s} fails, \texttt{\{s\}o} fails. 
Otherwise, the modifier \texttt{o} is applied to the resulting problems.
\end{itemize}

\paragraph{Configuration File} Since strategies can get quite complex, CSI provides the 
possibility to specify a config file. A config file consists of a sequence of 
abbreviations of the form \texttt{N = s} where \texttt{N} defines its name and \texttt{s} the strategy (in principle arbitrary text) it should expand to. The convention is 
to use all capital names for abbreviations. Each abbreviation has to be put on a 
separate line. You can spread a strategy over several lines by terminating 
each line with a \verb|\|. Last but not least, you can add comments to config files 
by putting a \texttt{\#} in front of each line.

\section{Default Strategy}
\label{sec:default}
This section explains the workflows of CSI's default competition strategy.
In the CoCo competition, CSI's execution command is roughly \texttt{csi -C CR -s AUTO -c x.trs}.
Here, \texttt{-C CR} denotes to prove (non-)confluence for a TRS. 
CSI is based on the termination tool \texttt{TTT2}~\cite{korp2009tyrolean} since some confluence techniques need to first prove termination. 
There are also other flags for \texttt{-C} to prove different properties of TRSs.
The flag \texttt{-s AUTO} means that CSI executes the \texttt{AUTO} strategy below in the default configuration document. 
Finally, \texttt{x.trs} means to prove (non-)confluence for the TRS $x$.
In ARI-COPS, the TRSs are represented by the \texttt{.ari} format. However,  CSI cannot receive \texttt{.ari} documents currently. We need to first execute a tool~\cite{ari-convert} to convert \texttt{.ari} documents to the corresponding \texttt{.trs} documents.
\begin{lstlisting}
AUTO = (if trs then (\
  sorted -order*;(AUTO_INNER \
    || (NOTCR | REDUNDANT_FC)3*!) \
 ) else fail)       
\end{lstlisting}

\texttt{AUTO} works as below:
\begin{enumerate}
    \item Check whether the given problem is a TRS problem. There are many different rewrite systems that can be input to CSI like higher-order rewrite systems, etc.  CSI fails if it is not a TRS problem.
    \item If it is, the default strategy uses ordered sorted decomposition~\cite{felgenhauer2015layer} to try to decompose the given TRS problem to a set of sub-problems. 
    % If it is not a TRS problem, CSI directly fails. 
    % Proving confluence for all decomposed problems equals proving confluence for 
    The original TRS is confluent if and only if all decomposed problems are confluent~\cite{felgenhauer2015layer}.
    \item \texttt{AUTO\_INNER} and \texttt{(NOTCR | REDUNDANT\_FC)3*!} are parallelly executed.  \texttt{AUTO\_INNER} mainly focuses on proving confluence, while  \texttt{(NOTCR | REDUNDANT\_FC)3*!} focuses on disproving confluence.
    \texttt{(NOTCR | REDUNDANT\_FC)3*!} first uses \texttt{NOTCR} to determine whether the given problem is nonconfluence, if it can be determined, the answer will be returned. Otherwise, it applies \texttt{REDUNDANT\_FC} which executes the redundant rule technique~\cite{nagele2015improving} to transform the given TRS. 
    \texttt{(NOTCR | REDUNDANT\_FC)} is repeatedly applied for at most three times to discover an answer.
    The specifier \texttt{!} is crucial since the transformation of  \texttt{REDUNDANT\_FC} may be successful, but it does not discover a nonconfluence proof. 
    According to the explanation in Section~\ref{sec:csi-grammar},
    CSI will return \texttt{YES} in this case if we do not use \texttt{!}. 
    % CSI will return \texttt{YES} if the 
    The details of \texttt{NOTCR} and \texttt{REDUNDANT\_FC} are explained in Section~\ref{sec:param}.
\end{enumerate}

% \begin{tcolorbox}[mycodebox]
\begin{lstlisting}
AUTO_INNER = \
  (AUTO_INNER0[30] | CPCS[5]2*)! | \
  ({AUTO_INNER0[30]}nono \
    | CPCS[5]2*)2*!
\end{lstlisting}
% \end{tcolorbox}

The definition of \texttt{AUTO\_INNER} is shown above. It first tries to discover a proof using \texttt{AUTO\_INNER0}. If it cannot discover a proof, the critical pair closing system (CPCS)~\cite{oyamaguchi2014confluence} transformation is applied. Then \texttt{AUTO\_INNER0} is executed again. Notice that the second \texttt{{AUTO\_INNER0[30]}nono} uses \texttt{nono} to ignore the proof for nonconfluence produced by \texttt{AUTO\_INNER0} because after the CPCS transformation, we cannot produce sound proofs for nonconfluence~\cite{oyamaguchi2014confluence}.

% \begin{tcolorbox}[mycodebox]
\begin{lstlisting}
AUTO_INNER0 = (GROUND || KB || AC ||\
  (REDUNDANT_DEL?;(CLOSED || DD \
     || SIMPLE || KB || AC \ 
     || {GROUND}nono))3*! \ 
  || (((CLOSED || DD) \ 
    | REDUNDANT_RHS)3*! \
  || (CLOSED || DD) \
    | REDUNDANT_JS)3*! || KH \
  || AT || SIMPLE || CPCS2 || \
 fail)
\end{lstlisting}
% \end{tcolorbox}

The above code explains the details of \texttt{AUTO\_INNER0}. It parallelly executes various techniques. The details of all techniques are explained in Section~\ref{sec:param}.

\section{Parameter Space}\label{sec:param}

\subsection{Overall Patterns}
The name of a parameter often follows one of the patterns:
\begin{itemize}
    \item \textit{abbreviation} such as \texttt{PRETRS}. It is a \emph{boolean-execution controlling parameter}. When it is set to \texttt{no}, we set \texttt{PRETRS = fail} in the generated configuration document.  The \texttt{fail} processor fails immediately. 
    CSI will not do any modification and discover any proof.
    \item \textit{abbreviation\_processor} such as \texttt{PRETRS\_matrix}.  It is a \emph{boolean-execution controlling parameter}. We use such names when a strategy contains several parallelly or sequentially executed processors.
    When it is set to \texttt{no}, we ignore the corresponding processor in the strategy.  
    \item \emph{abbreviation\_processor\_flag}.
    It belongs to \emph{processor flags},
    e.g.,  \texttt{PRETRS\_matrix\_ib}. 
    It chooses the flags for the processor. For instance, \texttt{PRETRS\_matrix\_ib} chooses the value for the \texttt{ib} flag.
    If \texttt{PRETRS\_matrix\_ib} chooses the value 6, we may obtain \texttt{matrix -ib 6}. 
    \item \textit{name\_time} or \textit{name\_loop}. 
    It belongs to \emph{iteration parameters}.
    They are used to control the execution time or the number of  
    repeated application times of a strategy. They control the values of the specifiers. For instance, if \texttt{PRETRS\_time} is set to 2, we generate the strategy \texttt{PRETRS = content of PRETRS[2]}.
\end{itemize}

% Replacing a definition such as \texttt{PRETRS} with \texttt{fail} is sound.
% There are different 

\subsection{Details of Patameters}
In this Section, we explain the strategies in CSI's competition strategy document, following its structure sequentially from the beginning to the end. 
The explanations of a flag and its soundness will be ignored if we have explained them in the previous strategy definitions.
After searching for the parameters, we convert the parameters into a configuration document which is used by CSI. 
The structures of strategies generally remain the same to confirm soundness, meaning that besides the newly discovered parameters, we do not change the processors and the order of their applications. 
We give examples to explain how to keep the structures in this section later via examples.
% We ignore the processor flags not searched by us since the number of them is indeed large and presenting all explanation is too verbose. We recommend to execute CSI's help function to understand the details. 

We ignore the processor flags not searched by us since the number of them is indeed intensive, and presenting all explanations is too verbose. We recommend executing CSI's help function to understand the details.

% \begin{tcolorbox}[mycodebox]
\begin{lstlisting}
PRETRS = ((\
 matrix -dim 1 -ib 3 -ob 5 | \
 matrix -dim 2 -ib 2 -ob 3 | \
 matrix -dim 3 -ib 1 -ob 1 | \
 matrix -dim 3 -ib 1 -ob 3)[2]*)    
\end{lstlisting}
% \end{tcolorbox}

It preprocesses the TRS before trying to discover a termination proof.
The preprocess tries to reduce the size of the original TRS by removing some rules from it.
The parameter space for \texttt{PRETRS} is presented below.
\begin{itemize}
    \item \texttt{PRETRS \{yes, no\}}. \texttt{PRETRS} can be chosen as \texttt{yes} or \texttt{no}. When it is set as \texttt{no}, we replace its definition with \texttt{fail}, i.e. \texttt{PRETRS = fail}. The processor \texttt{fail} means the processor simply fails.
    Replacing it with \texttt{fail} is sound since it is only used in \texttt{SN}.
    It is \texttt{(PRETRS;(...)}. According to the definition of \texttt{;}, replacing it with \texttt{fail} only makes the strategy for termination fail immediately.
    \item \texttt{PRETRS\_time} \{1, 2, 3, 4, 5\}. It is used to control the execution time of \texttt{PRETRS}. The default execution time is 2 seconds, controlled by \texttt{[2]} as in the definition. We set the execution time to be chosen from \texttt{\{1, 2, 3, 4, 5\}} seconds.
    \item \texttt{PRETRS\_matrix\_dim \{1, 2, 3, 4, 5, 6, 7, 8, 10, 12, 16\}}. The \texttt{matrix} processor means applying matrix interpretations~\cite{endrullis2008matrix}. 
    The parameter \texttt{PRETRS\_matrix\_dim} specifies the dimension of the matrices.
    Each matrix processor has several processor flags and associative values to choose from as below. Since there are several matrix processors in \texttt{PRETRS}, we actually associate each matrix processor with a set of parameters to search from. For simplicity, we only show the parameter space of a single matrix processor.
    The values \{yes, no\} indicate whether to use such flags. For instance, when \texttt{matrix\_real} is set to \texttt{yes}, we add the \texttt{real} flag to \texttt{matrix} and obtain \texttt{matrix -real}. In comparison, when it is set to \texttt{no}, we do not append the \texttt{real} flag and simply use \texttt{matrix}.
    The other flag values indicate the values of the flags. For instance, when \texttt{matrix\_ib} is set to \texttt{3}, we obtain a processor of \texttt{matrix -ib 3}.
    % The \texttt{matrix} processor also has many other flags to choose from. We do not use them.
    % The reasons for not using such flags are explained in \textcolor{blue}{TODO}
    \item \texttt{PRETRS\_matrix\_ib \{1, 2, 3, 4, 5, 6\}}. 
    Defines the number of bits that can be used to represent the smallest number that appears in the intermediate results
    % Defines the number of bits used to represent matrix entries.    
    \item \texttt{PRETRS\_matrix\_ob \{1, 2, 3, 4, 5, 6, 7, 8, 9, 10, 12, max\}}. 
    Defines the number of bits that can be used to represent the largest number that appears in the intermediate results.
    Actually \texttt{max} is not a valid value to the \texttt{ob} flag. However, Grackle has the mechanism to ignore a parameter if it equals the default value. We set \texttt{max} as \texttt{PRETRS\_matrix\_ob}'s default value. 
    Moreover, by default, CSI uses the largest 64-bit integers for \texttt{ob}.
    Hence, \texttt{max} reasonably represents the maximal value of \texttt{ob}.
    \item \texttt{PRETRS\_matrix\_rat \{1, 2, 3, 4\}}. Sets the denominator of integers. It makes integers become rational numbers. 
    The soundness of non-negative rational numbers for matrix interpretation is explained in in~\cite{gebhardt2007matrix,zankl2010satisfiability}. 
    % Since polynomial numbers belong to real numbers. The \texttt{-rat} flag is also sound. 
    The \texttt{matrix} processor also has a flag \texttt{neg}, which is set to \texttt{false} by default. Since we do not use \texttt{neg} in our parameter space, the interpretations are always non-negative and therefore sound.
    \item \texttt{PRETRS\_matrix\_db \{1, 2, 3, 4, max\}}. Specifies the bits after the decimal point. The reason for soundness is the same as  \texttt{PRETRS\_matrix\_rat}
    \item \texttt{PRETRS\_matrix\_real \{yes, no\}}. 
    The soundness of non-negative real numbers for matrix interpretation is explained in Lemma 23, Definition 24, and Lemma 25 in ~\cite{zankl2010satisfiability}.     
    Matrix interpretation over real numbers is unsound because negative numbers break the monotone constraints~\cite{endrullis2008matrix}. However, \texttt{matrix} also has a flag \texttt{neg}, which is set to \texttt{false} by default. Since we do not use \texttt{neg} in our parameter space, the real number interpretations are always non-negative and therefore sound.
    \item \texttt{PRETRS\_matrix\_triangle \{yes, no\}}. Use triangular matrices. It is sound since it only constrains the shape of the matrix. It is less expressive than the traditional matrix interpretation. Originally, it is invented for complexity analysis~\cite{moser2008complexity}.    
\end{itemize}

The structure of \texttt{PRETRS} is kept although we search for other parameters. This means the \texttt{matrix} processors are still connected by \texttt{|} instead of \texttt{||}. We may generate something like the strategy below.

% \begin{tcolorbox}[mycodebox]
% \begin{lstlisting}[breaklines]{text}
\begin{lstlisting}
PRETRS = ((\
 matrix -dim 1 -ib 4 -ob 5 | \
 matrix -dim 3 -ib 3 -ob 3 | \
 matrix -dim 3 -ib 3 -ob 7 | \
 matrix -dim 3 -ib 1 -ob 3)[4]*)    
\end{lstlisting}
% \end{tcolorbox}

Next, we explain the parameter space for \texttt{DIRECTTRS}.
% \begin{tcolorbox}[mycodebox]
\begin{lstlisting}
DIRECTTRS = ((\
 kbo || (lpo | (ref;lpo) \
 || (bounds -rfc -qc))*[7])!
\end{lstlisting}
% \end{tcolorbox}

\begin{itemize}
\item \texttt{DIRECTTRS \{yes, no\}}. \texttt{DIRECTTRS} and \texttt{DIRECTTRS\_time} works similar as \texttt{PRETRS} and \texttt{PRETRS\_time}. Hence, we ignore such explanations here and for the following sub-strategy definitions. 
It is sound as \texttt{DIRECTTRS} is only parallelly with other strategies in \texttt{SN}.
It makes one technique fail and does not affect other parallel executions.
\item \texttt{DIRECTTRS\_time \{1, 3, 5, 7, 9, 11\}}
\item \texttt{DIRECTTRS\_kbo \{yes, no\}}. The \texttt{kbo} processor 
means the application of the Knuth-Bendix order (KBO)~\cite{baader1998term}.
\item \texttt{DIRECTTRS\_kbo\_ep \{yes, no\}}. Demands an empty precedence (only for `-pbc').
According to the function \texttt{pbc\_aux} in \url{csi/src/processors/src/termination/orderings/kbo.ml},
it is only used when the PBC prover is invoked.
As explained in Definition 1 in~\cite{zankl2009kbo}, KBO has a set of precedences and a set of weight functions.
An empty set of precedences makes KBO weaker in discovering termination proofs.
Empty precedences have also been used in Example 3 in the paper. 
Section 9 in~\cite{zankl2009kbo} shows empty precedences are sometimes useful for the PBC backend.
\item \texttt{DIRECTTRS\_kbo\_ib \{1, 2, 3, 4, 5, 6\}}. 
Defines the number of bits that can be used to represent the smallest number that appears in the intermediate results
% Defines the number of bits used to represent weights. 
\item \texttt{DIRECTTRS\_kbo\_minp \{yes, no\}}. Minimizes the precedence comparisons (only for `-pbc').
According to the function \texttt{solve2} in \url{csi/src/logic/src/solver.ml} and the function \texttt{context} in \url{csi/src/processors/src/termination/orderings/kbo.ml}, it is only used when the PBC prover is invoked.
According to the function \texttt{solve} in \url{csi/src/logic/src/miniSatP.ml}, the usage of \texttt{-pbc} is sound without 
\texttt{-minp} or \texttt{-minw}.
According to Section 9 in~\cite{zankl2009kbo}, it is sound and is helpful for generating human-readable proofs.
KBO has a set of precedence and a set of weight functions.
This flag tries to discover a termination proof with a small set of precedence.
% Section 9 in~\cite{zankl2009kbo} shows its soundness.
\item \texttt{DIRECTTRS\_kbo\_minw \{yes, no\}}. Minimize the sum of weights (only for `-pbc').
According to the function \texttt{solve2} in \url{csi/src/logic/src/solver.ml} and the function \texttt{context} in \url{csi/src/processors/src/termination/orderings/kbo.ml}, it is only used when the PBC prover is invoked.
According to the function \texttt{solve} in \url{csi/src/logic/src/miniSatP.ml}, the usage of \texttt{-pbc} is sound without \texttt{-minp} or \texttt{-minw}.
According to Section 9 in~\cite{zankl2009kbo}, it is sound and is helpful for generating human-readable proofs.
Small weights are more readable.
KBO has a set of precedence and a set of weight functions.
This flag tries to discover a termination proof with weight functions that map symbols to small weights.
\item \texttt{DIRECTTRS\_kbo\_ob \{1, 2, 3, 4, 5, 6, 7, 8, 9, 10, 12, max\}}. Defines the number of bits that can be used to represent the largest number that appears in the intermediate results.
\item \texttt{DIRECTTRS\_kbo\_pbc \{yes, no\}}. Uses PBC backend. PBC, SAT, or SMT are just solvers to search for interpretations. The choice does not affect the soundness.
\item \texttt{DIRECTTRS\_kbo\_quasi \{yes, no\}}. Allows quasi-precedences. It is sound according to Definition 3.1 in~\cite{sternagel2013formalizing}.
\item \texttt{DIRECTTRS\_kbo\_rat \{1, 2, 3, 4\}}. Sets the denumerator (only in combination with `-sat' or `-smt').
\item \texttt{DIRECTTRS\_kbo\_sat \{yes, no\}}.  Uses SAT backend (default).
\item \texttt{DIRECTTRS\_kbo\_smt \{yes, no\}}. Uses SMT backend.
\item \texttt{DIRECTTRS\_lpo \{yes, no\}}. Applies Lexicographic Path Order~\cite{baader1998term}. There are two \texttt{lpo} processors in \texttt{DIRECTTRS}. 
We employ the same parameters for them to reduce the size of the parameter space.
Since \texttt{lpo} are weak techniques for proving termination, assigning different parameters to two \texttt{lpo} processors should only have little influence on the performance of proving termination.
\item \texttt{DIRECTTRS\_lpo\_direct \{yes, no}\}. Try to finish the termination proof. The \texttt{lpo} processor can each time prove the termination of a single rule of a TRS and remove the rule from the TRS. Afterwards, it can be repeatedly applied to prove the termination of another rule in the smaller TRS. It shows the termination of the entire TRS by proving the termination of all rules. Notice that \texttt{*[7]} in the definition means repeatedly applying \texttt{lpo} and \texttt{lpo} as much as possible in seven seconds.
Using \texttt{direct}, it tries to show termination for all rules; thus, it only makes the searching difficult.
\item \texttt{DIRECTTRS\_lpo\_quasi \{yes, no\}}. Allows quasi-precedences (currently not supported together with -dp flag). The proofs of the soundness are presented in Theorem 2.37 and Theorem 2.26 in Chapter 2 Preliminaries in~\cite{hirokawa2006automated}. 
\item \texttt{DIRECTTRS\_lpo\_sat \{yes, no\}}. 
\item \texttt{DIRECTTRS\_lpo\_smt \{yes, no\}}
\item \texttt{DIRECTTRS\_bounds \{yes, no\}}. This processor proves termination of a given problem by using the match-bound technique~\cite{korp2009match}.
\item \texttt{DIRECTTRS\_bounds\_qc \{yes, no\}}. Computes quasi-compatible tree automata instead of compatible     tree automata.
    Different values of the parameter are sound according to~\cite{korp2009match}. 
    Moreover, it is used in \texttt{DIRECTTRS}.
\item \texttt{DIRECTTRS\_bounds\_rc \{explicit, implicit\}}.
    Defines the algorithm that is used to construct raise-consistent tree automata. Possible values are \texttt{explicit} and \texttt{implicit} where. Per default \texttt{implicit} is used.  
Different values of the parameter are sound according to~\cite{korp2009match}. 
% Different values of the parameter are It is sound according to~\cite{korp2009match}.
\item \texttt{DIRECTTRS\_bounds\_rfc \{yes, no\}}. Uses right-hand sides of forward closures.
    Different values of the parameter are sound according to~\cite{korp2009match}.
    Moreover, it is used in \texttt{DIRECTTRS}.
% hence, \texttt{yes} is sound.
\item \texttt{DIRECTTRS\_bounds\_steps \{-1, 1, 2, 3, 4, 5, 6, 7, 8, 10, 12, 16, 32\}}. Specifies the maximum number of compatibility violations that should be solved. This guarantees that the procedure always terminates. Otherwise it might happen that the graph approximation does not terminate.
The match-bound technique tries to build a tree automata using the tree automata completion technique. 
The tree automata completion tries to solve all compatibility violations. 
However, sometimes there are infinite violations. Meanwhile solving all violations may take too much time.
According to \url{csi/src/processors/src/termination/bounds/bounds.ml}, CSI uses -1 by default, meaning to solve all violations. 
Other values make the match-bound technique fail earlier.
This flag only controls the size of the search space and therefore sound.
\end{itemize}

\begin{tcolorbox}[mycodebox]
\begin{lstlisting}{text}
ARCTICTRS = arctic -dp -ur \
  -dim 2 -ib 2 -ob 2[2]

ARCTICBZTRS = arctic -bz -dp -ur \
  -dim 2 -ib 2 -ob 2[2]
\end{lstlisting}
\end{tcolorbox}

Use arctic interpretation~\cite{koprowski2008arctic}.
The parameter space for \texttt{ARCTICTRS} and \texttt{ARCTICBZTRS} is presented below.
\begin{itemize}
\item \texttt{ARCTICTRS \{yes, no\}}. Notice that \texttt{ARCTICTRS} and \texttt{ARCTICBZTRS} use the flag \texttt{ur} because they are employed in \texttt{MAINTRS}, which first applies the transformation using the \texttt{ur} processor. 
The \texttt{ur} processor removes all rules of the given dependency pair (DP) problem which are not usable~\cite{suzuki2011argument}.
The \texttt{-ur} flag for \texttt{arctic} uses usable rules with respect to interpretation.
If we remove \texttt{-ur} here, we will produce unsoundness. 
Therefore, we always choose to use the \texttt{-ur} flag for \texttt{arcitc} and only search for other parameters.
It is sound as \texttt{ARCTICTRS} is only parallelly executed with other strategies in \texttt{MAINTRS}.
It makes one technique fail and does not affect other parallel executions.
\item \texttt{ARCTICTRS\_time \{1, 2, 3, 4\}}
\item \texttt{ARCTICTRS\_arctic\_dim \{1, 2, 3, 4, 5, 6, 7, 8, 10, 12, 16\}}. The reasons for the soundness of \texttt{dim, ib, ob, direct, rat, real} have been explained for the previous strategy definitions.
\item \texttt{ARCTICTRS\_arctic\_ib \{1, 2, 4, 8, 16\}}.
\item \texttt{ARCTICTRS\_arctic\_ob \{1, 2, 4, 8, max\}}.
Defines the number of bits that can be used to represent the largest number that appears in the intermediate results.
\item \texttt{ARCTICTRS\_arctic\_bz \{yes, no\}}. Since both \texttt{yes} and \texttt{no} are empolyed in \texttt{ARCTICTRS} and \texttt{ARCTICBZTRS}, and they are used in the same strategy definition \texttt{MAINTRS}, we conclude either the choice of \texttt{yes} or \texttt{no} is sound. Moreover, its soundness is explained in Section 7~\cite{koprowski2008arctic}.
\item \texttt{ARCTICTRS\_arctic\_direct \{yes, no\}}
\item \texttt{ARCTICTRS\_arctic\_dp \{yes, no\}}.  Allows non-monotone interpretations, i.e., `0' as a coefficient. In the original definition of \texttt{ARCTICTRS}, \texttt{-dp} is used; hence, \texttt{yes} is sound. Choosing \texttt{no} merely makes the interpretation monotone.
Basically, termination techniques require monotone interpretations. The \texttt{-dp} flag is an exception since it works in the dependency pair (DP) frameworks~\cite{giesl2005dependency}.
The soundness of  non-monotone interpretations is explained in Section 6~\cite{koprowski2008arctic}.
\item \texttt{ARCTICTRS\_arctic\_rat \{1, 2, 3, 4\}}. Use rational numbers for arctic interpretations. The soundness is explained in 
Section 5~\cite{sternagel2014formalizing}.
\item \texttt{ARCTICTRS\_arctic\_real \{yes, no\}}. Use real numbers for arctic interpretations. The soundness is explained in 
Section 6~\cite{sternagel2014formalizing}.
\item \texttt{ARCTICBZTRS \{yes, no\}}. The parameter space for \texttt{ARCTICBZTRS} is the same as that for \texttt{ARCTICTRS}; hence, we ignore it here.
It is sound as \texttt{ARCTICBZTRS} is only parallelly executed with other strategies in \texttt{MAINTRS}.
It makes one technique fail and does not affect other parallel executions.

\end{itemize}

\begin{tcolorbox}[mycodebox]
\begin{lstlisting}{text}
BOUNDS = (bounds -dp -rfc -qc \
  || bounds -dp -all -rfc -qc \
  || bounds -rfc -qc)    
\end{lstlisting}
\end{tcolorbox}

\begin{itemize}
\item \texttt{BOUNDS \{yes, no\}}. Prove termination of a given problem by using the match-bound technique~\cite{korp2009match}.
It is sound as \texttt{ARCTICBZTRS} is only parallelly executed with other strategies in \texttt{MAINTRS}.
It makes one technique fail and does not affect other parallel executions.

\item \texttt{BOUNDS\_bounds \{yes, no\}}. There are three \texttt{bounds}, we only explain the parameter space for one for simplicity.
\item \texttt{BOUNDS\_bounds\_all \{yes, no\}}. This flag is only effective if a DP, critical pair, or relative termination problem is given. In that case, all rewrite rules are proved to be finite (relative terminating) instead of a single rewrite rule. It is sound since it works like the \texttt{-direct} flag that proves a certain property for all rewrite rules at the same time.
Moreover, both \texttt{yes} and \texttt{no} are used for \texttt{bounds} in \texttt{BOUNDS}.

\item \texttt{BOUNDS\_bounds\_dp \{yes, no\}}.  Uses the enrichments \texttt{match-DP} and \texttt{top-DP} instead of \texttt{match} and \texttt{top} if a DP problem is given. Make sure that as enrichment either \texttt{top} or \texttt{match} has been chosen because the soundness of \texttt{roof-DP} is unknown. The enrichments are determined by the flag \texttt{-e}. Since we do not search for \texttt{-e} here, it uses the default value \texttt{match} and hence is sound.
\item \texttt{BOUNDS\_bounds\_qc \{yes, no\}}, \texttt{BOUNDS\_bounds\_rc \{explicit, implicit\}}, \texttt{BOUNDS\_bounds\_rfc \{yes, no\}}, \texttt{BOUNDS\_bounds\_steps \{-1, 1, 2, 4, 8, 16, 32\}}. Their soundness has been explained for \texttt{DIRECTTRS}
\end{itemize}

\begin{tcolorbox}[mycodebox]
\begin{lstlisting}{text}
MAINTRS = (dp;edg[0.5]?;(sccs | \
 (sc || sct || \
 {ur?;( \
  (matrix -dp -ur -dim 1 \
    -ib 3 -ob 5 | \ 
   matrix -dp -ur -dim 2 \
    -ib 2 -ob 3 | \
   matrix -dp -ur -dim 3 \ 
    -ib 1 -ob 1 | \
   matrix -dp -ur -dim 3 \ 
    -ib 1 -ob 3) || \
  (kbo -ur -af | lpo -ur -af) || \
  ARCTICTRS || \
  ARCTICBZTRS ) \
 }restore || \
 BOUNDS[1]
 ))*[6])!    
\end{lstlisting}
\end{tcolorbox}

The \texttt{MAINTRS} is the main strategy for proving termination.
It first transforms TRS problems into dependency pair (DP) problems using the \texttt{dp} processor. Next, the \texttt{edg} processor reduces the sizes of the DP problems. Afterwards, the \texttt{sccs} processors tries to decompose each DP problems into several smaller DP problems. Finally, many processors are executed to solve the DP problems. In particular, the \texttt{ur} processor removes all rules of the given DP Problem which are not usable. Note that this processor is not sound if the given DP problem is duplicating.
To confirm the soundness, the termination processors after the application of the transformation processor \texttt{ur} must use the flag \texttt{-ur}.
The processor \texttt{restore}  remains unchanged to confirm soundness.
It restores the original TRS within the given DP problem.
In CSI, TRS rules and DP rules are stored in different data structures. Each execution of processors like \texttt{matrix} in \texttt{MAINTRS} may remove TRS rules or DP rules, but the TRS rule will be restored afterward.
The details of the parameter space are presented below.
\begin{itemize}
\item \texttt{MAINTRS \{yes, no\}}. The main technique to prove termination.
It is sound as \texttt{MAINTRS} is only parallelly executed with other strategies in \texttt{SN}.
It makes one technique fail and does not affect other parallel executions.

\item \texttt{MAINTRS\_time \{2, 4, 6, 8\}}
\item \texttt{MAINTRS\_edg\_time \{0.2, 0.5, 1\}}.
The \texttt{edg} processor~\cite{giesl2005proving} removes all edges from the current dependency graph (DG) that are not contained in the EDG (approximation of DG based on recursive unification and symmetry).
Here, the parameter controls its execution time.
% \item \texttt{MAINTRS\_edg\_i \{yes, no\}}.  Computes the innermost EDG, if possible.
\item \texttt{MAINTRS\_edg\_gtcap \{yes, no\}}. Use a general tcap-like non-reachability analysis. Sound as explained in Theorem 13 of~\cite{giesl2005proving}. 
\item \texttt{MAINTRS\_edg\_nl \{yes, no\}}. Try to exploit nonlinearity for \texttt{-gtcap}. Non-linear order is more expressive than linear order. The soundness is explained in Section 3 in~\cite{giesl2005proving}.
\item \texttt{MAINTRS\_BOUNDS\_time \{0.5, 1, 2, 3\}}
\item \texttt{MAINTRS\_sc \{yes, no\}}. Applies the subterm criterion processor~\cite{sternagel2016generalized}.
\item \texttt{MAINTRS\_sc\_sat \{yes, no\}}. Uses SAT backend (default).
\item \texttt{MAINTRS\_sc\_smt \{yes, no\}}. Uses Yices backend (default).
\item \texttt{MAINTRS\_sc\_rec \{yes, no\}}.  Allow recursive simple projections. 
It is sound because in~\cite{sternagel2016generalized}, it is explained below Definition 5 and is used in the proving termination as shown in Table 1.
% The \texttt{rec}, \texttt{mulex}, and \texttt{defs} flags are sound according to~\cite{sternagel2016generalized}.
\item \texttt{MAINTRS\_sc\_mulex \{yes, no\}}. Allow projections to multisets of terms.
It is sound because in~\cite{sternagel2016generalized}, it is explained in Definition 5 and is used in the proving termination as shown in Table 1.
\item \texttt{MAINTRS\_sc\_defs \{yes, no\}}. Allow projection of defined symbols (only relevant 
for \texttt{-rec} and \texttt{-mulex}; default \texttt{false}). It is sound according to the explanation below Definition 5 in~\cite{sternagel2016generalized}.
\item \texttt{MAINTRS\_sc\_mbits \{1, 2, 3, 4, 5\}}. Bits used for multiplicity of terms in multisets corresponding to left- and right-hand sides (default 2). 
The soundness is explained in Section 3~\cite{sternagel2016generalized}.
% The flag determines 
% It is sound since it only changes the size of the search space.
\item \texttt{MAINTRS\_sc\_wbits \{1, 2, 3, 4, 5\}}. Bits used for multiplicity (weight) of arguments in projections (default 2). 
The soundness is explained in Section 3~\cite{sternagel2016generalized}.
\item \texttt{MAINTRS\_sc\_nsteps \{0, 1, 2, 3, 4\}}. Number of rewrite steps before checking for subterms (default 0). It only uses the rules to rewrite the TRS. The properties of the original TRS remain the same.
\item \texttt{MAINTRS\_sct \{yes, no\}}.  Applies the size-change termination processor to a DP problem.
\item \texttt{MAINTRS\_matrix}. The parameter space for the four \texttt{matrix} processors is similar to that for
\texttt{PRETRS}. Hence, we ignore its details here. The only difference is that we always use flags \texttt{-ur} and \texttt{-dp} for every  \texttt{matrix} processor here. 
The usage \texttt{-ur} is for soundness, while the usage of \texttt{-dp} aims at making the matrix interpretation stronger in discovering termination.
\item \texttt{MAINTRS\_kbo \{yes, no\}}. Applies Knuth-Bendix order~\cite{baader1998term}. Always use the \texttt{-ur -af} flags.
\item \texttt{MAINTRS\_kbo\_ib \{1, 2, 3, 4, 5, 6\}}. 
Defines the number of bits that can be used to represent the smallest number that appears in the intermediate results
\item \texttt{MAINTRS\_kbo\_ob \{1, 2, 3, 4, 5, 6, 7, 8, 9, 10, 12, max}\}. Defines the number of bits that can be used to represent the largest number that appears in the intermediate results.
Actually \texttt{max} is not a valid value to the \texttt{ob} flag. However, Grackle has the mechanism to ignore a parameter if it equals the default value. We set \texttt{max} as \texttt{MAINTRS\_kbo\_ob}'s default value. 
Moreover, by default, CSI uses the largest 64-bit integers for \texttt{ob}.
Hence, \texttt{max} reasonably represents the maximal value of \texttt{ob}.
\item \texttt{MAINTRS\_kbo\_quasi \{yes, no\}}. Allow quasi-precedences. It is sound as explained in~\cite{sternagel2013formalizing}
\item \texttt{MAINTRS\_lpo \{yes, no\}}. The soundness has been explained previously.  Always use the \texttt{-ur -af} flags.
\item \texttt{MAINTRS\_lpo\_sat \{yes, no\}}.
\item \texttt{MAINTRS\_lpo\_smt \{yes, no\}}.
\item \texttt{MAINTRS\_lpo\_direct \{yes, no\}}.
\end{itemize}

The structures are also kept unchanged, meaning that we first execute \texttt{dp} and then \texttt{edg}, etc.
We may probably generate a strategy like below. The flags to \texttt{matrix} processors and the time limit of \texttt{edg} are changed. However, the overall structure remains the same, and all termination processors always employ the \texttt{-ur} and \texttt{-dp} flags.
\begin{tcolorbox}[mycodebox]
\begin{lstlisting}{text}
MAINTRS = (dp;edg[1]?;(sccs | \
 (sct || \
 {ur?;( \
  (matrix -dp -ur -dim 2 \
    -ib 4 -ob 5 | \
   matrix -dp -ur -dim 2 \
     -ib 2 -ob 3 | \
   matrix -dp -ur -dim 4 \
     -ib 2 -ob 2 | \
   matrix -dp -ur -dim 3 \ 
     -ib 1 -ob 3) || \
  (kbo -ur -af | lpo -ur -af) || \
  ARCTICTRS || \
  ARCTICBZTRS ) \
 }restore || \
 BOUNDS[2]
 ))*[8])!    
\end{lstlisting}
\end{tcolorbox}

We will subsequently explain the parameter space for \texttt{SN}. It is a sub-strategy for proving termination, which is also called \emph{strong normalization}.
\begin{tcolorbox}[mycodebox]
\begin{lstlisting}{text}
SN = (PRETRS;(MAINTRS \
  || DIRECTTRS || \
  (rev;(MAINTRS || DIRECTTRS))))
\end{lstlisting}
\end{tcolorbox}
The only parameter is \texttt{SN\_rev \{yes, no\}}. If it is \texttt{no}, \texttt{(rev;(MAINTRS || DIRECTTRS))} is not used in \texttt{SN}. Otherwise, it remains in \texttt{SN}.
It is sound as \texttt{SN\_rev} is only parallelly executed with other strategies in \texttt{SN}.
It makes one technique fail and does not affect other parallel executions.

\begin{tcolorbox}[mycodebox]
\begin{lstlisting}{text}
SNRELATIVE_STEP = ( \
 lpo -quasi || \
 (matrix -dim 1 -ib 3 -ob 4 | \
  matrix -dim 2 -ib 2 -ob 2 | \
  matrix -dim 3 -ib 1 -ob 2 | \
  arctic -dim 2 -ib 2 -ob 2) || \
 (if duplicating then fail else 
   (bounds -rt || \
     bounds -rt -qc))[1] || \
 poly -ib 2 -ob 4 -nl2 -heuristic 1)

SNRELATIVE = (SNRELATIVE_STEP[5]*)
\end{lstlisting}
\end{tcolorbox}
\texttt{SNRELATIVE\_STEP} tries to prove the relative termination of a rule.
\texttt{SNRELATIVE} repeatedly apply \texttt{SNRELATIVE\_STEP} in five seconds until all rules are proven to be relatively terminated. 
\begin{itemize}
\item \texttt{SNRELATIVE\_STEP\_time \{1, 2, 3, 4, 5, 6, 7, 8\}}. Assign an execution time for \texttt{SNRELATIVE\_STEP}. Notice that CSI's default strategy assigns five seconds. 
\item \texttt{SNRELATIVE\_lpo \{yes, no\}}. The parameters of \texttt{lpo} are the same as those of \texttt{lpo} in \texttt{DIRECTTRS}.
\item \texttt{SNRELATIVE\_matrix \{yes, no\}}.
The parameters of \texttt{matrix} are the same as those of \texttt{matrix} in \texttt{PRETRS}.
\item \texttt{SNRELATIVE\_arctic \{yes, no\}}.
The parameters of \texttt{arctic} are \texttt{ib}, \texttt{ob}, \texttt{direct}, \texttt{rat}, \texttt{real}. Their soundness have been explained for \texttt{ARCTICTRS}.

\item \texttt{SNRELATIVE\_poly \{yes, no\}}.  Applies polynomial interpretations.
\item \texttt{SNRELATIVE\_poly\_dim \{1, 2, 3, 4, 5, 6, 7, 8, 10, 12, 16\}}. Specifies the dimension of the matrices.
\item \texttt{SNRELATIVE\_poly\_direct \{yes, no\}}.  Try to finish the termination proof.
\item \texttt{SNRELATIVE\_poly\_ib \{1, 2, 3, 4, 5, 6\}}. 
Defines the number of bits that can be used to represent the minimal weight that appears in the intermediate results.
\item \texttt{SNRELATIVE\_poly\_neg \{yes, no\}}.
Allow negative numbers (only for non-linear interpretations) for some coefficients.
It is sound for non-linear interpretation according to Corollary 3.9 in~\cite{neurauter2012termination}.
The combination of \texttt{neg} flag and the default linear interpretation may cause unsoundenss.
We avoid it by using Grackle's forbidden mechanism that disallows the combination of the \texttt{neg} flag and the default linear interpretation as shown in Section~\ref{app:sec-forbid}.
The combination of \texttt{neg}, the nonlinear interpretation (\texttt{nl} or \texttt{nl2}), and the real number interpretation (\texttt{real} or \texttt{rat > 1}) is also unsound. But CSI can detect it and forbade such combinations according to the function \texttt{context} in \url{csi/src/processors/src/termination/interpretations/polynomialInterpretation.ml}.
% Moreover, CSI seems to be able to confirm the interpretations are non-negative when we use \texttt{-neg} and the default linear interpretation.
\item \texttt{SNRELATIVE\_poly\_ob \{1, 2, 3, 4, 5, 6, 7, 8, 9, 10, 12, max\}}
Defines the number of bits that can be used to represent the largest number that appears in the intermediate results.
\item \texttt{SNRELATIVE\_poly\_rat \{1, 2, 3, 4\}}. 
Sets the denumerators for rational numbers. The soundness is explained in Section 2.1.2~\cite{neurauter2012termination}.
\item \texttt{SNRELATIVE\_poly\_real \{yes, no\}}. Uses reals. 
The soundness is explained in Section 2.1.2~\cite{neurauter2012termination}.
\item \texttt{SNRELATIVE\_poly\_nl \{yes, no\}}. Allow $x^2$ + $y^2$. 
By default, linear interpretation is used, which denotes the format of $f(x_1,...,x_n)=a_0 + a_1 x_1+ \dots + a_n x_n $ where $x_i \geq 0$. 
The \texttt{-nl} flag enables the interpretation in the format of $f(x_1,...,x_n)=a_0 + a_1 x_1+ \dots + a_n x_n + a_1' x_1^2 + \dots + a_n' x_n^2$ where $x_i \geq 0$ according to Section 3.2.2 in~\cite{neurauter2012termination} and the function \texttt{quadratic} in \url{csi/src/processors/src/termination/interpretations/polynomialInterpretation.ml}.
It is sound according to~\cite{neurauter2012termination}.
\item \texttt{SNRELATIVE\_poly\_nl2 \{yes, no\}}. Allow $x^2 + x*y + y^2$.
The \texttt{-nl2} flag enables the interpretation in the format of 
\[ f(x_1,...,x_n)=a_0 + \sum_{j=1}^{n} a_j x_j + \sum_{1 \leq j \leq k \leq n}^{n} a_{jk}x_jx_k \] where $x_i \geq 0$ according to Section 3.2.2 in~\cite{neurauter2012termination} and the function \texttt{quadratic} in \url{csi/src/processors/src/termination/interpretations/polynomialInterpretation.ml}.
It is sound according to~\cite{neurauter2012termination}.
\item \texttt{SNRELATIVE\_poly\_heuristic \{-1, 0, 1, 2, 3, 4}\}. 
-1 $\rightarrow$ all symbols (default); 
0 $\rightarrow$ no symbols ;
1 $\rightarrow$ symbols appearing at most once in each left-hand side (lhs)/ right-hand side (rhs); 
2 $\rightarrow$ symbols appearing at most twice in each lhs/rhs; 
3 $\rightarrow$ symbols appearing at parallel positions in each lhs/rhs; 
4 $\rightarrow$ defined symbols.
It decides which symbols should be interpreted by non-linear polynomials and is sound according to Section 5 in~\cite{neurauter2010monotonicity}.

\end{itemize}

\begin{tcolorbox}[mycodebox]
\begin{lstlisting}{text}
MATRIXSTAR=(( \
 matrix -dim 1 -ib 2 \
   -ob 2 -strict_empty -lstar | \
 matrix -dim 2 -ib 2 -ob 2 \
   -strict_empty -lstar)[2])
\end{lstlisting}
\end{tcolorbox}
\begin{itemize}
    \item \texttt{MATRIXSTAR\_time \{1, 2, 3, 4\}}
    \item The parameter space for each \texttt{matrix} processor is the same as that for a \texttt{matrix} processor in \texttt{PRETRS}. Notice that both processors \texttt{matrix} here employ the flags \texttt{-strict\_empty -lstar}. We fix the usage of \texttt{matrix -strict\_empty -lstar} and search for the other parameters.
\end{itemize}

\begin{tcolorbox}[mycodebox]
\begin{lstlisting}{text}
MATRIXREDEX=(( \
 matrix -dim 1 -ib 2 -ob 2 \
 -strict_empty -lredex)[2])
\end{lstlisting}
\end{tcolorbox}
\begin{itemize}
    \item \texttt{MATRIXREDEX\_time \{1, 2, 3, 4, 5\}} 
    \item The parameter space for each \texttt{matrix} processor is the same as that for a \texttt{matrix} processor in \texttt{PRETRS}. Notice that both processors \texttt{matrix} here employ the flags \texttt{-strict\_empty -lredex}. We fix the usage of \texttt{-strict\_empty -lredex} and search for the other parameters.
\end{itemize}

\begin{tcolorbox}[mycodebox]
\begin{lstlisting}{text}
LDH = (shift -dd;SNRELATIVE; \
  shift -ldh -force)
LDHF = (shift -dd -force; \
  SNRELATIVE;shift -ldh -force)
SSTAR = (cr M -star;MATRIXSTAR*; \
  shift -sstar)
DUP = (cr M -dup;SNRELATIVE; \
  shift -lstar)
REDEX = (cr M -redex;MATRIXREDEX*; \
  shift -lstar)
\end{lstlisting}
\end{tcolorbox}
The strategies in the above code block are very complicated, and we cannot understand them. Therefore, we keep them unchanged and do not search for the parameters.
We only have a parameter \texttt{REDEX \{yes, no\}}.
It is sound as it is only used in  \texttt{COR3 = (REDEX; ...)!}, and \texttt{COR3} is parallelly executed with other techniques in \texttt{DD}.
It only makes \texttt{COR3} immediately fail.

\begin{tcolorbox}[mycodebox]
\begin{lstlisting}{text}
GROUND = (if ground \
  then uncurry -curry?; \
  groundcr else fail)
\end{lstlisting}
\end{tcolorbox}
The decision procedure for ground systems~\cite{felgenhauer2012deciding}.
We only have \texttt{GROUND \{yes, no\}}.
It is sound as it is only parallel executed with other confluence techniques.

\begin{tcolorbox}[mycodebox]
\begin{lstlisting}{text}
NOTCR = ( \
 (nonconfluence -steps 0 \
   -tcap -fun | \
  nonconfluence -steps 2 \
   -tcap -fun | \
  nonconfluence -steps 25 \
   -width 1 -tcap -fun | \
  nonconfluence -steps 2 \
   -idem -fun) || \
 (nonconfluence -steps 2 \
   -tcap -var | \
  nonconfluence -steps 25 \
    -width 1 -tcap -var) || \
 (nonconfluence -steps 0 \
   -tree -fun | \
  nonconfluence -steps 0 \
    -tree -var | \
  nonconfluence -steps 1 \
    -tree -fun | \
  nonconfluence -steps 1 \
    -tree -var | \
  nonconfluence -steps 2 \
    -tree -fun | \
  nonconfluence -steps 2 \ 
    -tree -var | \
  nonconfluence -steps 25 \ 
    -width 1 -tree -fun | \
  nonconfluence -steps 25 \
    -width 1 -tree -var) \
)[10]
\end{lstlisting}
\end{tcolorbox}

The \texttt{NOTCR} strategy is used to disprove confluence. It uses the \texttt{||} combinator to parallelly execute three groups of nonconfluence techinuqes. Each group contains several \texttt{nonconfluence} processors employed with different parameters. These parameters determine the search space of the \texttt{nonconfluence} processors.
To improve the execution speed, the processors using smaller search spaces are invoked before those using larger 
search spaces in each group.
We only define the parameter space for one \texttt{nonconfluence} processor due to the following reasons.
First, although the sorted decomposition technique may decompose a TRS to several sub-TRSs, we only need to disprove confluence for a sub-TRS to disprove confluence for the original TRS ~\cite{felgenhauer2015layer}.
CSI's \href{https://cops.uibk.ac.at/results/2023-full-run/TRS/CSI/695.trs/615708971.txt}{solution} in CoCo for a nonconfluence problem also shows that we only need to prove nonconfluence for a sub-TRS to disprove confluence for the original TRS.
Second, we want to invent a set of complementary strategies and then combine them in the approach explained in Section Strategy Combination in our paper. CSI's default strategy combines sequential and parallel execution to try various nonconfluence techniques and increase the execution speed. 
In contrast, our goal of defining a parameter space is simply to invent a technique suitable for a set of problems. The combination of invented strategies will be considered later.
Therefore, our parameter space will produce a strategy like \texttt{NOTCR = nonconfluence -steps 0 -tcap -fun[10]} where the execution time of \texttt{nonconfluence} and its flags are decided by Grackle.

\begin{itemize}
\item \texttt{NOTCR \{yes, no\}}. Disprove confluence. It is sound because if we set it to \texttt{fail}, the strategy simply mainly tries to discover confluence proofs.
It will not cause unexpected transformations.
\item \texttt{nonconfluence\_time \{1, 2, 4, 6, 8, 10, 12, 14, 16, 18, 20, 25, 30\}}
\item \texttt{nonconfluence\_steps \{0, 1, 2, 3, 4, 5, 6, 7, 8, 9, 10, 11, 12, 16, 25, 32\}}.
Number of rewrite steps that are performed from critical pairs to test terms nonconfluent [default: 2].
Critical pairs are explained in Section~\ref{sec:rewriting}.
It is sound as it only changes the size of the search space.
\item \texttt{nonconfluence\_width \{-1, 1, 2, 3, 4, 5, 6, 7, 8, 9, 10, 11, 12, 16\}}.
Width of search tree for rewrite sequences; -1 means unbounded [default: -1].
It is sound as it only changes the size of the search space.
\item \texttt{nonconfluence\_fun \{yes, no\}}. Use overlaps at function positions only. 
As explained in Section~\ref{sec:rewriting}, an overlap roughly means that at a certain position, a sub-term can be applied with two rewrite rules~\cite{baader1998term}.
As explained in Section~\ref{sec:rewriting}, the basic idea for disproving confluence is to discover an overlap, discover a critical pair from the overlap, and check the non-joinability of the pair.
% discover an overlap of a term and rewrite the term to two disjoinable terms using the rules in the overlap~\cite{nagele2017csi}. 
This flag only determines where to find such an overlap and thereby is sound.
\item \texttt{nonconfluence\_var \{yes, no\}}. Use overlaps at variable positions only.
As explained in the last flag, this flag only determines where to find such an overlap and thereby is sound.
\item \texttt{nonconfluence\_iter \{-1, 1, 2, 3, 4, 5, 6, 7, 8, 9, 10, 11, 12, 16\}}. Specifies the maximum number of compatibility violations that should be solved. This guarantees that the procedure always terminates. 
Otherwise, it might happen that non-confluence check does not terminate.
It is only used for tree automata technique. According to Theorem 4 in~\cite{nagele2017csi}, for a critical pair $(s,t)$,
it first tries to respectively construct compatible tree automates $\mathcal{A}_1$ and $\mathcal{A}_2$.
Then, it checks the non-joinability of term reachable from $\mathcal{A}_1$ and $\mathcal{A}_2$.
The flag uses the tree automata completion technique~\cite{korp2009match} to build tree automata, which solves compatibility violations during the constructions. If it cannot solve all compatibility violations, the flag fails and cannot disprove confluence.
According to the \texttt{filter} function in \url{csi/src/processors/src/confluence/nonconfluence.ml}, when \texttt{-iter} is -1, it tries to solve all compatibility violations. However, the construction process may not terminate and fail.
Other values only make the tree automate completion fails earlier.
% We can discover a 
% However, one compatibility violation is enough to show nonconfluence.
% It is sound as it only changes the size of the search space.

\item \texttt{nonconfluence\_tcap \{yes, no\}}. Show nonconfluence by tcap (default on). It is 
sound as it is one of the nonconfluence techniques. It is explained in Lemma 1 of~\cite{zankl2011csi}.
\item \texttt{nonconfluence\_tree \{yes, no\}}. Show nonconfluence by tree automata (default off). It is sound as it is one of the nonconfluence techniques used in \texttt{NOTCR}..  It is explained in Theorem 4 of~\cite{zankl2011csi}
\item \texttt{nonconfluence\_idem \{yes, no\}}. Show nonconfluence by idem (default off).  It is sound as it is one of the nonconfluence techniques. 
Meanwhile, according to the function \texttt{idem} in \url{csi/src/processors/src/confluence/nonconfluence.ml}, it simply checks the nonjoinability of the reducts of the two terms in a critical pair.
The nonjoinability is approximated via defined symbols.
Moreover, it is used in the original \texttt{NOTCR}.
% \textcolor{blue}{which paper}
\item \texttt{nonconfluence\_nf \{yes, no\}}. Show no unique normal forms exist by finding distinct normal forms (default off). No unique normal forms imply nonconfluence~\cite{baader1998term}; hence, the flag is sound.
\end{itemize}

\begin{tcolorbox}[mycodebox]
\begin{lstlisting}{text}
KB = (cr -kb;SN)!
RL = (rule_labeling \
  | rule_labeling -left)
DECPAR = ((shift -par; \
  decreasing -par)| \
  (shift -par -m 2; decreasing -par))
DECWLL = ((rule_labeling -left \
  -persist;decreasing)|DECPAR)
DDLAB = (LDH;(decreasing | \
  RL?;decreasing))!
\end{lstlisting}
\end{tcolorbox}

\begin{itemize}
    \item \texttt{KB \{yes, no\}}. Denote the Knuth-Bendix criterion~\cite{knuth1983simple}. 
    It is sound because it is only a technique parallelly executed with other techniques.
    \item \texttt{DECPAR \{yes, no\}}. One technique in decreasing diagrams~\cite{aoto2014proving}. We cannot entirely understand it. Thus, there is only one boolean execution-controlling parameter.
    It is sound because \texttt{DECPAR} is simply used at the end of \texttt{DECWLL}.
    \texttt{DECWLL} is only used at the end of \texttt{DD}.
    \texttt{DD} is only a technique parallelly executed with other techniques.
    Setting \texttt{DECPAR} to \texttt{fail} only makes the strategy weaker in discovering proofs. 
    \item \texttt{DECPAR\_shift\_m \{0, 1, 2, 4, 6\}}. Search for (minimal+m)-length joins. It changes the value of \texttt{-m} of the second \texttt{shift} in \texttt{DECPAR}. 
     If $m <0$, it does not search for joins. Otherwise, it first searches for the $minimal$-length of joins. After that, according to the value of \texttt{-m}, it searches for joins of length $minimal+m$.
     It is sound as it only changes the size of the search space. Meanwhile, different values of \texttt{-m} are used in \texttt{DECPAR}.
    \item \texttt{DDLAB \{yes, no\}}. One technique in decreasing diagrams. We cannot entirely understand it.
    Thus, there is only one boolean execution-controlling parameter.
\end{itemize}

\begin{tcolorbox}[mycodebox]
\begin{lstlisting}{text}
COR1 = (DUP;DDLAB)!
COR2 = (SSTAR;DDLAB)!
COR3 = (REDEX;LDH;(decreasing \
 | rule_labeling?;decreasing))!
DDWLL = (LDHF;RL?;DECWLL)!
DD = (if left-linear then \
  (COR1 || COR2 || COR3 || \
  (cr -force;DDWLL)!) else fail)!
\end{lstlisting}
\end{tcolorbox}

\begin{itemize}
\item \texttt{COR1 \{yes, no\}}. One technique in decreasing diagrams~\cite{aoto2014proving}. We cannot entirely understand it. Thus, there is only one boolean execution-controlling parameter.
\item \texttt{COR2 \{yes, no\}}. One technique in decreasing diagrams~\cite{aoto2014proving}. We cannot entirely understand it. Thus, there is only one boolean execution-controlling parameter.
\item \texttt{COR3 \{yes, no\}}. One technique in decreasing diagrams~\cite{aoto2014proving}. We cannot entirely understand it. Thus, there is only one boolean execution-controlling parameter.
\item \texttt{DD \{yes, no\}}. Techniques in decreasing diagrams~\cite{aoto2014proving}.
\end{itemize}

\begin{tcolorbox}[mycodebox]
\begin{lstlisting}{text}
CLOSED_LINEAR = (if linear \
  then cr -closed -redundant -m -1; \
    (closed -feeble \
      | closed -strongly 7) \
  else fail)
CLOSED_LEFT = (if left-linear \
 then ((cr -closed -redundant -m -1;\
    (closed -feeble \ 
     | closed -development \
     | closed -upside \
     | closed -outside)) \ 
    | cr -okui) else fail)
CLOSED = (CLOSED_LINEAR || \
 CLOSED_LEFT)!
\end{lstlisting}
\end{tcolorbox}

\begin{itemize}
\item \texttt{CLOSED \{yes, no\}}. 
Test whether the critical pairs of a TRS are strongly or development closed~\cite{huet1980confluent,nagele2016certification}.
\item \texttt{CLOSED\_LINEAR \{yes, no\}}
\item \texttt{CLOSED\_LINEAR\_strongly \{-1, 1, 3, 5, 7, 9, 11\}}.  Check critical pairs strongly closed (in $\leq n$ steps). It is sound since it only changes the number of rewrite steps before checking whether critical pairs are strongly closed.
It determines the value of \texttt{-strongly 7} in \texttt{CLOSED\_LINEAR}.
\item \texttt{CLOSED\_LEFT \{yes, no\}}. Test whether the critical pairs of a left-linear system
are development closed~\cite{van1997developing}.
\item \texttt{CLOSED\_LEFT\_closed\_strongly \{-1, 1, 3, 5, 7, 9, 11\}}. There are four \texttt{closed} processors with this flag in \texttt{CLOSED\_LEFT}, we only explain one.
\end{itemize}

\begin{tcolorbox}[mycodebox]
\begin{lstlisting}{text}
CR_AUX = (sorted -order | \
    (KB || ((((CLOSED \
      || DD) | add)2*)!)))*
KH = (cr -rt;SNRELATIVE; \
  kh -mace;CR_AUX)!
\end{lstlisting}
\end{tcolorbox}

\begin{itemize}
    \item \texttt{{CR\_AUX\_loop} \{1, 2, 3, 4, 5\}}. It determines the times of the application of \texttt{CR\_AUX}. 
    \item \texttt{KH \{yes, no\}}. Perform the confluence test for associative communicative (AC) problems by using the theorem of Klein and Hirokawa~\cite{klein2012confluence}.
    \item \texttt{KH\_mace \{yes, no\}}. Use mace4 theorem prover if available. The \texttt{yes} value is sound since \texttt{kh -mace} is used in the default configuration.
    According to the \url{csi/src/processors/src/confluence/kleinHirokawa.ml} in CSI's source code, when it is set to \texttt{no}, no theorem prover will be invoked. According to the paper~\cite{klein2012confluence}, when no theorem prover is invoked, the method merely becomes weaker in constructing confluence proofs.
\end{itemize}

\begin{tcolorbox}[mycodebox]
\begin{lstlisting}{text}
AT1 = (at -theorem 1; SN)!
AT2 = (at -theorem 2; SN)!
AT3 = (at -theorem 3; SN)!
AT = (AT2 || AT3)
\end{lstlisting}
\end{tcolorbox}

\begin{itemize}
\item \texttt{AT  \{yes, no\}}. The confluence test for associative-communicative (AC) problems by using the theorems of Aoto and Toyama~\shortcite{aoto2012reduction}.

\item \texttt{AT2 \{yes, no\}} 
\item \texttt{AT2\_theorem \{1, 2, 3\}} Indicates which of the three theorems is used. By default, theorem 3 will be used. The three theorems are explained in~\cite{aoto2012reduction} and are all sound. The value \texttt{1}, \texttt{2}, and \texttt{3} respectively correspond to Theorem 3.8, Theorem 3.18, and Theorem 3.28 in~\cite{aoto2012reduction},
\item \texttt{AT2\_bound \{1, 2, 4, 8, 12, 14, 16, 24\}}.
Indicates an upper bound for the number of rewrite rules. If the number of rewrite rules is $>= b$, then the 
processor ends and fails. By default, $b = 12$ will be used.
It is sound as it only changes the size of the search space.
\item \texttt{AT3 \{yes, no\}}
\item \texttt{AT3\_theorem \{1, 2, 3\}}
\item \texttt{AT3\_bound \{1, 2, 4, 8, 12, 14, 16, 24\}}
\end{itemize}

\begin{tcolorbox}[mycodebox]
\begin{lstlisting}{text}
CPCS = (cr -cpcs; SNRELATIVE; \
 shift -lstar)
CPCS2 = (cr -cpcs2; SN)!
\end{lstlisting}
\end{tcolorbox}
\begin{itemize}
\item \texttt{CPCS \{yes, no\}}. 
The processor \texttt{cr -cpcs} computes the critical pair closing system by Theorem 2.4 in~\cite{oyamaguchi2014confluence}. 
It is sound since if it is set to \texttt{fail}, it simply does not transform the problem in 
\texttt{AUTO\_INNER}.
\item \texttt{CPCS2 \{yes, no\}}.
The processor \texttt{cr -cpcs2} computes the critical pair closing system by Theorem 2.11 in~\cite{oyamaguchi2014confluence}.
It is sound since it is parallelly executed with other techniques in \texttt{AUTO\_INNER0}.
\end{itemize}

\begin{tcolorbox}[mycodebox]
\begin{lstlisting}{text}
REDUNDANT_JS = (( \
 cr -force -redundant); \
 (redundant))
REDUNDANT_RHS = ( \
 (cr -m -1 -force -redundant); \
 (redundant -rhs))
REDUNDANT_DEL = ((cr -m -1 -force); \
 (redundant -remove 4))
\end{lstlisting}
\end{tcolorbox}

A group of redundant rule techniques. \texttt{REDUNDANT\_FC} is used in nonconfluence analysis, while the other three are mainly used for confluence analysis.
\begin{itemize}
\item \texttt{REDUNDANT\_JS\_cr\_m \{-1, 0, 1, 2, 3, 4, 5\}}. Search for (minimal+m)-length joins.
For a critical pair ($s,t$),  
According to \url{csi/src/processors/src/transformation/redundant.ml} and \url{csi/src/rewriting/src/rewrite.ml}, 
When $m=-1$, only critical pairs are returned, and no joins will be found. 
When $m = 0$, it tries to discover the minimal number $M$ of rewrite steps, such that $\exists u, s \rightarrow^M u \land t \rightarrow^M u$. Then joins reachable from $M$ steps are returned.
When $m > 0$, joins reachable from $M + m$ steps are returned.
Some redundant rule techniques use joins of critical pairs to generate redundant rules~\cite{nagele2015improving}.
This parameter is sound since it only controls the length of joins to be generated.
\item \texttt{REDUNDANT\_JS\_redundant\_size \{-1, 1, 2, 3, 4, 5, 6, 7, 8, 9, 10, 11, 12, 16, 32\}}.
Only add rules whose size is less than n (default: -1, i.e., unrestricted). It is sound since values other than the default merely limit the number of redundant rules to generate.
\item \texttt{REDUNDANT\_RHS\_cr\_m \{-1, 0, 1, 2, 3, 4, 5\}}
\item \texttt{REDUNDANT\_RHS\_redundant\_size \{-1, 1, 2, 3, 4, 5, 6, 7, 8, 9, 10, 11, 12, 16, 32\}}
\item \texttt{REDUNDANT\_DEL\_cr\_m \{-1, 0, 1, 2, 3, 4, 5\}}
\item \texttt{REDUNDANT\_DEL\_redundant\_js \{yes, no\}}. Add joining sequences of critical peaks as rules. It is sound as explained in Collary 6 and Section 5 in~\cite{nagele2015improving}.
\item \texttt{REDUNDANT\_DEL\_redundant\_development \{-1, 1, 2, 3, 4, 5, 6\}}. Add rules to make critical peaks development closed.
 It is sound as explained in Collary 6 and Section 5 in ~\cite{nagele2015improving}.
\item \texttt{REDUNDANT\_DEL\_redundant\_rhs \{yes, no\}}. Add rules by rewriting right-hand sides 1 step.
 It is sound as explained in Collary 6 and Section 5 in ~\cite{nagele2015improving}.
\item \texttt{REDUNDANT\_DEL\_redundant\_remove \{-1, 1, 2, 3, 4, 5, 6, 7, 8, 9, 10\}}. 
Remove rules whose left- and right-hand sides are joinable in $n$ steps. By default, it is -1, meaning no limitations on the number of rewrite steps.
The parameter is sound as other values are weaker than the default in removing rules.
% merely limit the number of rules to remove.
\item \texttt{REDUNDANT\_DEL\_redundant\_reverse \{-1, 1, 2, 3, 4, 5, 6\}}.  Add reversible rules.
 It is sound as explained in Collary 6 and Section 5 in ~\cite{nagele2015improving}.
\item \texttt{REDUNDANT\_DEL\_redundant\_size \{-1, 1, 2, 3, 4, 5, 6, 7, 8, 9, 10, 11, 12, 16, 32\}}
\end{itemize}

\begin{tcolorbox}[mycodebox]
\begin{lstlisting}{text}
REDUNDANT_FC = ((cr -m -1 -force); \
  (redundant -narrowfwd \
   -narrowbwd -size 7))
\end{lstlisting}
\end{tcolorbox}

% In the default strategy,  is used to
\begin{itemize}
\item \texttt{REDUNDANT\_FC \{yes, no\}}. It is sound because if it is set to \texttt{fail}, it simply does not do the redundant rule transformation for non-confluence analysis.
\item \texttt{REDUNDANT\_FC\_cr\_m \{-1, 0, 1, 2, 3, 4, 5\}}. The soundness has been explained before.
\item \texttt{REDUNDANT\_FC\_redundant\_js \{yes, no\}}. The soundness has been explained before.
\item \texttt{REDUNDANT\_FC\_redundant\_development \{-1, 1, 2, 3, 4, 5, 6\}}. The soundness has been explained before.
\item \texttt{REDUNDANT\_FC\_redundant\_rhs \{yes, no\}}. The soundness has been explained before.
\item \texttt{REDUNDANT\_FC\_redundant\_narrowfwd \{yes, no\}}.
Use narrowing forwards to generate new rules.
It is sound since it is in the original \texttt{REDUNDANT\_FC}.
\item \texttt{REDUNDANT\_FC\_redundant\_narrowbwd \{yes, no\}}.
Use narrowing backwards to generate new rules.
It is sound since it is in the original \texttt{REDUNDANT\_FC}.
\item \texttt{REDUNDANT\_FC\_redundant\_size \{-1, 1, 2, 3, 4, 5, 6, 7, 8, 9, 10, 11, 12, 16, 32\}}. 
The soundness has been explained before.
\end{itemize}

\begin{tcolorbox}[mycodebox]
\begin{lstlisting}{text}
SIMPLE = FULL( \
  (if right-linear \
   then if left-linear -ie \
    then if strongly-non-overlapping 
      then succ -reason \
       ToyamaOyamaguchi95Cor22 
      else fail else fail else fail)|\
  (if collapsing then fail else 
    if shallow -ws then 
      if strongly-non-overlapping then 
        succ -reason \
         SakaiOyamaguchiOgawa14 
      else fail else fail) | \
  fail)
\end{lstlisting}
\end{tcolorbox}
Simple criteria for proving confluence~\cite{sakai2015non,toyama1994church}.
The \texttt{FULL} keyword is a trick in the competition strategy for easily taking part in different categories of competitions in CoCo. 
For us, it means nothing and can be ignored.
The only parameter is \texttt{SIMPLE \{yes, no\}}. 

\begin{tcolorbox}[mycodebox]
\begin{lstlisting}{text}
AC_SN = ((acrpo \
 ||  ackbo -ib 3 -ob 5 -q -nt -sc  \
 || ackbo -ib 3 -ob 5 -kv2)*[10])
AC = (cr -ac;AC_SN)!
\end{lstlisting}
\end{tcolorbox}

\begin{itemize}
\item \texttt{AC \{yes, no\}}. Main techniques for proving confluence for AC problems.
\item \texttt{AC\_time \{1, 2, 4, 6, 8, 10, 15\}}
\item \texttt{AC\_acrpo \{yes, no\}}.  Applies AC-Recursive Path Order~\cite{yamada2016ac}.
\item \texttt{AC\_acrpo\_direct \{yes, no\}}
\item \texttt{AC\_acrpo\_sat \{yes, no\}}
\item \texttt{AC\_acrpo\_smt \{yes, no\}}
\item \texttt{AC\_ackbo \{yes, no\}}. Apply standard, Korovin/Voronkov's, and Steinbachs AC-KBO~\cite{yamada2016ac}.
There are two \texttt{ackbo} processors in \texttt{AC\_SN}, we only explain the parameter for one for simplification.
The flags \texttt{-kv}, \texttt{kv2}, and \texttt{st} respectively correspond to Korovin \&Voronkov, KV', and Steinbach methods in Table 1 of~\cite{yamada2016ac}.
When none of \texttt{-kv}, \texttt{kv2}, and \texttt{st} is used, by default,  \texttt{ackbo} uses the AC-KBO method in Table 1.
\item \texttt{AC\_ackbo\_ac0 \{yes, no\}}. In case of Steinbach's order, give AC symbols weight 0. Its soundness is explained in Theorem 3.3~\cite{yamada2016ac} and the related footnote.
\item \texttt{AC\_ackbo\_direct \{yes, no\}}. Try to finish the termination proof.
\item \texttt{AC\_ackbo\_ib \{1, 2, 3, 4, 5, 6\}}.
Defines the number of bits that can be used to represent the smallest number that appears in the intermediate results.
\item \texttt{AC\_ackbo\_kv2 \{yes, no\}}. Use corrected Korovin and Voronkov's ordering. 
The paper~\cite{yamada2016ac} finds a bug in the original Korovin and Voronkov's ordering and has corrected it.
\item \texttt{AC\_ackbo\_ob \{1, 2, 3, 4, 5, 6, 7, 8, 9, 10, 12, max\}}.
Defines the number of bits that can be used to represent the largest number that appears in the intermediate results.
\item \texttt{AC\_ackbo\_q \{yes, no\}}. Uses quasi-precedences. According to the annotation of the function \texttt{quasi\_adm} in \url{csi/src/processors/src/termination/orderings/ackbo.ml} in CSI's source code, here, quasi-precedences mean that AC-symbols never equal to non-AC symbols.
Using the flag only makes the order more strict and thereby is sound. Moreover, the \texttt{ackbo} processors have both use and do not use \texttt{-q} in \texttt{AC\_SN}.  
\item \texttt{AC\_ackbo\_sat \{yes, no\}}. Uses SAT backend (default).
\item \texttt{AC\_ackbo\_sc \{yes, no\}}. Uses subterm coefficients. It is sound since \texttt{ackbo} both employs it and ignores it in \texttt{AC\_SN}.  Moreover, its soundness is explained in Theorem 8.2 in~\cite{yamada2016ac}.
% It is sound since \texttt{sc} has both been used and ignored by \texttt{ackbo} in \texttt{AC\_SN}. 
\item \texttt{AC\_ackbo\_smt \{yes, no\}}. Uses SMT backend.
\item \texttt{AC\_ackbo\_st \{yes, no\}}. Use Steinbach's ordering. Its soundness is explained in Theorem 3.3 in~\cite{yamada2016ac}.
\item \texttt{AC\_ackbo\_nt \{yes, no\}}. Allow non-total precedences (not compatible with \texttt{-kv}). 
Precedences are defined between function symbols or constants.
It is essential as explained in Example 5.11, sometimes we need to make two function symbols incomparable to obtain a working order.
We use Grackle's forbidden mechanism to forbidde both \texttt{-kv} and \texttt{-nt} are employed by \texttt{ackbo}. According to \texttt{AC\_SN}, the combination of \texttt{-nt} and AC-KBO is sound. According to Definition 4.1 and Example 5.11,  the combination of \texttt{-nt} and \texttt{kv2} is sound.  According to Definition 3.1, the combination of \texttt{-nt} and \texttt{st} is sound.
\end{itemize}

\begin{tcolorbox}[mycodebox]
\begin{lstlisting}{text}
AUTO_INNER0_DEL = (REDUNDANT_DEL?; \
 (CLOSED || DD || SIMPLE || KB \
  || AC || {GROUND}nono))3*!

AUTO_INNER0_CLOSED_DD_REDUNDANT =
 (((CLOSED || DD) \
   | REDUNDANT_RHS)3*! || \
  (CLOSED || DD) \
   | REDUNDANT_JS)3*! 
   
AUTO_INNER0 = (GROUND || KB || AC || \
 AUTO_INNER0_DEL || \
 AUTO_INNER0_CLOSED_DD_REDUNDANT || \
 KH || AT || SIMPLE || CPCS2)
\end{lstlisting}
\end{tcolorbox}
As explained in Section~\ref{sec:param}, \texttt{AUTO\_INNER0} parallelly executes a set of techniques.
When searching for parameters, we group some techniques in \texttt{AUTO\_INNER0} into \texttt{AUTO\_INNER0\_DEL} and \texttt{AUTO\_INNER0\_CLOSED\_DD\_REDUNDANT}. 
We perform this modification since techniques inside the two groups have two common features that not exist in
the other techniques in \texttt{AUTO\_INNER0}. 
Grouping them is helpful for using a boolean execution-controlling flag to determine whether to execute the two groups of techniques.
The first distinct feature is that both groups utilize redundant rule techniques.
Moreover, both groups consist of multiple techniques which follow a specific invocation procedure.

\begin{itemize}
\item \texttt{AUTO\_INNER0 \{yes, no\}}.
\item \texttt{AUTO\_INNER0\_time \{2, 4, 6, 8, 12, 16, 30, 60\}}. Control the running time of \texttt{AUTO\_INNER0}.
\item \texttt{AUTO\_INNER0\_DEL \{yes, no\}}. 
% We group the sub-strategies following \texttt{REDUNDANT\_DEL} as \texttt{AUTO\_INNER0\_DEL}.
% This means \texttt{}
\item \texttt{AUTO\_INNER0\_DEL\_loop \{1, 2, 3, 4, 5, 6, 7\}}
\item \texttt{AUTO\_INNER0\_GROUND \{yes, no\}}. Control whether to execute \texttt{GROUND} in \texttt{AUTO\_INNER0}.
\item \texttt{AUTO\_INNER0\_KB \{yes, no\}}.
 Control whether to execute \texttt{KB} in \texttt{AUTO\_INNER0}.
\item \texttt{AUTO\_INNER0\_AC \{yes, no\}}.
 Control whether to execute \texttt{AC} in \texttt{AUTO\_INNER0}.
\item \texttt{AUTO\_INNER0\_CLOSED\_DD\_REDUNDANT \{yes, no\}}
\item \texttt{AUTO\_INNER0\_CLOSED\_DD\_REDUNDANT\_loop  \{1, 2, 3, 4, 5, 6, 7\}}. 
Control the repeated application times of \texttt{((CLOSED || DD) | REDUNDANT\_RHS)3*! || (CLOSED || DD) | REDUNDANT\_JS)}. 
\item \texttt{AUTO\_INNER0\_CLOSED\_DD\_REDUNDANT\_inner \_loop  \{1, 2, 3, 4, 5, 6, 7\}}.
Control the repeated application times of \texttt{(CLOSED || DD) | REDUNDANT\_RHS}.
\item \texttt{AUTO\_INNER0\_KH \{yes, no\}}.
 Control whether to execute \texttt{KH} in \texttt{AUTO\_INNER0}.
\item \texttt{AUTO\_INNER0\_AT \{yes, no\}}.
 Control whether to execute \texttt{AT} in \texttt{AUTO\_INNER0}.
\item \texttt{AUTO\_INNER0\_SIMPLE \{yes, no\}}.
 Control whether to execute \texttt{SIMPLE} in \texttt{AUTO\_INNER0}.
\item \texttt{AUTO\_INNER0\_CPCS2 \{yes, no\}}.
Control whether to execute \texttt{CPCS2} in \texttt{AUTO\_INNER0}.
\end{itemize}

\begin{tcolorbox}[mycodebox]
\begin{lstlisting}{text}
AUTO_INNER = (AUTO_INNER0[30] \
  | CPCS[5]2*)! \
    | ({AUTO_INNER0[30]}nono \
      | CPCS[5]2*)2*!
\end{lstlisting}
\end{tcolorbox}

\begin{itemize}
\item \texttt{AUTO\_INNER \{yes, no\}}
\item \texttt{AUTO\_INNER\_CPCS\_time1 \{1, 2, 3, 4, 5, 6, 7, 8, 9\}}. 
Control the running time of the first \texttt{CPCS}.
\item \texttt{AUTO\_INNER\_CPCS\_time2 \{1, 2, 3, 4, 5, 6, 7, 8, 9\}}
Control the running time of the second \texttt{CPCS}.
\item \texttt{AUTO\_INNER\_CPCS\_loop1 \{1, 2, 3, 4, 5, 6, 7\}}.
Control the number of repeated application times of the first \texttt{CPCS}.
\item \texttt{AUTO\_INNER\_CPCS\_loop2 \{1, 2, 3, 4, 5, 6, 7\}}.
Control the number of repeated application times of the second \texttt{CPCS}.
\item \texttt{AUTO\_INNER\_loop \{1, 2, 3, 4, 5, 6, 7\}}.
Control the number of repeated application times of the compound processor \texttt{(AUTO\_INNER0[30] | CPCS[5]2*)! | ({AUTO\_INNER0[30]}nono | CPCS[5]2*)}.

\end{itemize}

\begin{tcolorbox}[mycodebox]
\begin{lstlisting}{text}
AUTO = (if trs then (\
 sorted -order*;(AUTO_INNER \
 || (NOTCR | REDUNDANT_FC)3*!) \
 ) else fail)
\end{lstlisting}
\end{tcolorbox}

\begin{itemize}
\item \texttt{AUTO\_sorted\_order \{yes, no\}}. Decompose a problem due to sorted information~\cite{felgenhauer2015layer}. The flag \texttt{-order} tries order-sorted decomposition.
\item \texttt{AUTO\_sorted\_ms \{yes, no\}}. Try many-sorted decomposition. It is weaker than \texttt{-order} in decomposing TRSs~\cite{felgenhauer2015layer}. 
\item \texttt{NOTCR\_loop \{1, 2, 3, 4, 5, 6, 7, 8, 9, 10\}}. Control the number of repeated applications of 
\texttt{(NOTCR | REDUNDANT\_FC)3*!}, default 3.
\end{itemize}

\subsection{Forbidden Parameters}
\label{app:sec-forbid}
Grackle can forbid the occurrence of certain parameters.
We have listed the parameters we forbid below. We forbid them either to confirm soundness or to reduce the size of the parameter search space.
\begin{tcolorbox}[mycodebox]
\begin{lstlisting}{text}
{AC_ackbo1_nt=yes, AC_ackbo1_kv=yes}
{AC_ackbo2_nt=yes, AC_ackbo2_kv=yes}
{MAINTRS_edg_nl=yes, 
  MAINTRS_edg_gtcap=no}
{AUTO_INNER_sorted_order=no, 
  AUTO_INNER_sorted_ms=no}
{DIRECTTRS_kbo_pbc=no, 
  DIRECTTRS_kbo_eq=yes}
{DIRECTTRS_kbo_pbc=no, 
  DIRECTTRS_kbo_minp=yes}
{DIRECTTRS_kbo_pbc=no, 
  DIRECTTRS_kbo_minw=yes}
{DIRECTTRS_kbo_sat=no, 
  DIRECTTRS_kbo_smt=no, 
  DIRECTTRS_kbo_rat=2}
{DIRECTTRS_kbo_sat=no, 
  DIRECTTRS_kbo_smt=no, 
  DIRECTTRS_kbo_rat=3}
{DIRECTTRS_kbo_sat=no, 
  DIRECTTRS_kbo_smt=no, 
  DIRECTTRS_kbo_rat=4}
{MAINTRS_kbo_sat=no, 
  MAINTRS_kbo_smt=no,
  MAINTRS_kbo_rat=2}
{MAINTRS_kbo_sat=no, 
  MAINTRS_kbo_smt=no, 
  MAINTRS_kbo_rat=3}
{MAINTRS_kbo_sat=no, 
  MAINTRS_kbo_smt=no, 
  MAINTRS_kbo_rat=4}
{SNRELATIVE_poly_neg=yes, 
  SNRELATIVE_poly_nl=no, 
  SNRELATIVE_poly_nl2=no}
\end{lstlisting}
\end{tcolorbox}

\section{Examples of Invented Strategy}
% We show two examples of invented strategies this section.
Figure~\ref{fig:csi-0d232d} presents the invented strategy csi-0d232dbb588232c4fa2a8db3585ab8b2d0c28c44bdbcfb555-98ae901.
It follows the basic structure of the original competition strategy.
Boolean-execution controlling flags disable some sub-strategies be replacing their definitions to \texttt{fail}.
In \texttt{SNRELATIVE\_STEP}, the values of \texttt{-ib}, \texttt{-ob}, and \texttt{-dim} are chosen by Grackle, which differs from those in the original \texttt{SNRELATIVE\_STEP} in Seciton~\ref{sec:default}.

\begin{figure*}
\caption{The invented strategy csi-0d232dbb588232c4fa2a8db3585ab8b2d0c28c44bdbcfb55598ae901}%
\label{fig:csi-0d232d}%
\begin{lstlisting}{text}
AUTO = (if trs then (sorted -order*; \
  (AUTO_INNER || (NOTCR | REDUNDANT_FC)3*!)) else fail)

AUTO_INNER = (AUTO_INNER0[30] | CPCS[5]2*)! \
  | ({AUTO_INNER0[30]}nono | CPCS[5]2*)2*!

AUTO_INNER0 = (AUTO_INNER0_GROUND || AUTO_INNER0_KB || AUTO_INNER0_AC \
  || AUTO_INNER0_DEL || AUTO_INNER0_CLOSED_DD_REDUNDANT || AUTO_INNER0_KH \
  || AUTO_INNER0_AT || AUTO_INNER0_SIMPLE || AUTO_INNER0_CPCS2 || fail)

AUTO_INNER0_AC = fail

AUTO_INNER0_AT = fail

AUTO_INNER0_CLOSED_DD_REDUNDANT = fail

AUTO_INNER0_CPCS2 = fail

AUTO_INNER0_DEL = fail

AUTO_INNER0_GROUND = fail

AUTO_INNER0_KB = fail

AUTO_INNER0_KH = fail

AUTO_INNER0_SIMPLE = SIMPLE

CPCS = (cr -cpcs; SNRELATIVE; shift -lstar)

NOTCR = fail

REDUNDANT_FC = fail

SIMPLE = FULL((if right-linear then if left-linear -ie then \
    if strongly-non-overlapping then succ -reason ToyamaOyamaguchi95Cor22 \
  else fail else fail else fail) 
  | (if collapsing then fail else if shallow -ws then \
    if strongly-non-overlapping then succ -reason SakaiOyamaguchiOgawa14 \
    else fail else fail) | fail)

SNRELATIVE = (SNRELATIVE_STEP[5]*)

SNRELATIVE_STEP = (lpo -quasi \
  || (matrix -ib 6 -ob 6 | matrix -dim 2 -ib 2 -ob 2 \
    | matrix -dim 3 -ob 2 | arctic -dim 2 -ib 2 -ob 8) \
  || (if duplicating then fail else \
    (bounds -rt || bounds -rt -qc))[1] \
  || poly -heuristic 1 -ib 2 -nl2 -ob 4)
\end{lstlisting}
\end{figure*}

\section{Grackle's Initial Strategies}
As explained in Section~\ref{sec:default}, \texttt{AUTO\_INNER0} parallelly executes a set of strategies mainly for confluence. We separate each of them as an initial strategy. This means we use the default competition strategy except that we change \texttt{NOTCR} to \texttt{fail} and \texttt{REDUNDANT\_FC} to \texttt{fail} via parameters defined in Section~\ref{sec:param}. Moreover, we respectively change the definition of \texttt{AUTO\_INNER0} to one of the items below and obtain nine initial strategies.
\begin{itemize}
    \item \texttt{AUTO\_INNER0 = GROUND}
    \item \texttt{AUTO\_INNER0 = KB} 
    \item \texttt{AUTO\_INNER0 = AC} 
    \item \texttt{AUTO\_INNER0 = AUTO\_INNER0\_DEL}
    \item \texttt{AUTO\_INNER0 = AUTO\_INNER0\_CLOSED\_DD\_REDUNDANT}
    \item \texttt{AUTO\_INNER0 = KH} 
    \item \texttt{AUTO\_INNER0 = AT }
    \item \texttt{AUTO\_INNER0 = SIMPLE}
    \item \texttt{AUTO\_INNER0 = CPCS2}
\end{itemize}

We also separate each strategy in the nonconfluence analysis strategy \texttt{NOTCR} as an initial strategy. 
This means we use the default competition strategy except that we change \texttt{AUTO\_INNER} to \texttt{fail} using parameters defined in Section~\ref{sec:param}. 
Meanwhile, we respectively change the definition of \texttt{NOTCR} to one of the items below and obtain 14 initial strategies.

\begin{itemize}
\item \texttt{NOTCR = nonconfluence -steps 0 -tcap -fun[10]}
\item \texttt{NOTCR = nonconfluence -steps 2 -tcap -fun[10]}
\item \texttt{NOTCR = nonconfluence -steps 25 -width 1 -tcap -fun[10]}
\item \texttt{NOTCR = nonconfluence -steps 2 -idem -fun[10]}
\item \texttt{NOTCR = nonconfluence -steps 2 -tcap -var[10]}
\item \texttt{NOTCR = nonconfluence -steps 25 -width 1 -tcap -var[10]}
\item \texttt{NOTCR = nonconfluence -steps 0 -tree -fun[10]}
\item \texttt{NOTCR = nonconfluence -steps 0 -tree -var[10]}
\item \texttt{NOTCR = nonconfluence -steps 1 -tree -fun[10]}
\item \texttt{NOTCR = nonconfluence -steps 1 -tree -var[10]}
\item \texttt{NOTCR = nonconfluence -steps 2 -tree -fun[10]}
\item \texttt{NOTCR = nonconfluence -steps 2 -tree -var[10]}
\item \texttt{NOTCR = nonconfluence -steps 25 -width 1 -tree -fun[10]}
\item \texttt{NOTCR = nonconfluence -steps 25 -width 1 -tree -var[10]}
\end{itemize}

To generate the dataset, we further add two other strategies besides the initial strategies.
This means we use the default competition strategy except that we change \texttt{NOTCR} to \texttt{fail} and \texttt{REDUNDANT\_FC} to \texttt{fail}. Moreover, we respectively change the definition of \texttt{AUTO\_INNER0} to one of the items below and obtain two strategies. 
% \textcolor{blue}{reason}
\begin{itemize}
    \item \texttt{AUTO\_INNER0 = CLOSED}
    \item \texttt{AUTO\_INNER0 = DD} 
\end{itemize}
We particularly extract them since \texttt{CLOSED} and \texttt{DD} are combined with redundant rule techniques in \texttt{AUTO\_INNER0\_DEL} and \texttt{AUTO\_INNER0\_CLOSED\_DD\_REDUNDANT} among the initial strategies. 
Moreover, \texttt{AUTO\_INNER0\_DEL} and \texttt{AUTO\_INNER0\_CLOSED\_DD\_REDUNDANT} parallelly execute  \texttt{CLOSED} and \texttt{DD}.
If we do not use them for labeling, we will not be able to understand whether a TRS is mastered only by \texttt{CLOSED} or \texttt{DD}.

% The initial st
In the augmented dataset, we notice eight initial strategies can only prove confluence, and 14 initial strategies can only prove non-confluence. One can prove both.
\section{Strategy Combination}

To combine several strategies, we first assign a time limit for each using the method in Section 3.2. Each strategy is written into a document. 
We use a Python program to invoke CSI with each strategy document within the assigned time limit.
The advantage of writing each strategy into a document is avoiding naming conflict since our strategy invention does not change the name of strategies. We only change the parameters.
Every strategy document is still invoked from the definition \texttt{AUTO} via Python.
% avoid the naming conflicts of sub-strategies in the same document.

When we choose a strategy, we make sure this strategy can solve the largest number of problems in the training data unsolvable by previously chosen strategies.

When we run experiments on the ARI-COPS database and use four CPUs for each CSI execution,
the assigning time limit for each strategy is presented below in the format of \textit{strategy, time by seconds}. 
\begin{itemize}
    \item csi-0ccc287f955294ea83afdc800030b453a8e37b1ee371\-
    57d83dcf04eb, 0.5
    \item csi-9f19c01a538808477a8980f80d3d2face7303ce6f058\-
    f8b6170f9461, 0.5
    \item csi-224dc655e2c823145f6a545fa78601d91e647dcfeffb\-
    317d8f57b340 0.5
    \item csi-2ebf247f07a5706eaf7a7a399dcd47bb52bbb40694af\-
    808ec8624cf1, 0.5
    \item csi-5deb4704be2267c6724151a6417770ceb5ffc4799560\-
    7bbf3a169c28, 1
    \item csi-c3e1c6423a6b61ebf5d3b3c3999fca2d098d4461b23a4\-
    b50d77171b3, 2
    \item csi-af40c043b23e299af5c85347534fe62d2c963a8f7d6b0\-
    dac873b8846, 3
    \item csi-5deb4704be2267c6724151a6417770ceb5ffc47995607\-
    bbf3a169c28, 3
    \item csi-a4390ec528e4f6c2616765939087c14c1ca0bd47cf1dd\-
    d3b6d055ba0, 3
    \item csi-112ecf2d9f970dad13f1bb4e09c25c2b51385f7324062\-
    deb49084b46, 5
    \item csi-5c115df06bbafa913b50c0fd62a28f53e6739d1358646\-
    b7f3cea19bf, 6
    \item csi-14c86024498e9a3ce1b4b79795e5ab8f6c49a19f5df8c\-
    924ce82f577, 6
    \item csi-ec0fb3f8ce9f2d586b23f5cd9c4a399845157f2854ef5\-
    f8e2f9d3f3a, 6
    \item csi-af40c043b23e299af5c85347534fe62d2c963a8f7d6b0\-
    dac873b8846, 7
    \item csi-3f5a5b338144451e2d7e22bcc184940e796a74510ec87\-
    4966c88b93d, 8
    \item csi-fc9509efa9ad65cb01f89548200a9ae5480ef616e48df\-
    fb3e0d790a2, 8
\end{itemize}

When using one CPU, the time splits are presented below
\begin{itemize}
    \item csi-1789a05ac5304d685e89b710f75edd8a273fc474526-6a3c7faadba34, 0.5,
    \item csi-c3e1c6423a6b61ebf5d3b3c3999fca2d098d4461b23-a4b50d77171b3, 0.5,
    \item csi-075d10c4a77649791eac301cda263f977e1a66186e7-20906f67a2c34, 0.5,
    \item csi-ddc25446660daf33870ef6991406ec360404a5af909-928d5ea21794f, 0.5,
    \item csi-6c8b5e7b4af3c2970275389b01434b25395a13d8f72-4e4a8aba48e24, 0.5,
    \item csi-b3ab9764cd4f0717a3037a5849e47d41c56af511e84-d4a72659e0829, 0.5,
    \item csi-0ccc287f955294ea83afdc800030b453a8e37b1ee37-157d83dcf04eb, 0.5,
    \item csi-224dc655e2c823145f6a545fa78601d91e647dcfeff-b317d8f57b340, 0.5,
    \item csi-a8f7d2a391815f9c401550acbb694bae346f443a43a-098c5e072aae9, 0.5,
    \item csi-e2877b030e4118e870207e14ef1a44e240c33e16a0e-79c8f57adaacd, 0.5,
    \item csi-ddc25446660daf33870ef6991406ec360404a5af909-928d5ea21794f, 1,
    \item csi-af40c043b23e299af5c85347534fe62d2c963a8f7d6-b0dac873b8846, 4,
    \item csi-d56e502247dfc3854f0a5360649b5be5357e2b60693-fdfdc40dc5527, 8,
    \item csi-0e7e3ba5c042fabdf8151556dc8b26717d98bcd49c2-3193ec7f29d33, 9,
    \item csi-36920a8f429f37ae80839d40e3cd805ef096f1aafbb-2dd6d7b845531, 10,
    \item csi-5deb4704be2267c6724151a6417770ceb5ffc479956-07bbf3a169c28, 11,
    \item csi-33d868984681a7cc4455c15daaa16bf675ba3056c6f-f1f186057216e, 12
\end{itemize}

When using one CPU for the augmented dataset, the time splits are presented below
\begin{itemize}
    \item csi-04117237bacc588c78fade8bc184e29c2f22ce95f0f-3d8fc051128e4, 0.5
    \item csi-a4390ec528e4f6c2616765939087c14c1ca0bd47cf1-ddd3b6d055ba0, 0.5
    \item csi-379e7f8304b587a34081a91ae8a1624f0fb9a83ef83-2090b72713d9f, 0.5
    \item csi-931e7ced256711e95501f485d5b130953464871eebe-6e41dd72957e9, 0.5
    \item csi-c3e1c6423a6b61ebf5d3b3c3999fca2d098d4461b23-a4b50d77171b3, 0.5
    \item csi-af40c043b23e299af5c85347534fe62d2c963a8f7d6-b0dac873b8846, 0.5
    \item csi-224dc655e2c823145f6a545fa78601d91e647dcfeff-b317d8f57b340, 0.5
    \item csi-5d8720a14f25d32a3a4cecd80b5cc818cfafce6f753-b2acd33280292, 0.5
    \item csi-aecfb984dbfbfe0cb3ba80326dcc23b15a9c2f0a01c-1e9620916c9e4, 0.5
    \item csi-0ccc287f955294ea83afdc800030b453a8e37b1ee37-157d83dcf04eb, 0.5
    \item csi-f660fd4bbdf8ee1f7e501b27bc2a8223d96e0f47cda-55d550d72973c, 0.5
    \item csi-f660fd4bbdf8ee1f7e501b27bc2a8223d96e0f47cda-55d550d72973c, 0.5
    \item csi-f660fd4bbdf8ee1f7e501b27bc2a8223d96e0f47cda-55d550d72973c, 0.5
    \item csi-f660fd4bbdf8ee1f7e501b27bc2a8223d96e0f47cda-55d550d72973c, 0.5
    \item csi-aa24b5e64bf83e78a707b01ad35bc3b6e22e06f9402-a758631eee1bb, 1
    \item csi-af40c043b23e299af5c85347534fe62d2c963a8f7d6-b0dac873b8846, 6
    \item csi-f1d55b73677f250ea72db5a2658ffe0f92fc0197ea7-9717d372870d1, 7
    \item csi-1bccc2cf890e2f92f021b591ec34371ebdd2a68741d-08de7fff0924e, 8
    \item csi-5deb4704be2267c6724151a6417770ceb5ffc479956-07bbf3a169c28, 9
    \item csi-a4390ec528e4f6c2616765939087c14c1ca0bd47cf1-ddd3b6d055ba0, 10
    \item csi-36920a8f429f37ae80839d40e3cd805ef096f1aafbb-2dd6d7b845531, 12
\end{itemize}

When using four CPU for the augmented dataset, the time splits are presented below
\begin{itemize}
\item    (csi-1f0dd05ed5c19b06c8c11b6dda27bd743c3b691c8f-f22150acb2c5e3, 0.5),
\item    (csi-95e8deb6c1089354822fc19425ea6cab098fff49ea-9ec5afb65da9da, 0.5),
\item    (csi-f660fd4bbdf8ee1f7e501b27bc2a8223d96e0f47cd-a55d550d72973c, 0.5),
\item    (csi-36920a8f429f37ae80839d40e3cd805ef096f1aafb-b2dd6d7b845531, 0.5),
\item    (csi-042b74efcc92c96a15db35030eea542a97329bf422-cced8e7fc07453, 0.5),
\item    (csi-224dc655e2c823145f6a545fa78601d91e647dcfef-fb317d8f57b340, 0.5),
\item    (csi-a8f7d2a391815f9c401550acbb694bae346f443a43-a098c5e072aae9, 0.5),
\item    (csi-4c4f985d3c24d4e988879b93a20ba27aef4d1b5e43-1cff21ed9c304f, 0.5),
\item    (csi-007a111242d4e9cc17cbb487d338934ef25edfe677-bec379b08b5002, 0.5),
\item    (csi-007a111242d4e9cc17cbb487d338934ef25edfe677-bec379b08b5002, 0.5),
\item    (csi-042b74efcc92c96a15db35030eea542a97329bf422-cced8e7fc07453, 1),
\item    (csi-5704ae7a7f958a112c4ab6707b5b6708c839a7a23a-8927ff774d5a38, 1),
\item    (csi-af40c043b23e299af5c85347534fe62d2c963a8f7d-6b0dac873b8846, 4),
\item    (csi-535aa6a8940b19cbf86cda227104df4ff1f3c76e36-eb17aa89585b91, 4),
\item    (csi-5deb4704be2267c6724151a6417770ceb5ffc47995-607bbf3a169c28, 8),
\item    (csi-a4390ec528e4f6c2616765939087c14c1ca0bd47cf-1ddd3b6d055ba0, 9),
\item    (csi-3f5a5b338144451e2d7e22bcc184940e796a74510e-c874966c88b93d, 12),
\item    (csi-953d891df68f6a2fd85d53da81602b86b0f04a395d-a9aa547b02a4a3, 16),
\end{itemize}
\section{Certification}

% \begin{table}[t]
% \centering
% \begin{tabular}{|c|c|c|c|c|c|c|}
% % \hline
% \cline{1-7}
% & \multicolumn{3}{c|}{never by CSI} & \multicolumn{3}{c|}{never in CoCo} 
% % \multirow{2}{*}{
% % \begin{tabular}{@{}c@{}} CoCo \\ 2023\end{tabular}}
% \\
% \hline
% % \cline{1-7}
% CPU & yes & no & solved & yes & no & solved \\
% \hline
%  1 &  &  &  &  &  &  \\
%  % CeTA-1 & &  & & & &  \\
% 4 & 4 & 2 & 6 & 1 & 2 & 3 \\
%  % CeTA-4 &  &  &  & & &   \\ 
%  1\&4 &  &  &  &  &  &  \\
%  1-CeTA &  &  &  & 0 & 2 & 0 \\ 
%  4-CeTA & 1 & 0 & 1 & 0 & 0 & 0 \\ 
%  1\&4-CeTA &  &  &  & 0 & 2 & 0 \\ 
% \hline
% \end{tabular}
% \caption{Numbers of TRSs solved by all strategies in Grackle's final portfolio that have never been solved by all versions of CSI or any tool in CoCo. The suffix CeTA denotes the proofs can be certified by CeTA. The notion 1\&4 means the union of all strategies invented by employing one CPU and four CPUs per strategy execution. 
% }
% \label{tab:newly-solved}
% \end{table}

Besides carefully designing the parameter space of Grackle, 
we also perform various verification procedures to ensure the soundness of the invented strategies.
% we also verify the results produced by CSI using CeTA.

\subsection{Proof Consistence Checking}
One typical way to verify the correctness of proofs in CoCo is to check whether the proofs of a prover are consistent with other provers.
Here, the consistency means that we do not prove confluence (non-confluence) for a problem for which other provers prove its non-confluence (confluence).
We check whether the proofs found by invented strategies are consistent with all provers in CoCo.
The proofs found by Grackle are depicted in the final portfolio \emph{grackle.flee}.
The check is done by \url{stats/dif_coco_grackle.py}, which compares the difference between results in \emph{grackle.flee}
and the results obtained by CSI in CoCo2024. 
For proofs found in \emph{grackle.flee} but not by CSI in CoCo 2024, we manually check the consistency between them and proofs of all provers in the previous CoCo competitions.
We also confirm that the proofs obtained by the unified strategies are consistent with all provers in all CoCo competitions.
This is done by \url{stats/consistency.py} in our code.

\subsection{Certifying Newly Found Proofs}
We run CeTA for each problem solved by invented strategies but not by CSI in CoCo. If it can be certified by CeTA, we trust the results. Otherwise, we manually look at the error information to see whether it is really an error and try to reproduce the proof and the certification error using the strategy defined in CSI's competition strategy.
We aim to understand what changes they perform to the original strategy lead to the proofs.
From the analysis, we either slightly modify the sub-strategy defined in the competition strategy or directly use some existing sub-strategies to produce the same answers as the invented strategies. 
These modifications that lead to the answers are employed in the corresponding invented strategies, which are small and sound according to our knowledge of term rewriting.
We also check the certification errors output by CeTA to figure out whether they are indeed errors or just caused by limitations of CSI and CeTA.
% We aim to understand what changes they perform to the original strategy lead to the proofs.
% From the analysis, we either slightly modify the sub-strategy defined in the competition strategy or directly use some existing sub-strategies to produce the same answers as the invented strategies. 
% These modifications that lead to the answers are employed in the corresponding invented strategies, which are small and sound according to our knowledge of term rewriting.
The statistics of the certifications are shown in Table 2 of the paper.

When we use four CPUs per CSI execution on ARI-COPS, we prove the following problems that are unprovable by CSI in CoCo. 
We analyze each of them. The format is \textit{(strategy, newly proved problems in ARI-COPS, corresponding problems in COPS)}. The results \texttt{CERTIFIED} means the proof is certified by CeTA.
\begin{itemize}
    \item (csi-e5535657f8e54081f79c2291ebad9d81992f6e7248-49f0fa92a83cc9, YES, 1499.ari, 1652.trs). The output is \texttt{CERTIFIED}.
    \item (csi-9f19c01a538808477a8980f80d3d2face7303ce6f0-58f8b6170f9461, YES, 879.ari, 1024.trs. The output is \texttt{./csi: XML output is not supported for this method}. The reason why CeTA fails to certify is CeTA does not support the certification of Aoto-Toyama criteria.  It can be proved by
    \texttt{(at -bound 16; SN)!}. The strategy only changes the value of the \texttt{-bound} flag for \texttt{AT3}, which has been used in the original competition strategy.
    Using \texttt{(at -bound 16; SN)!} leads to the same certification error as the invented strategy.
    \item (csi-9f6172de97a8148d22ba4b910ba8b16dccd0196c276-c568d9ab2c0b5, NO, 852.ari (997.trs), 846.ari (991.trs))
    CeTA cannot support the verification of \texttt{nonconfluence -idem}. However, the essential for solving such two problems is the usage of \texttt{redundant -development 6}, which is discovered by Grackle. 
    Two problems can be solved if we change \texttt{redundant -narrowfwd -narrowbwd -size 7} in \texttt{REDUNDANT\_FC} to \texttt{redundant -development 6 -size 7}. Moreover, they are certifiable.
    \item csi-beb87b539aa3911f6c65d5e2a97ef40cb45898dfafc-7283f152a217a, YES, 794.ari,  939.trs, UNSUPPORTED.
    CeTA cannot certify AoTo-Toyama criteria. We can use \texttt{AT} defined in the competition strategy to prove it. 
    Using \texttt{AT} leads to the same certification error as the invented strategy.
    % Actually, we find it can also be proved by \texttt{REDUNDANT\_DEL?; KB}, which is also defined in the competition strategy.
    \item csi-ec0fb3f8ce9f2d586b23f5cd9c4a399845157f2854e-f5f8e2f9d3f3a, YES, 167.ari, 170.trs. UNSUPPORTED \texttt{Fatal: parse error on <acRuleRemoval> at [trsTerminationProof, wcrAndSN, crProof, redundantRules, crProof, proof, certificationProblem]}. 
    CeTA does not support AC confluence proving techniques; however, they are used in CSI's original competition strategy.
    From the invented strategy, we learn that it can be proven with two modifications to the original competition strategy.
    First, increase the number of repeated applications from two to five in \texttt{AUTO\_INNER}.
    It leads to \texttt{AUTO\_INNER = (AUTO\_INNER0[30] | CPCS[5]2*)! | ({AUTO\_INNER0[30]}nono | CPCS[5]2*)5*!}.
    Second, only run a subset of techniques in \texttt{AUTO\_INNER0}. It leads to
    \texttt{AUTO\_INNER0 = (REDUNDANT\_DEL?;(CLOSED || DD || SIMPLE || KB || AC || {GROUND}nono))3*!}
    If we only change \texttt{AUTO\_INNER0} and  \texttt{AUTO\_INNER} as mentioned above, CeTA can produce the same certification error as the invented strategy.
    \item (csi-b994d65167954d35cdd2b7a70646a4f8d550ec9e0de-bc650285a0d90, YES, 158.ari, 160.trs). 
    \texttt{Ackbo: no XML output for SCFs}. CSI cannot output a certificate for the \texttt{ackbo} processor with the flag \texttt{-sc}. 
    The soundness of it has been explained in Section~\ref{sec:param}.
    Moreover, CSI cannot produce a certificate for the CPCS transformation.
    % It can be solved by \texttt{CPCS*}, where \texttt{CPCS} is defined in the competition strategy. 
    We can prove it by changing the definition of \texttt{AUTO\_INNER0} to \texttt{AUTO\_INNER0 = (REDUNDANT\_DEL?;(AC))3*!}, which is used in the original \texttt{AUTO\_INNER0}.
    The new definition does not parallelly execute all techniques in \texttt{AUTO\_INNER0} and makes the execution faster.
    Using \texttt{AUTO\_INNER0 = (REDUNDANT\_DEL?;(AC))3*!} leads to the same certification error as the invented strategy.
    It is not solved by CSI in CoCo 2024 but was solved by CSI in the previous CoCo competitions.
    \item csi-b3ab9764cd4f0717a3037a5849e47d41c56af511e84-d4a72659e0829, YES, 1500.ari, 1653.trs. CPCS cannot be certified by CeTA. It can be proven by \texttt{CPCS*}.
    \texttt{CPCS} is defined in the competition strategy and the soundness is guaranteed.
    Using \texttt{CPCS*} leads to the same certification error as the invented strategy.
    It is not solved by CSI in CoCo 2024 but was solved by CSI in the previous CoCo competitions.
\end{itemize}

When we use one CPU per CSI execution on ARI-COPS, we prove the following problems that are unprovable by CSI in CoCo. 

\begin{itemize}
    \item csi-0d382fd14a431f6befe533e561bbaf68c777bf48b92-ccbbea5fd4346, NO, 449.ari, 540.trs, CERTIFIED.
    \item csi-9e016d5ab9730720dcd426c5368545f09f0f5b9b464-9cd6af41284ad, YES, 463.ari, 554.trs. 
    \texttt{the critical pair g(h(f(f(b, b), b))) <- . -> g(h(h(f(f(h(k(k(b, b), b)), h(k(k(b, b), b))), h(k(k(b, b), b)))))) is not (almost) parallel closed within None steps. hence the following TRS is not (almost) parallel closed.}
    CeTA cannot certify the \texttt{cr -okui} technique.
    We can prove it if we change \texttt{AUTO\_INNER0} to \texttt{AUTO\_INNER0 = (REDUNDANT\_DEL?;(CLOSED))3*! || ((CLOSED | REDUNDANT\_RHS)3*! || (CLOSED) | REDUNDANT\_JS)3*!}, which is used in the original definition of \texttt{AUTO\_INNER0}.
    The modified definition causes the same certification error as the invented strategy.
    \item csi-0e7e3ba5c042fabdf8151556dc8b26717d98bcd49c2-3193ec7f29d33, YES, 166.ari, 169.trs. \texttt{./csi: XML output is not supported for this method}. CSI cannot output certificates for Aoto-Toyama criteria.
    It can be proven by \texttt{(at -bound 16 -theorem 2; SN)!} where \texttt{SN} is defined in the original competition strategy. We only change \texttt{-bound 16} to increase the search space as explained in Section~\ref{sec:param}.
    The strategy \texttt{(at -bound 16 -theorem 2; SN)!} causes the same certification error as the invented strategy.
    \item csi-aabfb22bdd5b365b18568e462dd644b3a94146e63d2-c44ff35c98a2c, NO, 846.ari(991.trs), 852.ari(997.trs), CERTIFIED.
    \item (csi-d42ec2e614b9f4287137c1772a4a13176783da5195-2c630b016bc7c4, YES, 158.ari, 160.trs). 
    \texttt{Ackbo: no XML output for SCFs}. CSI cannot output a certificate for the \texttt{ackbo} processor with the flag \texttt{-sc}. 
    The soundness of it has been explained in Section~\ref{sec:param}.
    Moreover, CSI cannot produce a certificate for the CPCS transformation.
    % It can be solved by \texttt{CPCS*}, where \texttt{CPCS} is defined in the competition strategy. 
    We can prove it by changing the definition of \texttt{AUTO\_INNER0} to \texttt{AUTO\_INNER0 = (REDUNDANT\_DEL?;(AC))3*!}, which is used in the original \texttt{AUTO\_INNER0}.
    The new definition does not parallelly execute all techniques in \texttt{AUTO\_INNER0} and makes the execution faster.
    It is not solved by CSI in CoCo 2024 but was solved by CSI in the previous CoCo competitions.
    The strategy \texttt{AUTO\_INNER0 = (REDUNDANT\_DEL?;(AC))3*!} causes the same certification error as the invented strategy.
    \item csi-b3ab9764cd4f0717a3037a5849e47d41c56af511e84-d4a72659e0829, YES, 1500.ari, 1653.trs. CPCS cannot be certified by CeTA. It can be proven by \texttt{CPCS*}.
    \texttt{CPCS} is defined in the competition strategy and the soundness is guaranteed.
    It is not solved by CSI in CoCo 2024 but was solved by CSI in the previous CoCo competition.
    \texttt{CPCS*} causes the same certification error as the invented strategy.
\end{itemize}
% never in CoCo

\subsection{Certifying Strategies on Mastered Problems}
For every invented strategy in the ARI-COPS dataset, we run it on the problems  it matsered and try to certify the proofs.
The problems mastered by each strategy is calculated by Grackle.
Since the outputs of CeTA on such problems are indeed lengthy, we do not present them in the technical appendix. The outputs exist in the attached code.
We refer readers to read such logs for details.
As explained in the main paper, CeTA may fail to certify the proofs due to several reasons. We manually check whether the outputs indeed denote errors. We have not found any unsoundness. The typical reasons why CeTA's rejection information does not indicate unsoundness are shown below.

% We explain the typical reasons causing CeTA to fail in the verification of a proof and the output message.
\begin{itemize}
\item \texttt{Fatal: parse error on <ac> at [statusPrecedenceEntry, statusPrecedence, pathOrder, redPair, orderingConstraintProof, ruleRemoval, trsTerminationProof, wcrAndSN, crProof, proof, certificationProblem].} CeTA cannot verify AC processors.
\item \texttt{Fatal: parse error on <acRuleRemoval> at [trsTerminationProof, wcrAndSN, crProof, redundantRules, crProof, proof, certificationProblem]}.  CeTA does not support AC confluence proving techniques
\item \texttt{Error in checking parallel closedness for the rewrite system ... The critical pair XXX is not (almost) parallel closed within None steps. hence the following TRS is not (almost) parallel closed.}
CSI outputs an empty certificate for the CPCS transformation and Church Rosser Transformation Processor (okui), confusing CeTA there are no proof steps.
\item \texttt{Fatal: parse error on <unknownAssumption> at [proof, certificationProblem].}
CSI uses a theorem that is not supported by CeTA.
% e.g., ToyamaOyamaguchi95Cor22.
\item \texttt{./csi: order-sorted decomposition: xml proof not supported}.
CeTA does not support order-sorted decomposition.
\item \texttt{Fatal: parse error on text element "-1" at [stronglyClosed, crProof, redundantRules, crProof, proof, certificationProblem]}.
CSI implements some development closedness techniques that cannot be verified by CeTA.
\item \texttt{Fatal: parse error on <uncurry> at [proof, certificationProblem]}.
Uncurry is not supported by CeTA.
\item \texttt{Could not infer that X and Y are not joinable, could not ensure closure under rewriting for first automaton, problem when ensuring (state-)compatibility of TRS with TA}.
The processor \texttt{nonconfluence -tree} cannot be verified. Need to use \texttt{nonconfluence -tree -cert}.
\item \texttt{Fatal: parse error on <magic> at [nonJoinableFork, crDisproof, proof, certificationProblem]}.
The technique \texttt{nonconfluence -idem} is not supported by CeTA.
\item \texttt{Error when closing critical pairs of rules, C not a subsystem of R, hence the following TRS is not critical pair closing
rewrite system.}
The technique \texttt{cr -cpcs2} is not certifiable. It should be changed to \texttt{cr -cpcs2 -cpcscert} for certification. But \texttt{cr -cpcs2} is used in the original competition strategy.
\item \texttt{Error below strong normalization + wcr; R is not empty in the following termination-problem.}
The usage of AC processors makes CSI output an empty proof; thus, confusing CeTA.
\item \texttt{./csi: MatrixInterpretation.fprintfx: XML output not supported expecting "<", but found: ''}
CPCS is not supported, and sometimes outputs entire empty certificates.

\item \texttt{./csi: XML output is not supported for this method}. CSI does not implement the functions to output the certificates for some processors.

\item \texttt{parser error : Excessive depth in document}. The certificate is too large for CeTA to parse.
\item \texttt{Ackbo: no XML output for SCFs} CSI cannot output a certificate for \texttt{ackbo -sc}. However it is used in the original competition strategy, and its soundness has been explained in Section~\ref{sec:param}.
\item \texttt{./csi: not an integer}. CSI cannot generate a certificate if \texttt{kbo} uses rational weights. The soundness of rational weights is explained in Section~\ref{sec:param}
\item \texttt{could not apply the reduction pair processor with the following polynomial interpretation over polynomial interpretation}. CeTA cannot certify the flag \texttt{-heuristic 1} for the \texttt{poly} processor. But \texttt{poly -heuristic 1} is used in the original competition strategy, and its soundness has been explained in Section~\ref{sec:param}. From the proofs, we know we can use \texttt{DD} to prove the problems, which are defined in the competition strategy. We can also reproduce the certification error if we only use \texttt{DD} and CeTA.
\end{itemize}

% \section{Additional Statistics of the Novelty}
% The fundamental difficulty of discovering the proofs that are previously unprovable in CoCo 
% is the significant complexity of CSI's competition strategy. 
% For instance, the development redundant rule technique was designed in ~\cite{nagele2015improving}. 
% Their evaluation shows no improvement over other redundant rule techniques in COPS at the time. 
% Thus, CSI's developers decided not to use it in the competition strategy. 
% As COPS grows, it becomes helpful in some new problems such as \texttt{846.ari} and \texttt{852.ari}. 
% However, the default strategy has only slightly changed over the past years, and the development redundant rule technique has never been tried. 
% One reason is that appropriately choosing sound parameters is challenging even for rewriting experts. 
% Meanwhile, competition strategy is highly complicated and has a prohibitively large configuration space both in the number of parameters and structures of the strategy itself.
% We leverage Grackle to do the tedious strategy search. It can automatically optimize the strategies better than experts as the dataset grows.
% Other rewriting tools do not discover the proof perhaps because they do not implement the essential techniques for solving the problems.

% \bibliographystyle{named}

%% file: ijcai-arxiv.bbl
\begin{thebibliography}{}

\bibitem[\protect\citeauthoryear{Abate \bgroup \em et al.\egroup }{2021}]{abate2021learning}
Alessandro Abate, Mirco Giacobbe, and Diptarko Roy.
\newblock Learning probabilistic termination proofs.
\newblock In {\em Computer Aided Verification: 33rd International Conference, CAV 2021, Virtual Event, July 20--23, 2021, Proceedings, Part II 33}, pages 3--26. Springer, 2021.

\bibitem[\protect\citeauthoryear{Aleksandrova \bgroup \em et al.\egroup }{2024}]{aleksandrova2024prover9}
Kristina Aleksandrova, Jan Jakubuv, and Cezary Kaliszyk.
\newblock Prover9 unleashed: Automated configuration for enhanced proof discovery.
\newblock In {\em Proceedings of 25th Conference on Logic for Pro}, volume 100, pages 360--369, 2024.

\bibitem[\protect\citeauthoryear{Aoto and Toyama}{2012}]{aoto2012reduction}
Takahito Aoto and Yoshihito Toyama.
\newblock A reduction-preserving completion for proving confluence of non-terminating term rewriting systems.
\newblock {\em Logical Methods in Computer Science}, 8, 2012.

\bibitem[\protect\citeauthoryear{Aoto \bgroup \em et al.\egroup }{2014}]{aoto2014proving}
Takahito Aoto, Yoshihito Toyama, and Kazumasa Uchida.
\newblock Proving confluence of term rewriting systems via persistency and decreasing diagrams.
\newblock In {\em Rewriting and Typed Lambda Calculi: Joint International Conference, RTA-TLCA 2014, Held as Part of the Vienna Summer of Logic, VSL 2014, Vienna, Austria, July 14-17, 2014. Proceedings 25}, pages 46--60. Springer, 2014.

\bibitem[\protect\citeauthoryear{Baader and Nipkow}{1998}]{baader1998term}
Franz Baader and Tobias Nipkow.
\newblock {\em Term rewriting and all that}.
\newblock Cambridge university press, 1998.

\bibitem[\protect\citeauthoryear{Bachmair and Ganzinger}{1994}]{bachmair1994rewrite}
Leo Bachmair and Harald Ganzinger.
\newblock Rewrite-based equational theorem proving with selection and simplification.
\newblock {\em Journal of Logic and Computation}, 4(3):217--247, 1994.

\bibitem[\protect\citeauthoryear{Bezem \bgroup \em et al.\egroup }{2003}]{bezem2003term}
Marc Bezem, Jan~Willem Klop, and Roel de~Vrijer.
\newblock {\em Term rewriting systems}.
\newblock Cambridge University Press, 2003.

\bibitem[\protect\citeauthoryear{Brown and Kaliszyk}{2022}]{brown2022lash}
Chad~E Brown and Cezary Kaliszyk.
\newblock Lash 1.0 (system description).
\newblock In {\em International Joint Conference on Automated Reasoning}, pages 350--358. Springer, 2022.

\bibitem[\protect\citeauthoryear{De~Moura and Bj{\o}rner}{2008}]{de2008z3}
Leonardo De~Moura and Nikolaj Bj{\o}rner.
\newblock Z3: An efficient smt solver.
\newblock In {\em International conference on Tools and Algorithms for the Construction and Analysis of Systems}, pages 337--340. Springer, 2008.

\bibitem[\protect\citeauthoryear{Endrullis \bgroup \em et al.\egroup }{2008}]{endrullis2008matrix}
J{\"o}rg Endrullis, Johannes Waldmann, and Hans Zantema.
\newblock Matrix interpretations for proving termination of term rewriting.
\newblock {\em Journal of Automated Reasoning}, 40:195--220, 2008.

\bibitem[\protect\citeauthoryear{Felgenhauer \bgroup \em et al.\egroup }{2015}]{felgenhauer2015layer}
Bertram Felgenhauer, Aart Middeldorp, Harald Zankl, and Vincent Van~Oostrom.
\newblock Layer systems for proving confluence.
\newblock {\em ACM Transactions on Computational Logic (TOCL)}, 16(2):1--32, 2015.

\bibitem[\protect\citeauthoryear{Felgenhauer}{2012}]{felgenhauer2012deciding}
Bertram Felgenhauer.
\newblock Deciding confluence of ground term rewrite systems in cubic time.
\newblock In {\em 23rd International Conference on Rewriting Techniques and Applications (RTA'12)(2012)}. Schloss-Dagstuhl-Leibniz Zentrum f{\"u}r Informatik, 2012.

\bibitem[\protect\citeauthoryear{Freund and Schapire}{1997}]{freund1997decision}
Yoav Freund and Robert~E Schapire.
\newblock A decision-theoretic generalization of on-line learning and an application to boosting.
\newblock {\em Journal of computer and system sciences}, 55(1):119--139, 1997.

\bibitem[\protect\citeauthoryear{Gebhardt \bgroup \em et al.\egroup }{2007}]{gebhardt2007matrix}
Andreas Gebhardt, Dieter Hofbauer, and Johannes Waldmann.
\newblock Matrix evolutions.
\newblock In {\em Proc. Workshop on Termination, Paris}, 2007.

\bibitem[\protect\citeauthoryear{Giacobbe \bgroup \em et al.\egroup }{2022}]{giacobbe2022neural}
Mirco Giacobbe, Daniel Kroening, and Julian Parsert.
\newblock Neural termination analysis.
\newblock In {\em Proceedings of the 30th ACM Joint European Software Engineering Conference and Symposium on the Foundations of Software Engineering}, pages 633--645, 2022.

\bibitem[\protect\citeauthoryear{Giesl \bgroup \em et al.\egroup }{2005a}]{giesl2005dependency}
J{\"u}rgen Giesl, Ren{\'e} Thiemann, and Peter Schneider-Kamp.
\newblock The dependency pair framework: Combining techniques for automated termination proofs.
\newblock In {\em International Conference on Logic for Programming Artificial Intelligence and Reasoning}, pages 301--331. Springer, 2005.

\bibitem[\protect\citeauthoryear{Giesl \bgroup \em et al.\egroup }{2005b}]{giesl2005proving}
J{\"u}rgen Giesl, Ren{\'e} Thiemann, and Peter Schneider-Kamp.
\newblock Proving and disproving termination of higher-order functions.
\newblock In {\em International Workshop on Frontiers of Combining Systems}, pages 216--231. Springer, 2005.

\bibitem[\protect\citeauthoryear{Gramlich}{1996}]{gramlich1996confluence}
Bernhard Gramlich.
\newblock Confluence without termination via parallel critical pairs.
\newblock In {\em Colloquium on Trees in Algebra and Programming}, pages 211--225. Springer, 1996.

\bibitem[\protect\citeauthoryear{He \bgroup \em et al.\egroup }{2023}]{he2023mcts}
Guoliang He, Zak Singh, and Eiko Yoneki.
\newblock Mcts-geb: Monte carlo tree search is a good e-graph builder.
\newblock In {\em Proceedings of the 3rd Workshop on Machine Learning and Systems}, pages 26--33, 2023.

\bibitem[\protect\citeauthoryear{Hirokawa}{2006}]{hirokawa2006automated}
Nao Hirokawa.
\newblock {\em Automated termination analysis for term rewriting}.
\newblock Citeseer, 2006.

\bibitem[\protect\citeauthoryear{Huet}{1980}]{huet1980confluent}
G{\'e}rard Huet.
\newblock Confluent reductions: Abstract properties and applications to term rewriting systems: Abstract properties and applications to term rewriting systems.
\newblock {\em Journal of the ACM (JACM)}, 27(4):797--821, 1980.

\bibitem[\protect\citeauthoryear{H{\r u}la and Jakub{\r u}v}{2022}]{huula2022targeted}
Jan H{\r u}la and Jakub{\r u}v.
\newblock Targeted configuration of an smt solver.
\newblock In {\em International Conference on Intelligent Computer Mathematics}, pages 256--271. Springer, 2022.

\bibitem[\protect\citeauthoryear{Hutter \bgroup \em et al.\egroup }{2009}]{hutter2009paramils}
Frank Hutter, Holger~H Hoos, Kevin Leyton-Brown, and Thomas St{\"u}tzle.
\newblock Paramils: an automatic algorithm configuration framework.
\newblock {\em Journal of artificial intelligence research}, 36:267--306, 2009.

\bibitem[\protect\citeauthoryear{James and Fabian}{2023}]{ari-convert}
Fox James and Mitterwallner Fabian, 2023.

\bibitem[\protect\citeauthoryear{Kerschke \bgroup \em et al.\egroup }{2019}]{kerschke2019automated}
Pascal Kerschke, Holger~H Hoos, Frank Neumann, and Heike Trautmann.
\newblock Automated algorithm selection: Survey and perspectives.
\newblock {\em Evolutionary computation}, 27(1):3--45, 2019.

\bibitem[\protect\citeauthoryear{Klein and Hirokawa}{2012}]{klein2012confluence}
Dominik Klein and Nao Hirokawa.
\newblock Confluence of non-left-linear trss via relative termination.
\newblock In {\em International Conference on Logic for Programming Artificial Intelligence and Reasoning}, pages 258--273. Springer, 2012.

\bibitem[\protect\citeauthoryear{Knuth and Bendix}{1983}]{knuth1983simple}
Donald~E Knuth and Peter~B Bendix.
\newblock Simple word problems in universal algebras.
\newblock {\em Automation of Reasoning: 2: Classical Papers on Computational Logic 1967--1970}, pages 342--376, 1983.

\bibitem[\protect\citeauthoryear{Koprowski and Waldmann}{2008}]{koprowski2008arctic}
Adam Koprowski and Johannes Waldmann.
\newblock Arctic termination... below zero.
\newblock In {\em International Conference on Rewriting Techniques and Applications}, pages 202--216. Springer, 2008.

\bibitem[\protect\citeauthoryear{Korp and Middeldorp}{2009}]{korp2009match}
Martin Korp and Aart Middeldorp.
\newblock Match-bounds revisited.
\newblock {\em Information and Computation}, 207(11):1259--1283, 2009.

\bibitem[\protect\citeauthoryear{Korp \bgroup \em et al.\egroup }{2009}]{korp2009tyrolean}
Martin Korp, Christian Sternagel, Harald Zankl, and Aart Middeldorp.
\newblock Tyrolean termination tool 2.
\newblock In {\em Rewriting Techniques and Applications: 20th International Conference, RTA 2009 Bras{\'\i}lia, Brazil, June 29-July 1, 2009 Proceedings 20}, pages 295--304. Springer, 2009.

\bibitem[\protect\citeauthoryear{Lindauer \bgroup \em et al.\egroup }{2022}]{lindauer2022smac3}
Marius Lindauer, Katharina Eggensperger, Matthias Feurer, Andr{\'e} Biedenkapp, Difan Deng, Carolin Benjamins, Tim Ruhkopf, Ren{\'e} Sass, and Frank Hutter.
\newblock Smac3: A versatile bayesian optimization package for hyperparameter optimization.
\newblock {\em Journal of Machine Learning Research}, 23(54):1--9, 2022.

\bibitem[\protect\citeauthoryear{McCune}{2005}]{prover9-mace4}
W.~McCune.
\newblock Prover9 and {M}ace4.
\newblock \url{http://www.cs.unm.edu/~mccune/prover9/}, 2005.
\newblock Accessed: 2025-05-21.

\bibitem[\protect\citeauthoryear{Meseguer}{2003}]{meseguer2003software}
Jos{\'e} Meseguer.
\newblock Software specification and verification in rewriting logic.
\newblock {\em Nato Science Series Sub Series III Computer and Systems Sciences}, 191:133--194, 2003.

\bibitem[\protect\citeauthoryear{Moser \bgroup \em et al.\egroup }{2008}]{moser2008complexity}
Georg Moser, Andreas Schnabl, and Johannes Waldmann.
\newblock Complexity analysis of term rewriting based on matrix and context dependent interpretations.
\newblock In {\em IARCS Annual Conference on Foundations of Software Technology and Theoretical Computer Science (2008)}. Schloss-Dagstuhl-Leibniz Zentrum f{\"u}r Informatik, 2008.

\bibitem[\protect\citeauthoryear{Nagele and Middeldorp}{2016}]{nagele2016certification}
Julian Nagele and Aart Middeldorp.
\newblock Certification of classical confluence results for left-linear term rewrite systems.
\newblock In {\em International Conference on Interactive Theorem Proving}, pages 290--306. Springer, 2016.

\bibitem[\protect\citeauthoryear{Nagele \bgroup \em et al.\egroup }{2015}]{nagele2015improving}
Julian Nagele, Bertram Felgenhauer, and Aart Middeldorp.
\newblock Improving automatic confluence analysis of rewrite systems by redundant rules.
\newblock In {\em 26th International Conference on Rewriting Techniques and Applications (RTA 2015)}. Schloss Dagstuhl-Leibniz-Zentrum fuer Informatik, 2015.

\bibitem[\protect\citeauthoryear{Nagele \bgroup \em et al.\egroup }{2017}]{nagele2017csi}
Julian Nagele, Bertram Felgenhauer, and Aart Middeldorp.
\newblock Csi: New evidence--a progress report.
\newblock In {\em International Conference on Automated Deduction}, pages 385--397. Springer, 2017.

\bibitem[\protect\citeauthoryear{Neurauter \bgroup \em et al.\egroup }{2010}]{neurauter2010monotonicity}
Friedrich Neurauter, Aart Middeldorp, and Harald Zankl.
\newblock Monotonicity criteria for polynomial interpretations over the naturals.
\newblock In {\em Automated Reasoning: 5th International Joint Conference, IJCAR 2010, Edinburgh, UK, July 16-19, 2010. Proceedings 5}, pages 502--517. Springer, 2010.

\bibitem[\protect\citeauthoryear{Neurauter}{2012}]{neurauter2012termination}
Friedrich Neurauter.
\newblock {\em Termination analysis of term rewriting by polynomial interpretations and matrix interpretations}.
\newblock na, 2012.

\bibitem[\protect\citeauthoryear{Oyamaguchi and Hirokawa}{2014}]{oyamaguchi2014confluence}
Michio Oyamaguchi and Nao Hirokawa.
\newblock Confluence and critical-pair-closing systems.
\newblock {\em Proc. 3rd IWC}, pages 29--33, 2014.

\bibitem[\protect\citeauthoryear{Ramírez \bgroup \em et al.\egroup }{2016}]{7814606}
Nicolás~Gálvez Ramírez, Youssef Hamadi, Eric Monfroy, and Frédéric Saubion.
\newblock Evolving smt strategies.
\newblock In {\em 2016 IEEE 28th International Conference on Tools with Artificial Intelligence}, 2016.

\bibitem[\protect\citeauthoryear{Sakai \bgroup \em et al.\egroup }{2015}]{sakai2015non}
Masahiko Sakai, Michio Oyamaguchi, and Mizuhito Ogawa.
\newblock Non-e-overlapping, weakly shallow, and non-collapsing trss are confluent.
\newblock In {\em Automated Deduction-CADE-25: 25th International Conference on Automated Deduction, Berlin, Germany, August 1-7, 2015, Proceedings 25}, pages 111--126. Springer, 2015.

\bibitem[\protect\citeauthoryear{Sternagel and Thiemann}{2013}]{sternagel2013formalizing}
Christian Sternagel and Ren{\'e} Thiemann.
\newblock Formalizing knuth-bendix orders and knuth-bendix completion.
\newblock In {\em 24th International Conference on Rewriting Techniques and Applications (RTA 2013)}. Schloss-Dagstuhl-Leibniz Zentrum f{\"u}r Informatik, 2013.

\bibitem[\protect\citeauthoryear{Sternagel and Thiemann}{2014a}]{sternagel2014certification}
Christian Sternagel and Ren{\'e} Thiemann.
\newblock The certification problem format.
\newblock {\em arXiv preprint arXiv:1410.8220}, 2014.

\bibitem[\protect\citeauthoryear{Sternagel and Thiemann}{2014b}]{sternagel2014formalizing}
Christian Sternagel and Ren{\'e} Thiemann.
\newblock Formalizing monotone algebras for certification of termination and complexity proofs.
\newblock In {\em International Conference on Rewriting Techniques and Applications}, pages 441--455. Springer, 2014.

\bibitem[\protect\citeauthoryear{Sternagel}{2016}]{sternagel2016generalized}
Christian Sternagel.
\newblock The generalized subterm criterion in ttt2.
\newblock {\em arXiv preprint arXiv:1609.03432}, 2016.

\bibitem[\protect\citeauthoryear{Stump \bgroup \em et al.\egroup }{2014}]{stump2014starexec}
Aaron Stump, Geoff Sutcliffe, and Cesare Tinelli.
\newblock Starexec: A cross-community infrastructure for logic solving.
\newblock In {\em International joint conference on automated reasoning}, pages 367--373. Springer, 2014.

\bibitem[\protect\citeauthoryear{Suzuki \bgroup \em et al.\egroup }{2011}]{suzuki2011argument}
Sho Suzuki, Keiichirou Kusakari, and Fr{\'e}d{\'e}ric Blanqui.
\newblock Argument filterings and usable rules in higher-order rewrite systems.
\newblock {\em IPSJ Online Transactions}, 4:114--125, 2011.

\bibitem[\protect\citeauthoryear{Thiemann and Sternagel}{2009}]{thiemann2009certification}
Ren{\'e} Thiemann and Christian Sternagel.
\newblock Certification of termination proofs using ceta.
\newblock In {\em International Conference on Theorem Proving in Higher Order Logics}, pages 452--468. Springer, 2009.

\bibitem[\protect\citeauthoryear{Toyama and Oyamaguchi}{1994}]{toyama1994church}
Yoshihito Toyama and Michio Oyamaguchi.
\newblock Church-rosser property and unique normal form property of non-duplicating term rewriting systems.
\newblock In {\em International Workshop on Conditional Term Rewriting Systems}, pages 316--331. Springer, 1994.

\bibitem[\protect\citeauthoryear{Van~Oostrom}{1994}]{van1994confluence}
Vincent Van~Oostrom.
\newblock Confluence by decreasing diagrams.
\newblock {\em Theoretical computer science}, 126(2):259--280, 1994.

\bibitem[\protect\citeauthoryear{Van~Oostrom}{1997}]{van1997developing}
Vincent Van~Oostrom.
\newblock Developing developments.
\newblock {\em Theoretical Computer Science}, 175(1):159--181, 1997.

\bibitem[\protect\citeauthoryear{Willsey \bgroup \em et al.\egroup }{2021}]{willsey2021egg}
Max Willsey, Chandrakana Nandi, Yisu~Remy Wang, Oliver Flatt, Zachary Tatlock, and Pavel Panchekha.
\newblock Egg: Fast and extensible equality saturation.
\newblock {\em Proceedings of the ACM on Programming Languages}, 5(POPL):1--29, 2021.

\bibitem[\protect\citeauthoryear{Winkler and Moser}{2019}]{winkler2019smarter}
Sarah Winkler and Georg Moser.
\newblock Smarter features, simpler learning?
\newblock In {\em Proceedings of the Second International Workshop on Automated Reasoning: Challenges, Applications, Directions, Exemplary Achievements}, 2019.

\bibitem[\protect\citeauthoryear{Xu \bgroup \em et al.\egroup }{2010}]{xu2010hydra}
Lin Xu, Holger Hoos, and Kevin Leyton-Brown.
\newblock Hydra: Automatically configuring algorithms for portfolio-based selection.
\newblock In {\em Proceedings of the AAAI Conference on Artificial Intelligence}, volume~24, pages 210--216, 2010.

\bibitem[\protect\citeauthoryear{Yamada \bgroup \em et al.\egroup }{2016}]{yamada2016ac}
Akihisa Yamada, Sarah Winkler, Nao Hirokawa, and Aart Middeldorp.
\newblock Ac-kbo revisited.
\newblock {\em Theory and Practice of Logic Programming}, 16(2):163--188, 2016.

\bibitem[\protect\citeauthoryear{Zankl and Middeldorp}{2010}]{zankl2010satisfiability}
Harald Zankl and Aart Middeldorp.
\newblock Satisfiability of non-linear (ir) rational arithmetic.
\newblock In {\em Logic for Programming, Artificial Intelligence, and Reasoning: 16th International Conference, LPAR-16, Dakar, Senegal, April 25--May 1, 2010, Revised Selected Papers 16}, pages 481--500. Springer, 2010.

\bibitem[\protect\citeauthoryear{Zankl \bgroup \em et al.\egroup }{2009}]{zankl2009kbo}
Harald Zankl, Nao Hirokawa, and Aart Middeldorp.
\newblock Kbo orientability.
\newblock {\em Journal of Automated Reasoning}, 43(2):173--201, 2009.

\bibitem[\protect\citeauthoryear{Zankl \bgroup \em et al.\egroup }{2011}]{zankl2011csi}
Harald Zankl, Bertram Felgenhauer, and Aart Middeldorp.
\newblock Csi--a confluence tool.
\newblock In {\em Automated Deduction--CADE-23: 23rd International Conference on Automated Deduction, Wroc{\l}aw, Poland, July 31-August 5, 2011. Proceedings 23}, pages 499--505. Springer, 2011.

\bibitem[\protect\citeauthoryear{Zantema}{2004}]{zantema2004relative}
Hans Zantema.
\newblock Relative termination in term rewriting.
\newblock In {\em WST’04 7th International Workshop on Termination}, page~51, 2004.

\end{thebibliography}
